\documentclass{elsart}
\newcounter{bla}

\usepackage{graphicx,amsmath}
\newcommand{\newc}{\newcommand}
\newc{\TAUOLA}{\textsf{TAUOLA}}
\newc{\HWPP}{\textsf{Herwig++}}
\newc{\HW}{\textsf{HERWIG}}
\newc{\fortran}{\textsf{FORTRAN}}
\newc{\decayer}{\textsf{Decayer}}
\newc{\ie}{{\it i.e.\/}\ }

\begin{document}

\begin{frontmatter}


\title{Simulation of Tau Decays in the Herwig++ Event Generator}


\author{David Grellscheid}

\address{Institute of Particle Physics Phenomenology, Department of Physics \\ 
        University of Durham, Durham, DH1 3LE, UK.\ead{david.grellscheid@durham.ac.uk}}

\author{Peter Richardson} 

\address{Institute of Particle Physics Phenomenology, Department of Physics \\ 
        University of Durham, Durham, DH1 3LE, UK; and \\
	Theoretical Physics Group, CERN, CH-1211, Geneva 23, Switzerland.\ead{peter.richardson@durham.ac.uk}}

\begin{abstract}
We describe the simulation of tau decays in the \HWPP\ event generator, which 
includes sophisticated modelling of the hadronic currents
and full treatment of spin correlation effects.
The structure of the simulation makes it easy to add new models of tau decay,
change the parameters of the existing models, and use the 
models from tau decay for the decay of other particles.
The results are compared in detail with an existing simulation, and
the benefits of the new structure are illustrated by considering 
the decay \mbox{$\chi^\pm_1\to\chi^0_1+{\rm hadrons}$} in 
Anomaly-Mediated SUSY-Breaking~(AMSB) models.
\end{abstract}

\end{frontmatter}

\section{Introduction}

\vspace{-21cm}
\begin{flushright}
IPPP/07/64\\
DCPT/07/128\\
CERN-PH-TH/2006-183
\end{flushright}
\vspace{19.0cm}

  The use of Monte Carlo event generators is an essential part
  of all experimental analyses, both in interpreting data from existing
  experiments and in the design and planning of future experiments.
  The crucial role that Monte Carlo simulations play in
  experimental studies mean it is imperative these simulations are as
  accurate as possible.

  While the existing Monte Carlo event generators have been highly successful
  over the last twenty years, a new generation
  of programs is necessary for the LHC. The reasons for this are twofold:
  a number of new ideas to improve the accuracy of the simulations
  have been suggested, 
  \emph{e.g.}~\cite{Richardson:2001df,Catani:2001cc,Frixione:2002ik,Gieseke:2003rz,Nason:2004rx}, and
  the existing code structures required major redesign to allow new theoretical
  developments to be incorporated and to make long-term maintenance easier.
  Given the changing nature of computing in high energy
  physics, the natural choice is to write these new programs in C++.
  In preparation for the LHC a major effort is therefore underway
  to produce new versions of established
  simulations~\cite{Bertini:2000uh,Gieseke:2003hm},
  as well as completely new event generators~\cite{Gleisberg:2003xi} in C++.

  As part of the process of writing the new \HWPP\ event
  generator~\cite{Gieseke:2003hm} we wish to improve
  many aspects of the simulation process. One area where major improvements
  are needed is in the simulation of tau lepton decays.

  In \fortran\ \HW\ the $\tau$ decays were treated in the same way as the decays of 
  the hadrons. The weak $V-A$ matrix element was used for the leptonic decays,
  and the other decays were generated using a phase-space distribution for the 
  decay products. In addition, there were two interfaces to the specialised 
  \TAUOLA\cite{Jadach:1993hs,Golonka:2003xt} package for $\tau$ decays. The first
  interface~\cite{Richardson:2001df}, which was part of \HW, was capable of generating
  longitudinal correlations in $\tau$ decays for all taus produced in the 
  perturbative part of the event, while the second external 
  interface~\cite{Golonka:2003xt} was capable of generating the longitudinal spin
  correlations of taus produced in $W$, $Z/\gamma^*$ and $H^\pm$ decays and the
  full correlation effects in neutral Higgs decays.

  In \HWPP\ we wanted to:
\begin{enumerate}
\item include the matrix elements for $\tau$ decay as an integral part of the simulation,
      in order to both have a unified treatment of all decays and to remove
      the inherent problems in interfacing to external 
      packages.\footnote{In the \fortran\ simulation there were 
                         problems with the \HW\ interface
                         due to changes in the \TAUOLA\ program.}
\item include the absent transverse spin correlations for all $\tau$ decays;
\item make it easy to change parameters in the hadronic
  currents used in the tau decays and the modelling of individual decay modes.
\end{enumerate}
  In this paper we will describe the simulation of $\tau$ decays in \HWPP.
  The aim of the simulation is to give a good description of tau decays with
  all the experimentally observed decay modes with branching ratio 
  above the 5~per~mille level included
  together with a reasonable model of the kinematics of the decay. Where possible
  we have used models which have been compared with, or tuned to, experimental
  data. In all cases the parameters of the models can be easily 
  adjusted so that they can be tuned to new experimental data as it becomes
  available.

  First we describe the factorization of the 
  matrix element for $\tau$ decays and the structure of the code. This is
  followed by a description of the different models of the hadronic current which
  are used for the decay modes, together with comparisons with
  previous results. We then discuss our choice of models for the various $\tau$
  decay modes. One major advantage of the new structure is that the hadronic currents
  can be used for applications other than $\tau$ decays which we illustrate
  by considering the weak decay of charginos to neutralinos in 
  Anomaly-Mediated-SUSY Breaking~(AMSB) models, where
  there is a small mass difference between the neutralino and chargino.
  Finally we present our conclusions.

\section{Formalism}  
\label{sect:tau}

  The matrix element for the decay of the $\tau$ lepton can be written as
\begin{equation}
\mathcal{M} = \frac{G_F}{\sqrt{2}}\,L_\mu\,J^\mu,\qquad
L_\mu       = \bar{u}(p_{\nu_\tau})\,\gamma_\mu(1-\gamma_5)\,
        u(p_{\tau}),
\label{eqn:taudecay}
\end{equation} 
  where $p_\tau$ is the momentum of the $\tau$ and $p_{\nu_\tau}$ is the momentum of the
  neutrino produced in the decay. The information on the decay products of the 
  virtual W boson is contained in the hadronic current, $J^\mu$.
  This factorization allows us to implement
  the leptonic current $L_\mu$ for the decaying tau and the hadronic current separately and then 
  combine them to calculate the $\tau$ decay matrix element.

  In  \HWPP\ the most important part of the simulation of hadronic decays
  is handled by the \decayer\ class. This class is responsible for:
\begin{enumerate}
\item generating the kinematics of the decay products;
\item inserting the spin-unaveraged matrix element used to perform the 
      decay and the wavefunctions\footnote{In \HWPP\ these are the spinors and polarization vectors of the fermions and vector mesons.}
      for the decay products into the \HWPP\ structure.  
\end{enumerate}
  All of the $\tau$ decays are handled by the \textsf{TauDecayer} class. This
  class
  inherits from the \textsf{DecayIntegrator} class. The \textsf{DecayIntegrator} 
  includes a sophisticated multi-channel integrator to perform the phase-space
  integration of the decay modes, leaving the implementation
  of the matrix element calculation and definition of the resonance
  structure of a multi-body decay mode the only tasks which need to be 
  performed when implementing a new decay model.

  The hadronic currents for a large range of modes are implemented as
  described in Section~\ref{sect:weakcurrent}. The classes implementing
  these currents all inherit from \textsf{WeakDecayCurrent}. The \textsf{TauDecayer}
  combines the hadronic current with its internal calculation of the
  leptonic current to compute the matrix element, and generates
   the momenta of the decay products using the features of its parent 
  \textsf{DecayIntegrator} class for the generation of the phase
  space.

  Given the spin-unaveraged matrix elements, the algorithm described 
  in~\cite{Knowles:1988vs,Knowles:1988hu,Collins:1987cp,Richardson:2001df}
  is used in \HWPP\ to include spin correlation 
  effects in all stages of the event generation process.
  The \HWPP\ structure, in particular the \textsf{DecayHandler} class, is
  responsible for passing the correct information between the \decayer\ objects
  in order to generate the spin correlations.

\begin{figure}
\includegraphics[width=0.45\textwidth,angle=90]{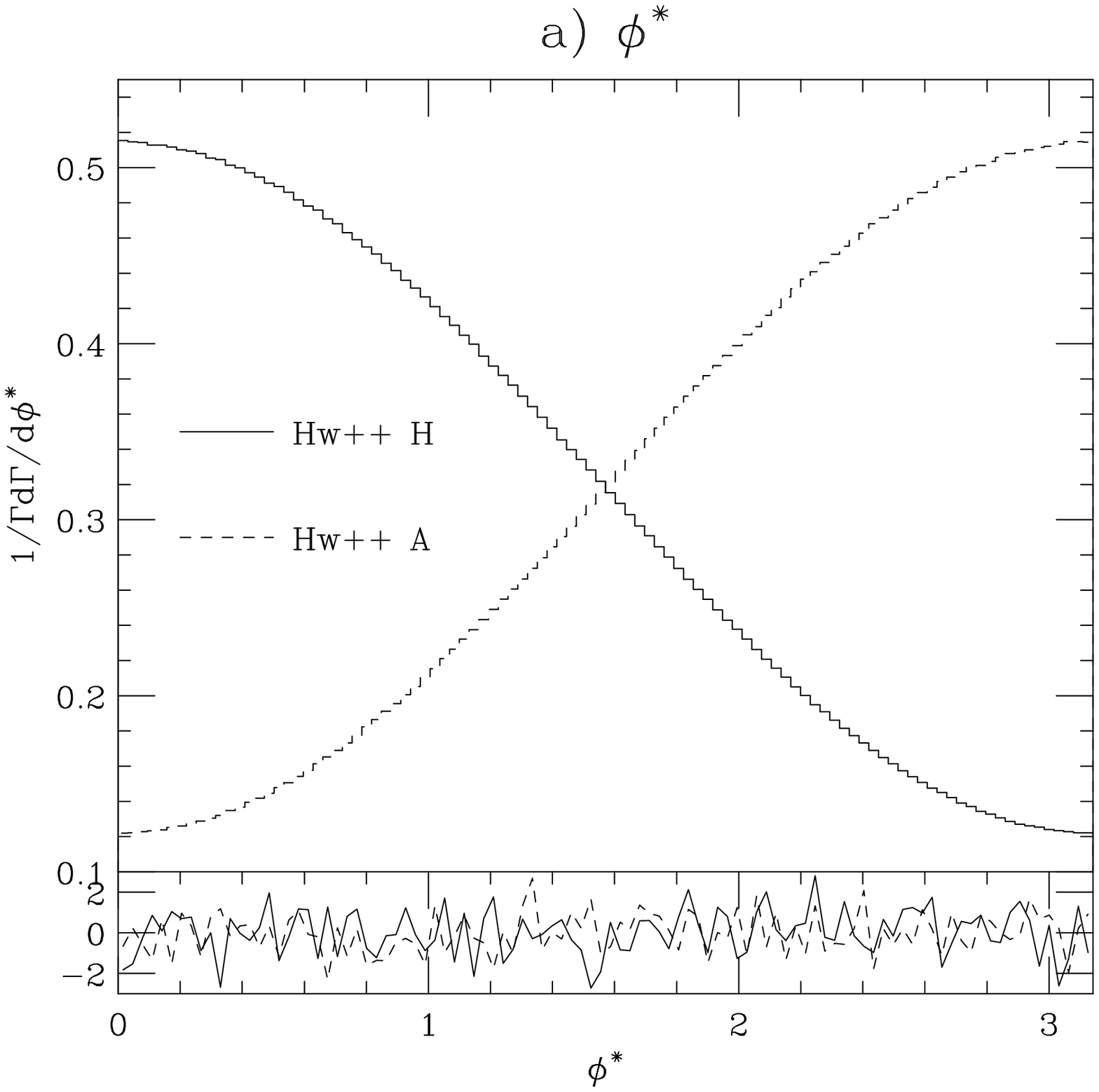}\hfill
\includegraphics[width=0.45\textwidth,angle=90]{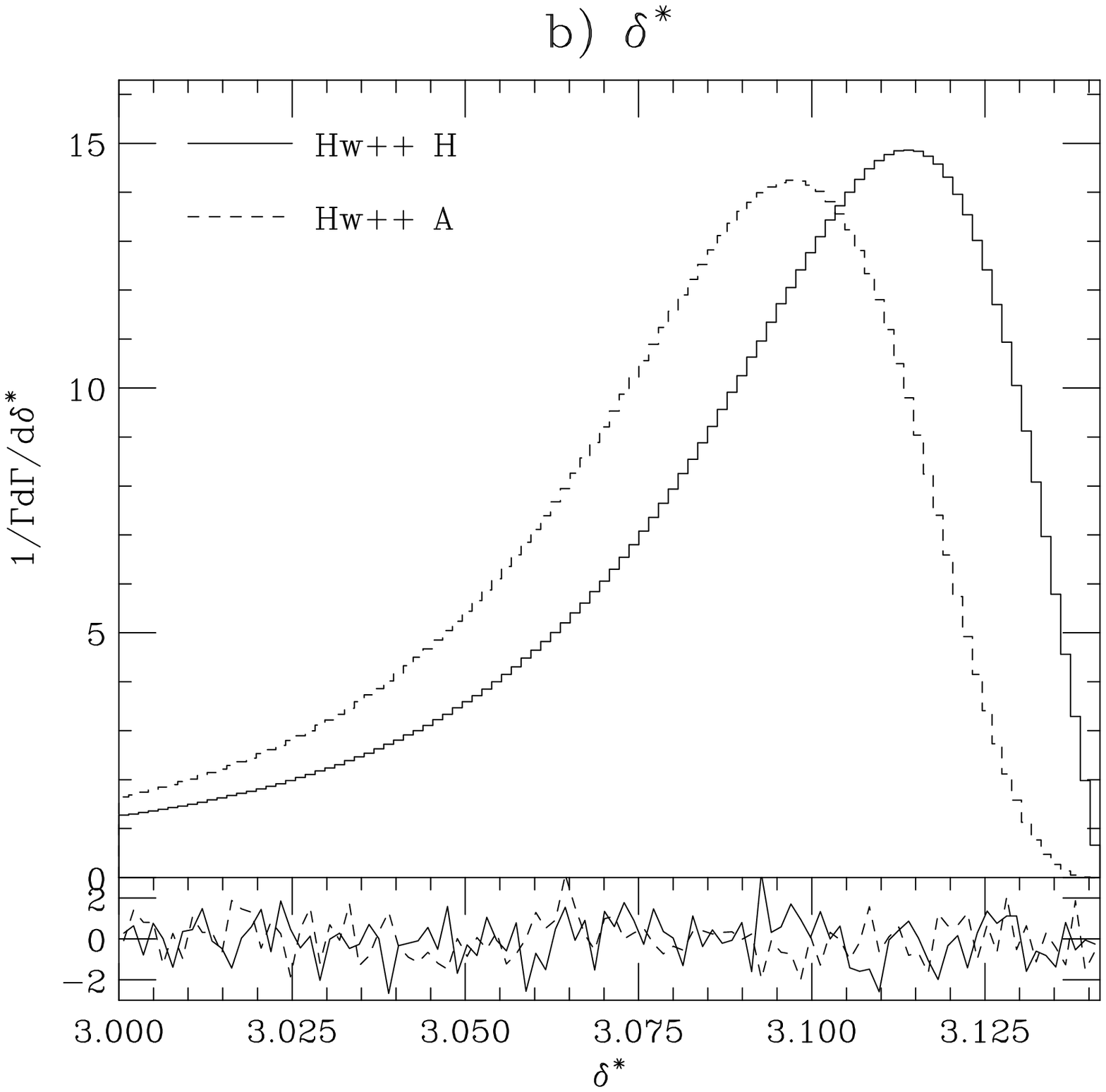}\\
\caption{Correlations between the decay products for $\{H,A\}\to\tau^+\tau^-$
         followed by $\tau\to\pi\nu_\tau$ for a Higgs boson mass of 120\,GeV.
         This plot shows the variables considered in \cite{Was:2002gv}.
         Figure (a) shows the angle between the
         planes of the two tau decays and (b) shows the angle between the pions
         in the rest frame of the Higgs boson.}
\label{fig:picorrelation}
\end{figure}

\begin{figure}
\includegraphics[width=0.45\textwidth,angle=90]{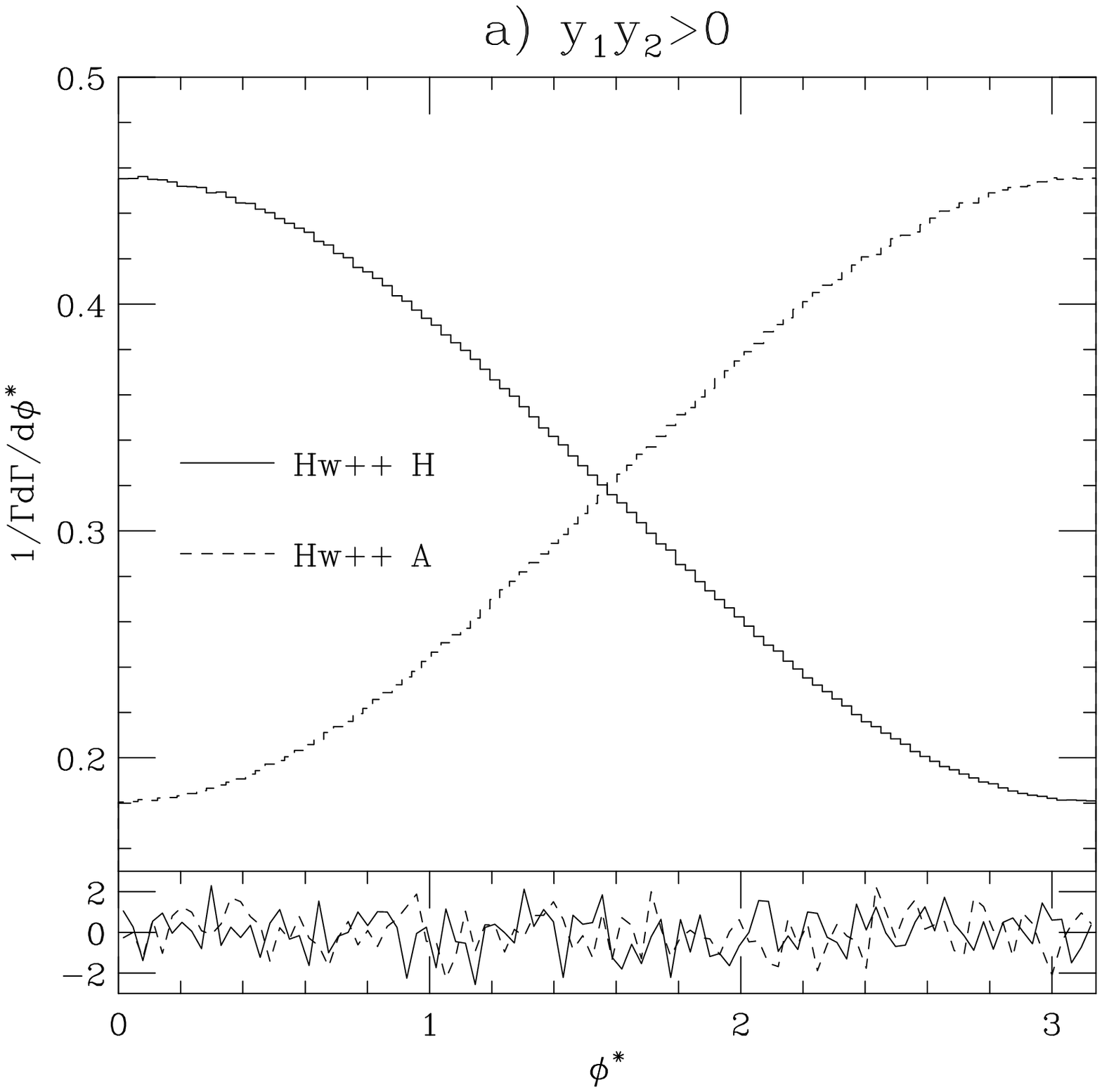}\hfill
\includegraphics[width=0.45\textwidth,angle=90]{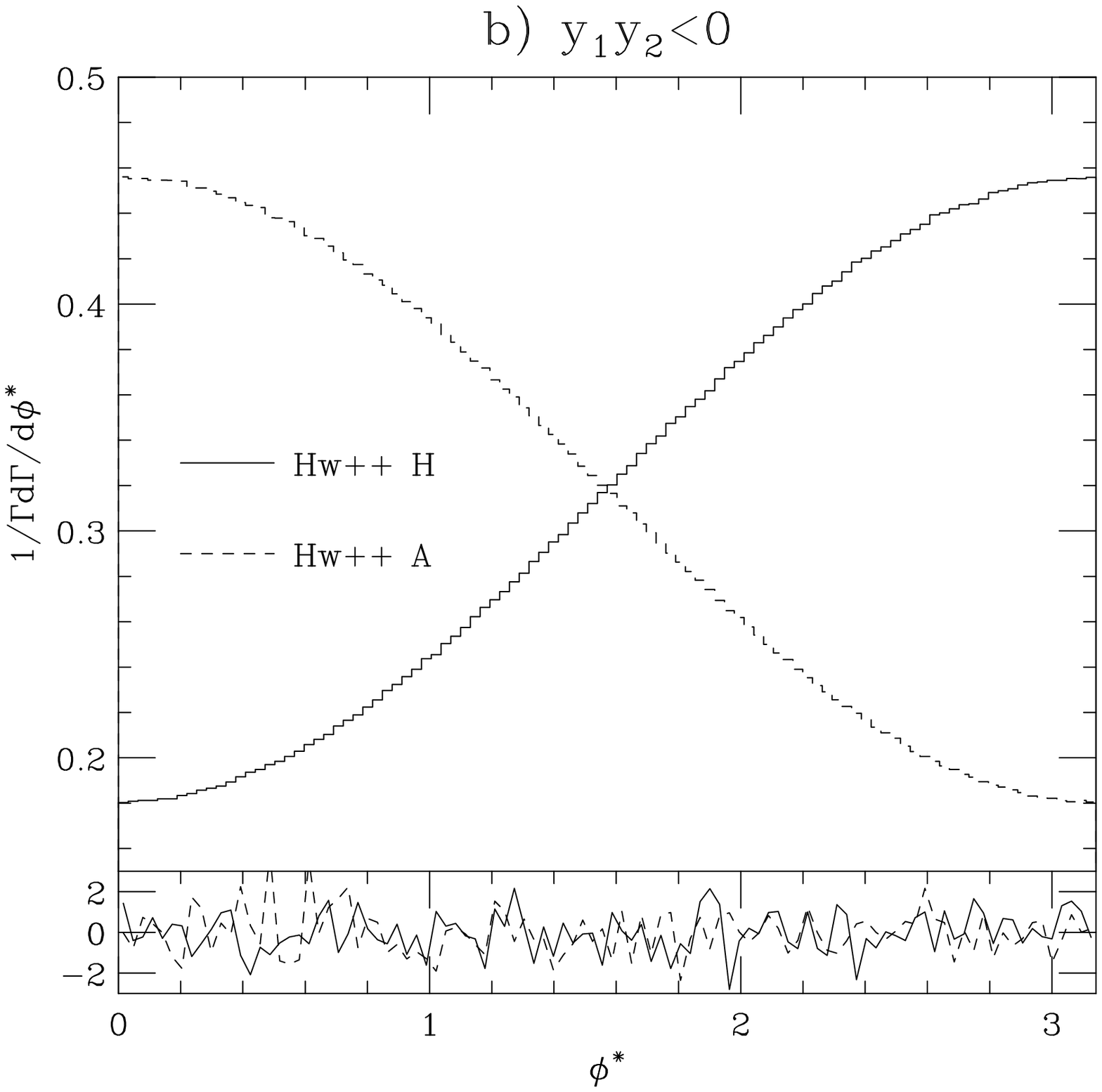}\\
\caption{Correlations between the decay products for $\{H,A\}\to\tau^+\tau^-$
         followed by $\tau\to\rho\nu_\tau$ for a Higgs boson mass of 120\,GeV.
         This plot shows the variables considered in \cite{Desch:2003mw}.
         The  angle between the decay planes of the two $\rho$ mesons 
         in the $\rho^+\rho^-$ rest frame is shown for (a) $y_1y_2>0$ and (b) $y_1y_2<0$.
         The variables $y_{1,2}$ are defined in terms of the energies of the particles in
         the respective $\tau$ rest frames, 
         $y_1=\frac{E_{\pi^+}-E_{\pi^0}}{E_{\pi^+}+E_{\pi^0}}$
         for the $\rho^+$ decay and $y_2=\frac{E_{\pi^-}-E_{\pi^0}}{E_{\pi^-}+E_{\pi^0}}$
         for the $\rho^-$ decay.}
\label{fig:rhocorrelation}
\end{figure}

  Examples of these correlations for the decay of scalar and pseudoscalar Higgs 
  bosons to $\tau^+\tau^-$ followed by the decays of $\tau\to\pi\nu_\tau$ 
  and $\tau\to\rho\nu_\tau$ are shown in Figures~\ref{fig:picorrelation}
  and~\ref{fig:rhocorrelation} 
  respectively.\footnote{In all the plots version 2.1 of \HWPP\ is compared with
  the \TAUOLA-\textsf{PHOTOS} package of October 2005.} In the absence of the correct
  correlations the $\phi^*$ distributions in both cases would be flat.
  The approach of~\cite{Knowles:1988vs,Knowles:1988hu,Collins:1987cp,Richardson:2001df}
  is in good agreement with the results of \TAUOLA\ where the two methods can be
  compared\footnote{The full effects can only be compared in neutral Higgs
	 decay as this is the only case in which \TAUOLA\ implements the full
	 correlations. However, there are  many cases where the transverse correlations
         are not important and for which the two approaches are in good agreement, 
	 see for example Section~4.4 of \cite{Gigg:2007cr}.}
  but can generate the full correlations regardless of how the tau is
  produced.

  In Figures~\ref{fig:picorrelation} and~\ref{fig:rhocorrelation}, and all subsequent,
  figures the \TAUOLA\ result is not shown as it is virtually indistinguishable
  from the \HWPP\ result due to the high statistics used. Similarly where present
  the lower panel shows the difference between the \TAUOLA~\cite{Jadach:1993hs} 
  and \HWPP\ results in terms of the statistical error for mass distributions, and the 
  fractional difference for running widths.

%%%%%%%%%%%%%%%%%%%%%%%%%%%%%%%%%%%%%%%%%%%%%%%%%%%%%%%%%%%%%%%%%%%%%%%%%%%%%%%%
%   Weak Hadronic currents                                                     %
%%%%%%%%%%%%%%%%%%%%%%%%%%%%%%%%%%%%%%%%%%%%%%%%%%%%%%%%%%%%%%%%%%%%%%%%%%%%%%%%
\section{Hadronic Currents}
\label{sect:weakcurrent}
  We have implemented a wide range of models for the weak hadronic currents.
  The various models are described in this section together with comparisons
  with the results of \TAUOLA, where appropriate.

\begin{table}
\begin{center}
\begin{tabular}{|c|c|c|c|}
\hline
Mode & \HWPP & \TAUOLA & Difference \\
     & $\Gamma_{\rm partial}/{\rm 10^{-13}\,GeV}$
     & $\Gamma_{\rm partial}/{\rm 10^{-13}\,GeV}$ & $/{\rm 10^{-17}\,GeV}$ \\
\hline
\multicolumn{4}{|c|}{PseudoScalar Meson}\\
\hline
 $\pi^-$ & $2.4334$ & $2.4334$ &   $\phantom{-}0$\\[-1mm]
 $K^-$   & $0.16611$ & $0.16611$ & $\phantom{-}0$\\
\hline
\multicolumn{4}{|c|}{Two PseudoScalar Mesons via Intermediate Vector Mesons}\\
\hline
 $\pi^-\pi^0$     & $5.3998 \pm0.0002 $ & $5.3998 \pm0.0002$  & $\phantom{-}0\pm3$\\[-1mm]
 $K^-      \pi^0$ &  $0.081295\pm0.000002$& $0.081293\pm0.000001$ & $\phantom{-}0.02\pm0.03$\\[-1mm]
 $\bar{K}^0\pi^-$ & $0.156196\pm0.000004$ & $0.156195\pm0.000003$ & $\phantom{-}0.01\pm0.04$\\[-1mm]
 $K^-K^0$         & $0.0024248 \pm0.000001$  & $0.0024251 \pm0.000001$  & $-0.03\pm0.02$\\
\hline
\multicolumn{4}{|c|}{$K\pi$ via Intermediate Scalar and Vector Mesons}\\
\hline
 $K^-      \pi^0$ & $0.081292\pm0.000003$ & $0.081293\pm0.000001$ & $-0.01\pm0.04$\\[-1mm]
 $\bar{K}^0\pi^-$ & $0.156200\pm0.000005$ & $0.156195\pm0.000003$ & $\phantom{-}0.05\pm0.06$\\
\hline
\multicolumn{4}{|c|}{Charged Lepton and Neutrino}\\
\hline
 $e^-\nu_e$     & $4.0491\pm0.0002$ &$4.0492\pm0.0002$ & $-1\pm3$\\[-1mm]
 $\mu^-\nu_\mu$ & $3.9380\pm0.0002$ &$3.9380\pm0.0002$ & $ \phantom{-}0\pm3$\\
\hline
\end{tabular}
\end{center}
\caption{Partial widths for two- and three-body decay modes of the $\tau$ calculated with
         \textsf{Herwig++} and \textsf{TAUOLA}.
        In order to compare the results of \HWPP\ with \TAUOLA, the masses of the
        decay products in \TAUOLA\ were adjusted to the \HWPP\ values, as were
        the pion and kaon decay constants and the Cabibbo angle. For the 
	two meson, via intermediate vector mesons, decays the parameters of 
	the resonances in 
        the form-factors in \HWPP\ were set to those used in \TAUOLA, the \textsf{CPC}
        version was used for the $\pi^-\pi^0$ and $K^-K^0$ modes and the
        \textsf{CLEO} version for the $K\pi$ modes.
        For the $K\pi$ modes including scalar resonances
	a transverse form of the projection operator
        was used together with the resonance parameters from the \textsf{CLEO}
        version of \TAUOLA\ in \HWPP. In the leptonic decays the radiative corrections in
        \TAUOLA\ were switched off. For the $K\pi$ and $K^-K^0$ modes our choice of 
        the normalisation was used.}
\label{tab:tautwothree}
\end{table}

%
%  Scalar Current
%
\subsection{Pseudoscalar Meson}
\label{sect:scalarmesoncurrent}
  The simplest hadronic current is that for the production of a pseudoscalar
  meson, {\it e.g.}~the current for the production of $\pi^\pm$ in the decay of the 
  tau.
  The hadronic current can be written as
\begin{equation}
J^\mu = f^{}_P\, p^\mu_P,
\end{equation}
  where $p^\mu_P$ is the momentum
  of the pseudoscalar meson and $f_P$ is the pseudoscalar meson decay
  constant.

  The partial widths for the decays $\tau^-\to\pi^-\nu_\tau$ and 
  $\tau^-\to K^-\nu_\tau$ are given in Table\,\ref{tab:tautwothree} and 
  agree with those from \TAUOLA. Another important check 
  is that the spin correlation effects are
  correctly implemented. The correlations in Higgs decays 
  followed by the decay of the tau to a single pion are shown in
  Fig.\,\ref{fig:picorrelation} and are in excellent agreement with the results
  of \TAUOLA.

%
%  Vector Current
%
\subsection{Vector Meson}
\label{sect:vectormesoncurrent}

  The current for the production of a vector meson is given by
\begin{equation}
J^\mu = \sqrt{2}g_{V}\epsilon^{*\mu}_V,
\end{equation}
  where $\epsilon^{*\mu}_V$ is the polarization vector for the outgoing meson and
        $g_V$ is the decay constant of the vector meson. 

 This current was included to test some aspects of the spin correlations, test the 
 treatment of off-shell effects and for future extensions, but is not used in
 tau decays because all such decays are better modeled by a complete description of the
 the production and decay of the vector meson as implemented in the various models described
 below.

\begin{figure}
\includegraphics[width=0.49\textwidth,angle=90]{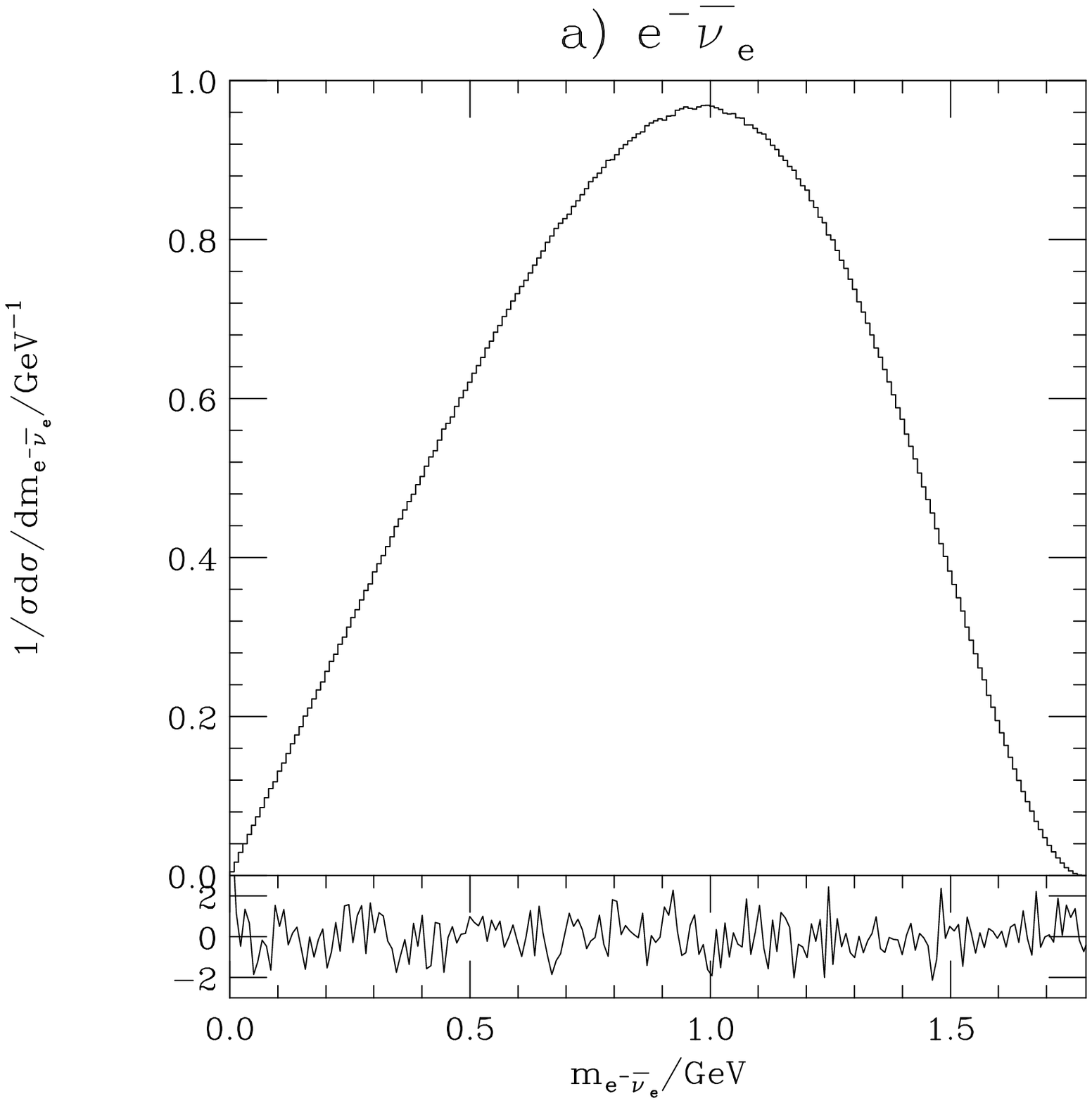}\hfill
\includegraphics[width=0.49\textwidth,angle=90]{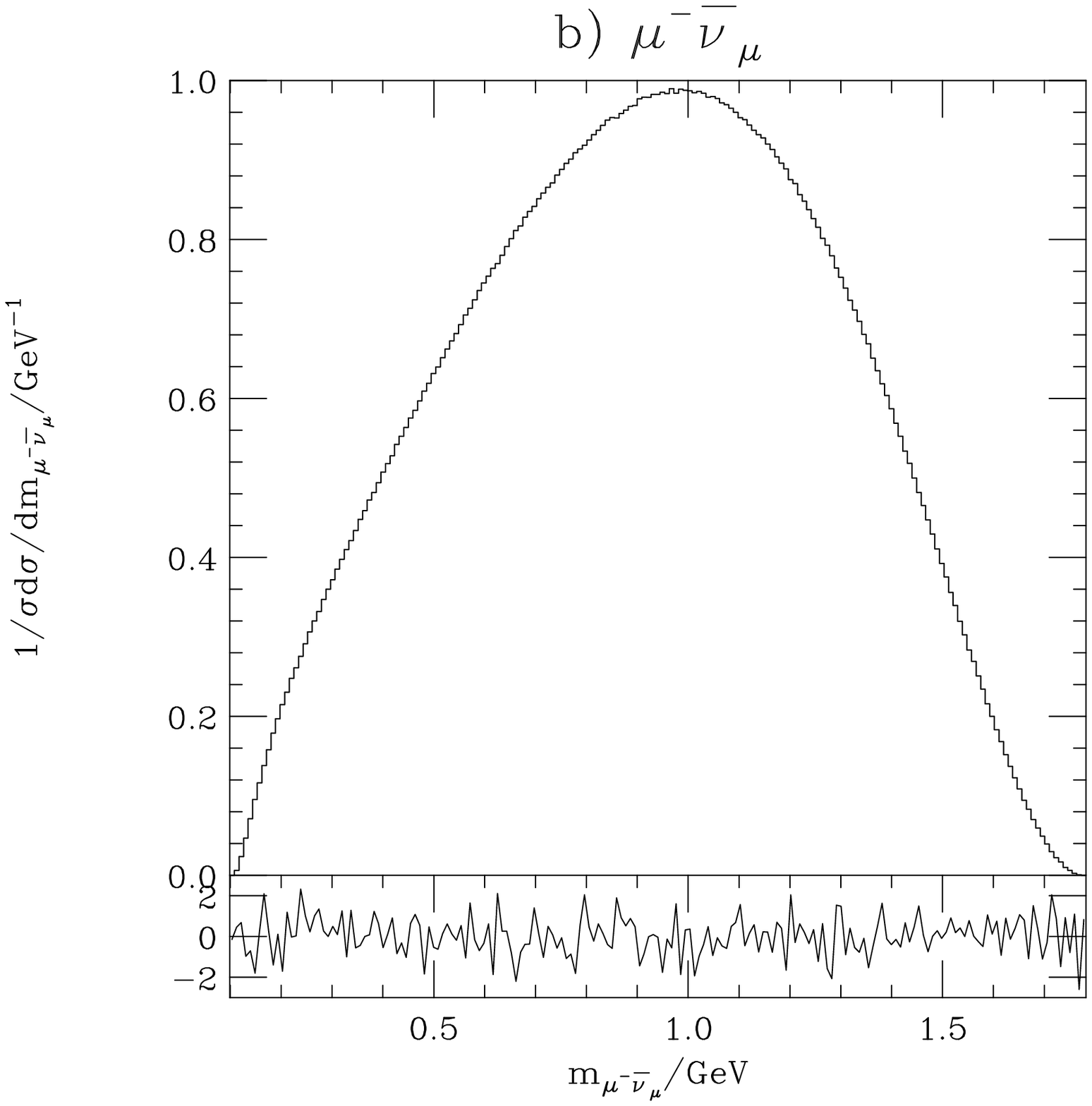}\\
\caption{The differential decay rate with respect to the mass of the lepton-neutrino
        pair for $\tau^-\to\ell^-\bar{\nu}_\ell\nu_\tau$.
        The \TAUOLA\ result was generated with the radiative corrections
        switched off.}
\label{fig:lnu}
\end{figure}
%
%  Lepton Neutrino Current
%
\subsection{Charged Lepton and Neutrino}
\label{sect:leptonneutrinocurrent}
 The current for weak decay to a lepton and the associated anti-neutrino is 
 given by 
\begin{equation}
J^\mu = \bar{u}(p_\ell)\gamma^\mu(1-\gamma_5)v(p_{\bar{\nu}}),
\end{equation} 
  where $p_{\bar{\nu}}$  is the momentum of the anti-neutrino and $p_\ell$
  is the momentum of the charged lepton.

  The mass distribution of the $\ell^-\bar{\nu}_\ell$ produced in the decay 
  \mbox{$\tau^-\to\ell^-\bar{\nu}_\ell\nu_\tau$} is shown in Fig.\,\ref{fig:lnu}.
  As can be seen there is good agreement between \textsf{Herwig++} and
  \textsf{TAUOLA} in terms of both
  the differential distributions and the partial widths given in 
  Table~\ref{tab:tautwothree}. For these comparisons the masses of the 
  particles in \TAUOLA\ were set to the \HWPP\ values and the radiative
  corrections switched off. 

  In general we have not yet considered radiative corrections to tau decays
  in \HWPP. However, we do have a treatment of electromagnetic radiation in 
  \mbox{$1\to2$}~decays~\cite{Hamilton:2006xz} based on the YFS 
  formalism~\cite{Yennie:1961ad} which resums the dominant soft radiation to 
  all orders and treats the universal large collinear logarithms.
  This can be applied to $\tau$ decays
  in the approximation that the decay is treated as a series of $1\to2$
  processes, which is reasonable for many of the decay modes.
  The YFS formalism can be systematically improved 
  to incorporate exact, process specific, higher order corrections so
  in the future we can extend this simulation to include the full radiative
  corrections in tau decay.

%
%  Two Meson current
%
\subsection{Two Pseudoscalar Mesons via Intermediate Vector Mesons}
\label{sect:twomeson}

\begin{figure}
\includegraphics[width=0.49\textwidth,angle=90]{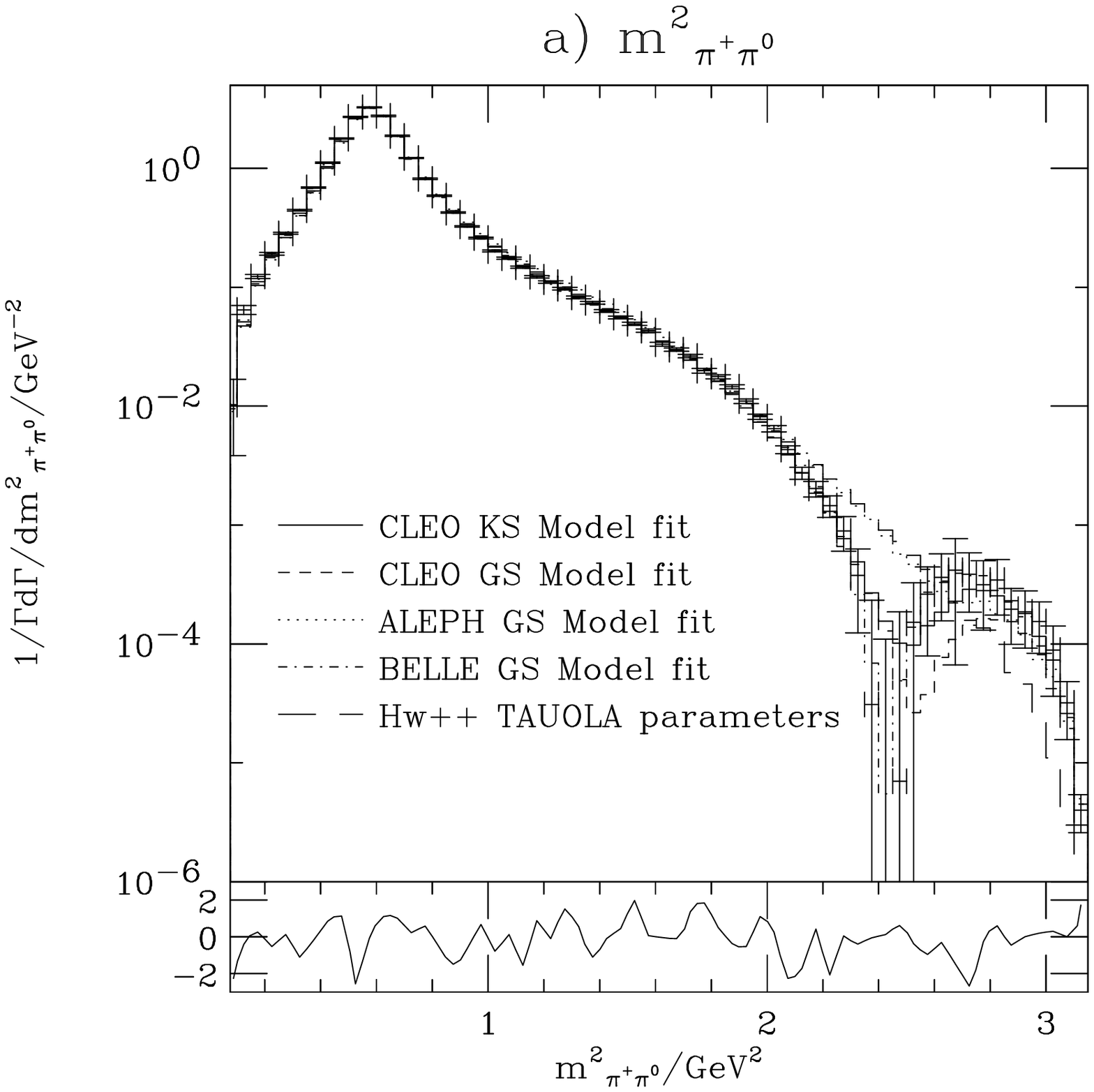}\hfill
\includegraphics[width=0.49\textwidth,angle=90]{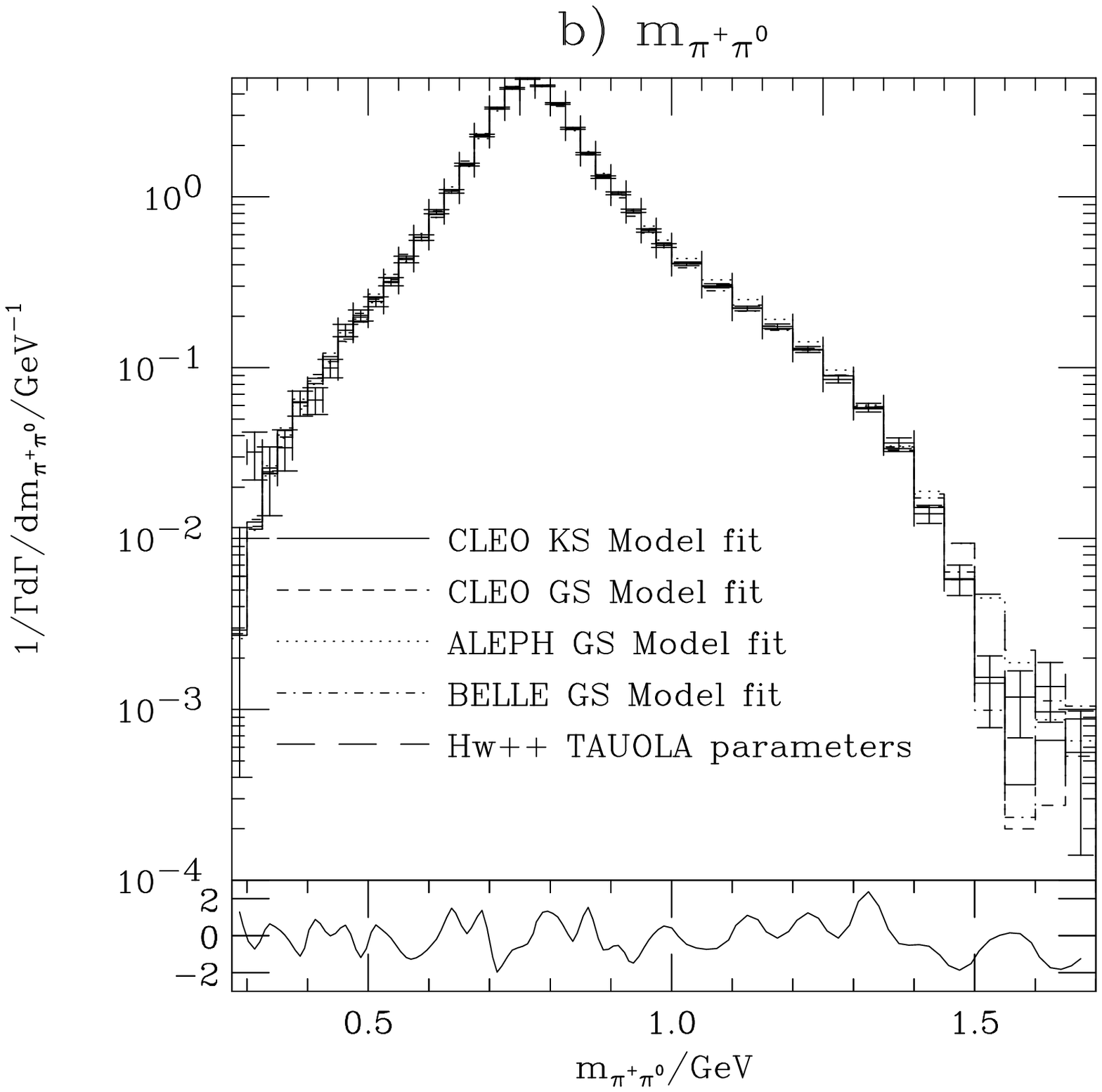}\\
\caption{The mass spectrum for $\pi^-\pi^0$ in the decay $\tau\to\pi^-\pi^0\nu_\tau$.
        a) shows $m^2_{\pi^-\pi^0}$ compared with CLEO data taken 
        from~\cite{Anderson:1999ui} and b) shows $m_{\pi^-\pi^0}$ compared with
        Belle data from \cite{Abe:2005ur}. The results of \HWPP\ using the 
        form-factor parameters from the fits of CLEO~\cite{Anderson:1999ui}, 
        Belle~\cite{Abe:2005ur} and ALEPH~\cite{Schael:2005am} are shown
        together with the result from \HWPP\ using
        the \TAUOLA\ parameters.}
\label{fig:pimpi0}
\end{figure}

  The weak current for the production of two mesons via the $\rho$ or $K^*$
  resonances has the form
\begin{equation}
J^\mu =(p_1-p_2)_\nu\left(g^{\mu\nu}-\frac{q^\mu q^\nu}{q^2}\right)
 \frac{W}{\sum_k\alpha_k}\sum_k \alpha_k B_k(q^2),
\end{equation}
  where $p_{1,2}$ are the momenta of the outgoing mesons, $q=p_1+p_2$,
  $B_k(q^2)$ is the Breit-Wigner distribution for the
  intermediate vector meson $k$ and $\alpha_k$ is the weight for the resonance.
  The Breit-Wigner terms are summed over the $\rho$ or $K^*$ resonances that
  can contribute to a given decay mode.

  The models of either K\"{u}hn and Santamaria~\cite{Kuhn:1990ad}, which uses
  a Breit-Wigner distribution with a p-wave running width Eqn.\,\ref{eqn:runningBW},
  or Gounaris and Sakurai~\cite{Gounaris:1968mw},
  which uses the form given in Eqn.\,\ref{eqn:GSBW}, are supported for the
  shape of the Breit-Wigner distribution.

\begin{figure}
%\vspace{0.5cm}
\includegraphics[width=0.49\textwidth,angle=90]{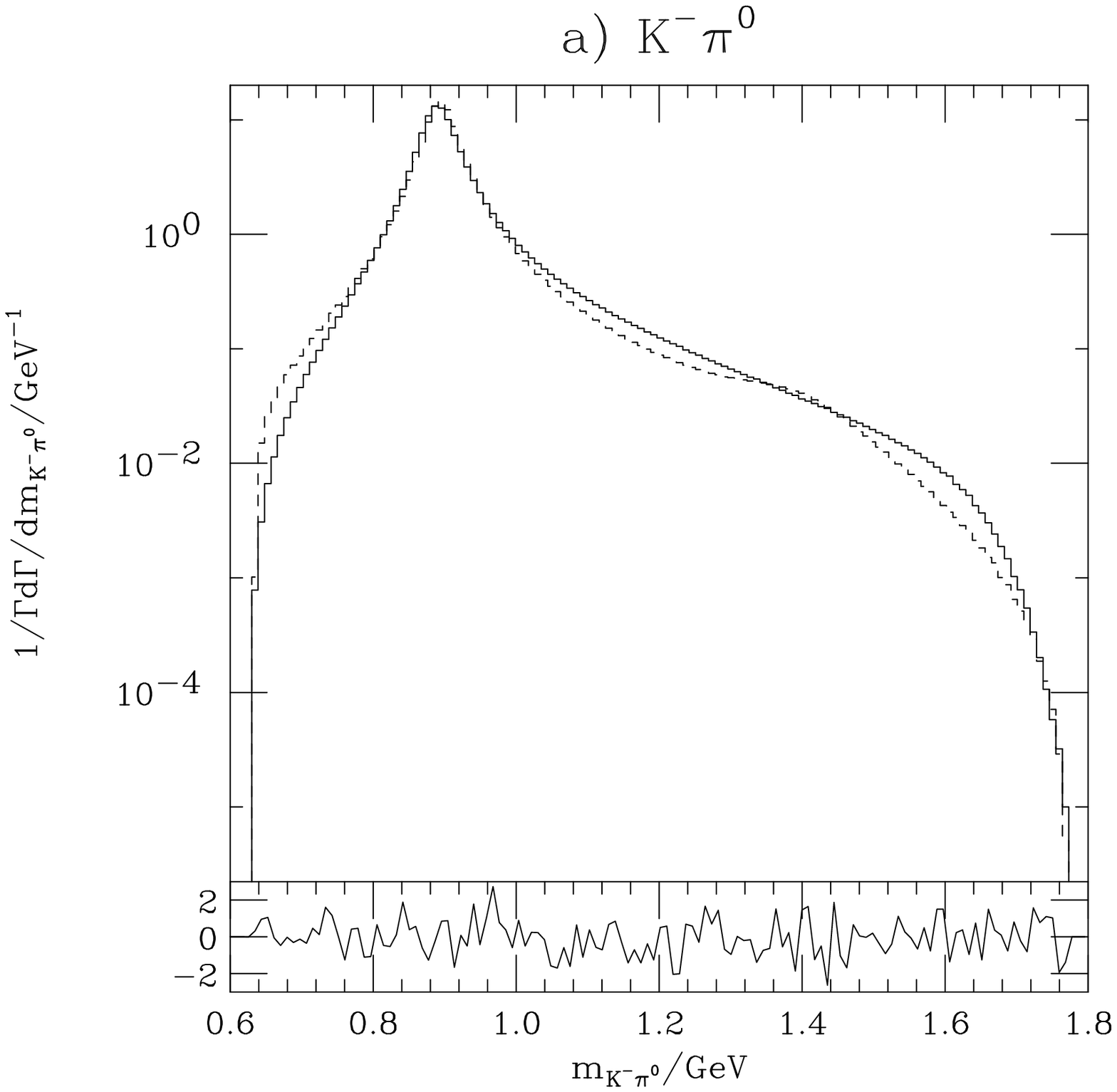}\hfill
\includegraphics[width=0.49\textwidth,angle=90]{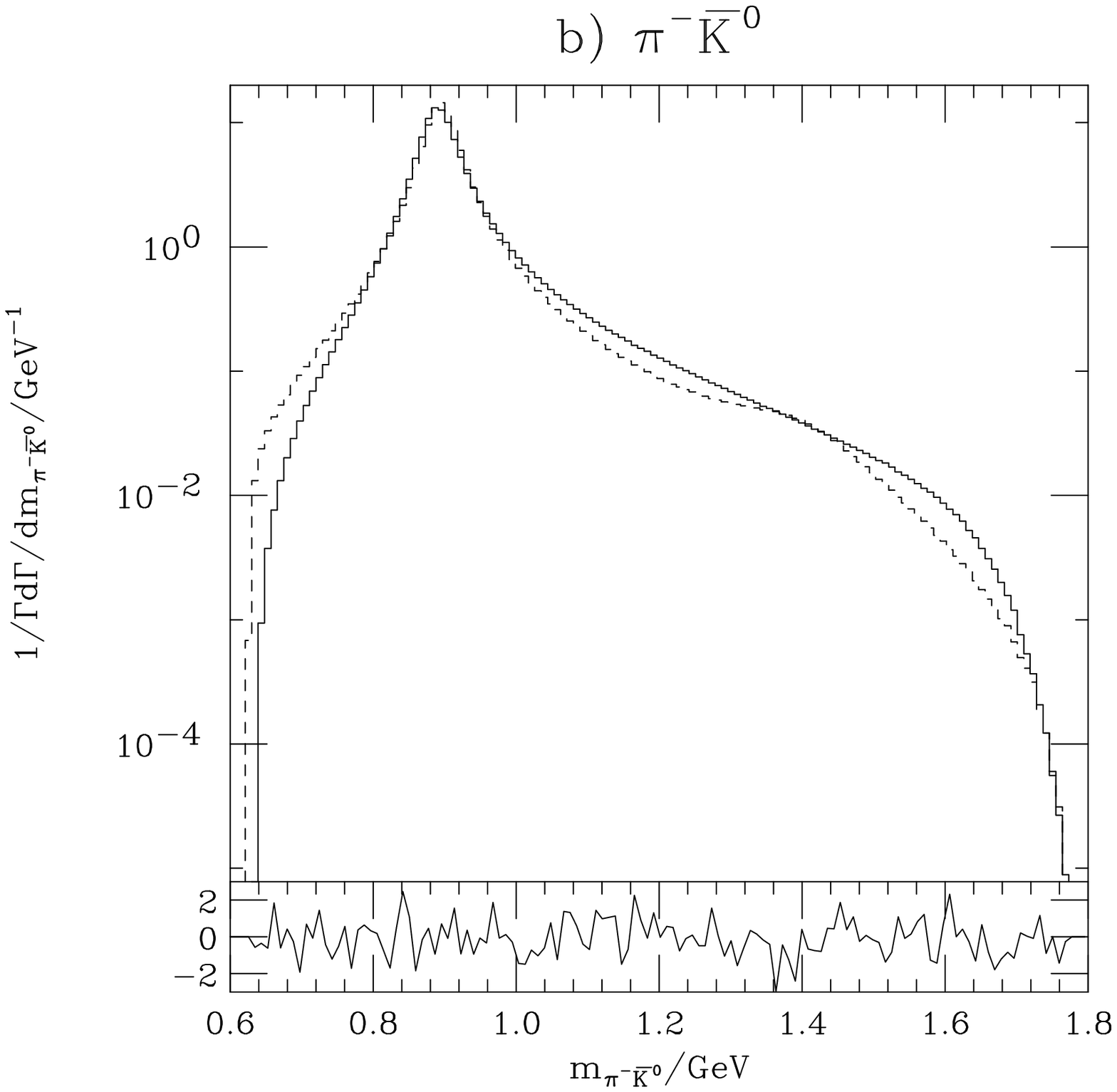}\\
\caption{The mass spectrum of a) $K^-\pi^0$ and b) $\pi^-\bar{K}^0$. In a) the solid
        line is the result of the model described in Section~\protect\ref{sect:twomeson} 
	using the \TAUOLA\
        parameters for the form-factor
	and in b) the solid line is the result of the model described in 
 	Section~\protect\ref{sect:kpicurrent} with no scalar component,
        using with a transverse form of the projection operator 
	and the parameters of the vector form factor set to the \TAUOLA\ values.
	In both plots the dashed line was generated
        using the model from Section~\protect\ref{sect:kpicurrent} using the fit of \cite{Epifanov:2007rf}.}
\label{fig:kpi}
%\vspace{-1cm}
\end{figure}

  The following decay modes: $\rho^-\to\pi^-\pi^0$; $K^{*-}\to K^-\pi^0$;
  $K^{*-}\to \bar{K}^0\pi^-$; $\rho^-\to K^-K^0$; $K^{*-}\to K^-\eta$
  are supported with weights, $W$: $\sqrt{2}$; $\frac1{\sqrt{2}}$; $1$; $1$; 
  $\sqrt{\frac32}$. 
  respectively.\footnote{It should be noted that this normalisation for the $K\pi$
                         and $KK$ modes is different from that in 
                         \textsf{TAUOLA}~\cite{Jadach:1993hs,Golonka:2003xt}.
                         However, it agrees with that of \cite{Finkemeier:1995sr} for the
                         $K\pi$ modes and \cite{Pich:1987qq} for the $K\eta$ mode.}

  There has been a number of recent studies of the $\pi^\pm\pi^0$ 
  mode~\cite{Anderson:1999ui,Schael:2005am,Abe:2005ur}. The fits from these
  studies are compared with the data from CLEO~\cite{Anderson:1999ui} and 
  Belle~\cite{Abe:2005ur} in Fig.\,\ref{fig:pimpi0} together with a comparison
  of the results of \HWPP\ and \TAUOLA. The partial width using the 
  \TAUOLA\ parameters is compared with that from \TAUOLA\ 
  in Table\,\ref{tab:tautwothree}.
  The CLEO, ALEPH and Belle fits differ in the amount of the $\rho(1700)$ resonance which 
  is present, giving destructive interference in the high mass region.
  We have therefore chosen to 
  use the parameters of the CLEO fit using the model of 
  K\"{u}hn and Santamaria~\cite{Kuhn:1990ad}, which lies between the ALEPH
  and Belle fits, as our default choice.

  There have been a number of recent studies of the $K\pi$ modes
  \cite{Barate:1999hj,Lyon:2004sp,Epifanov:2007rf} which favour a low-mass
  enhancement and little contribution of the $K^*(1410)$. We have therefore
  chosen use a model which includes scalar resonances, 
  as described in Section~\ref{sect:kpicurrent}, by default for the $K\pi$ modes. 
  For this model we use same parameters as the 
  \textsf{CLEO} version of \TAUOLA, which includes the
  $K^*(892)$ and $K^*(1680)$ resonances. The mass spectrum for the $K^+\pi^0$ mode 
  shown in Fig.\,\ref{fig:kpi}a shows good agreement
  between \HWPP\ and \TAUOLA, as does the partial width for the $K\pi$ modes given
  in Table~\ref{tab:tautwothree}.

\begin{figure}
\includegraphics[width=0.49\textwidth,angle=90]{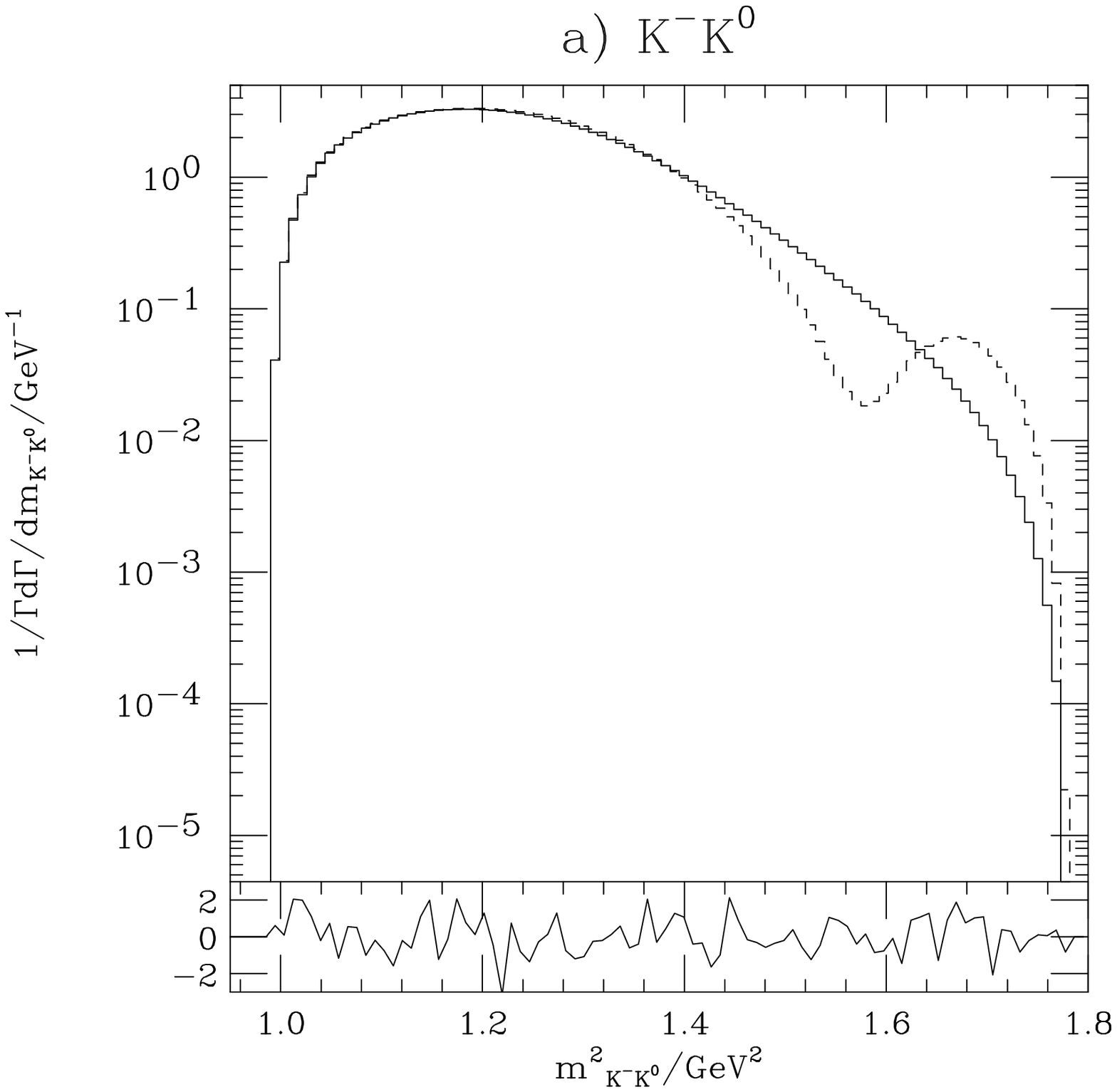}\hfill
\includegraphics[width=0.49\textwidth,angle=90]{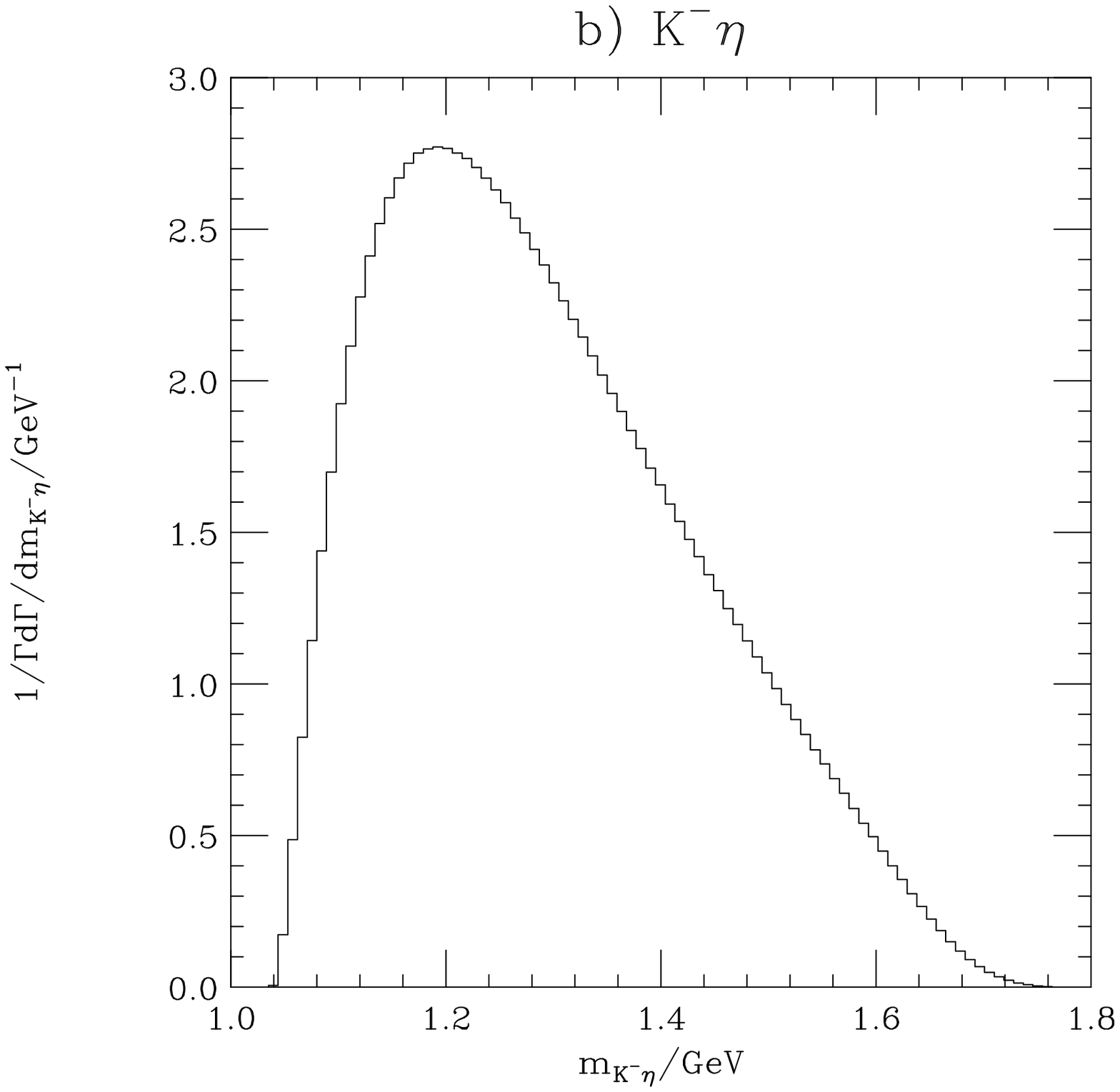}\\
\caption{The mass spectrum for a) $K^-K^0$ and b) $K^-\eta$.
        For the $K^-K^0$ spectrum the solid lines were generated using the 
        parameters from the \textsf{CPC} version of \TAUOLA\ for the form factor
        while the dashed line uses the default \HWPP\ parameters for the rho form
        factor.}
\label{fig:kk}
%\vspace{0.5cm}
\end{figure}

  The mass spectrum for $K^-K^0$ generated by \HWPP\ is compared with that generated
  by \TAUOLA\ in Fig.\,\ref{fig:kk}a, again there is good agreement between \HWPP\ 
  and \TAUOLA\ for both the shape of the distribution and the partial width for the
  decay mode given in Table~\ref{tab:tautwothree}. The spectrum for the $K\eta$ mode 
  is shown in Fig.\,\ref{fig:kk}b.

\subsection[$K\pi$ via Intermediate Scalar and Vector Mesons]
{$K\pi$ via Intermediate Scalar and Vector Mesons}
\label{sect:kpicurrent}
  Unlike the $\pi^+\pi^0$ decay of the tau the $K\pi$ decay mode can occur
  via either intermediate scalar or vector mesons. We therefore include 
  a model for the current for the $K\pi$ decay mode including
  the contribution of both vector and scalar resonances based on the model 
  of~\cite{Finkemeier:1996dh}. The current is given by
\begin{equation}
J^\mu = c_V(p_1-p_2)_\nu\frac1{\sum_k\alpha_k}\sum_k\alpha_k{\rm BW}_k(q^2)
        \left(g^{\nu\mu}-\frac{q^\nu q^\mu}{M^2_k}\right)+c_Sq^\mu\frac1{\sum_k\beta_k}\sum_k\beta_k{\rm BW}_k(q^2),\ \ \ 
\end{equation}
where $p_{1,2}$ are the momenta of the outgoing mesons, $q=p_1+p_2$,
${\rm BW}_k(q^2)$ is the Breit-Wigner distribution for the intermediate mesons 
and $\alpha_k$ is the weight for the resonance. The sum over the resonances
is over the vector $K^*$ states in the first, vector, part of the current and
the excited scalar $K^*$ resonances in the second, scalar, part of the current.
By default the vector part of of the current includes the $K^*(892)$ and $K^*(1410)$
states and the scalar part of the current includes the $K^*_0(1430)$ together with
the option of including the $\kappa(800)$ to model any low-mass enhancement in the 
mass of the $K\pi$ system, although additional resonances can be included if necessary.

  The mass spectrum  for the $\pi^-\bar{K}^0$ mode is
  shown in Fig.\,\ref{fig:kpi}b, where the parameters have been chosen to give the same
  form factor as \TAUOLA. There is good agreement with \TAUOLA\ for both the shape of the distribution and
the partial widths, which are given in Table~\ref{tab:tautwothree}.

\begin{figure}
\begin{center}
\includegraphics[width=0.58\textwidth,angle=90]{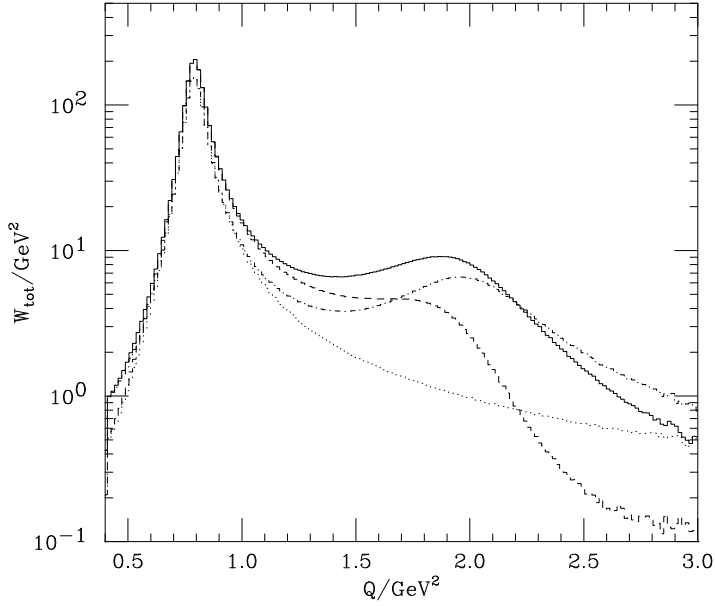}\\
\caption {
        The hadronic structure function $W_{\rm tot}$ for $\tau\to K\pi\nu_\tau$.
        The solid black line shows the result with the default parameters of 
        \cite{Finkemeier:1996dh}, the short 
        dashed line shows the result without the scalar
        contribution, the dot-dashed line shows the result without
        the $K^*(1410)$ and the dotted line shows the result with neither
        the $K^*(1410)$ or scalar contribution.
        This plot is the equivalent of 
        Fig.1a of \cite{Finkemeier:1996dh}.}
\label{fig:kpiQ}
\end{center}
\end{figure}

  The results for the hadronic structure function defined in \cite{Finkemeier:1996dh}
  are shown in Fig.\,\ref{fig:kpiQ} for different values of the parameters. This
  shows good agreement with Fig.1a of \cite{Finkemeier:1996dh}, upon which it is based.
  The branching ratios for the different parameters values are also in good agreement
  with the results of \cite{Finkemeier:1996dh}.

  The most recent study \cite{Epifanov:2007rf} of the $K\pi$ mass spectrum 
  sees a low-mass enhancement which is modelled using the scalar $\kappa$ resonance.
  By default we use a set of parameters this model which reproduces
  the fit of~\cite{Epifanov:2007rf} to the mass spectrum with the $\kappa$, 
  $K^*(892)$ and $K^*(1410)$ resonances, as shown in Fig.\,\ref{fig:kpi}
  for both the $K^-\pi^0$ and $\pi^-\bar{K}^0$ mass distributions. The low-mass
  enhancement due to the $\kappa$ can clearly be seen in these distributions.

\subsection{Three Pseudoscalar Mesons}
\label{sect:threemesoncurrentbase}
  In order to simply the implementation of a number of standard currents for the 
  production of three pseudoscalar mesons we define the current in terms of 
  a number of form factors. The current is defined to be~\cite{Jadach:1993hs}
\begin{eqnarray} 
J^\mu &=& \left(g^{\mu\nu}-\frac{q^\mu q^\nu}{q^2}\right)
   \left[F_1(p_2-p_3)^\mu +F_2(p_3-p_1)^\mu+F_3(p_1-p_2)^\mu\right]\\
 &&  +q^\mu F_4
   +iF_5\epsilon^{\mu\alpha\beta\gamma}p_1^\alpha p_2^\beta p_3^\gamma,
\nonumber
\end{eqnarray}
  where $p_{1,2,3}$ are the momenta of the mesons in the order given below and
  $F_{1\to5}$ are the form factors.
  We use this approach for a number of three meson  modes
  which occur in $\tau$ decays: 
  $    \pi^-  \pi^-    \pi^+ $;
  $    \pi^0  \pi^0    \pi^- $;
  $    K^-   \pi^-    K^+ $;
  $    K^0   \pi^-    \bar{K}^0$;
  $    K^-   \pi^0    K^0 $;
  $    \pi^0  \pi^0    K^- $;
  $    K^-   \pi^-    \pi^+ $;
  $    \pi^-  \bar{K}^0  \pi^0 $;
  $    \pi^-  \pi^0    \eta $;
  $ K^0_S\pi^-K^0_S$;
  $ K^0_L\pi^-K^0_L$;
  $ K^0_S\pi^-K^0_L$.
  The current is implemented in terms of these form factors in a base class 
  so that any model for these currents can be implemented by inheriting
  from this class and specifying the form factors.

  We currently implement three models for these decays, the general model
  of~\cite{Kuhn:1990ad,Decker:1992kj,Jadach:1993hs} which treats all the 
  decay modes, the model of CLEO~\cite{Asner:1999kj} for the three-pion modes
  and the model of \cite{Finkemeier:1995sr} for the kaon modes. 

\subsubsection{General Model}
\label{sect:threemesondefaultcurrent}

\begin{figure}
\begin{center}
\includegraphics[width=0.45\textwidth,angle=90]{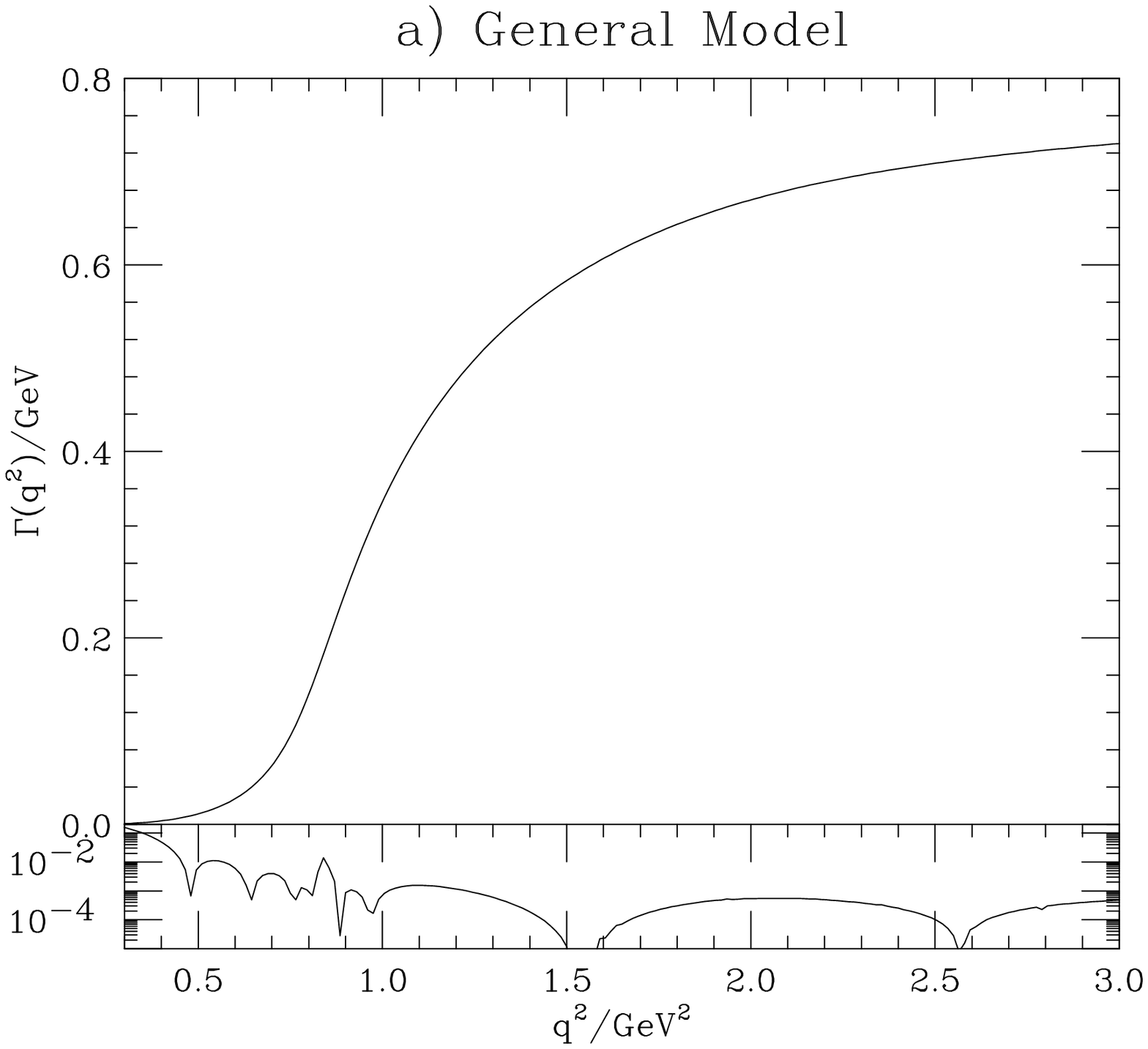}\hfill
\includegraphics[width=0.45\textwidth,angle=90]{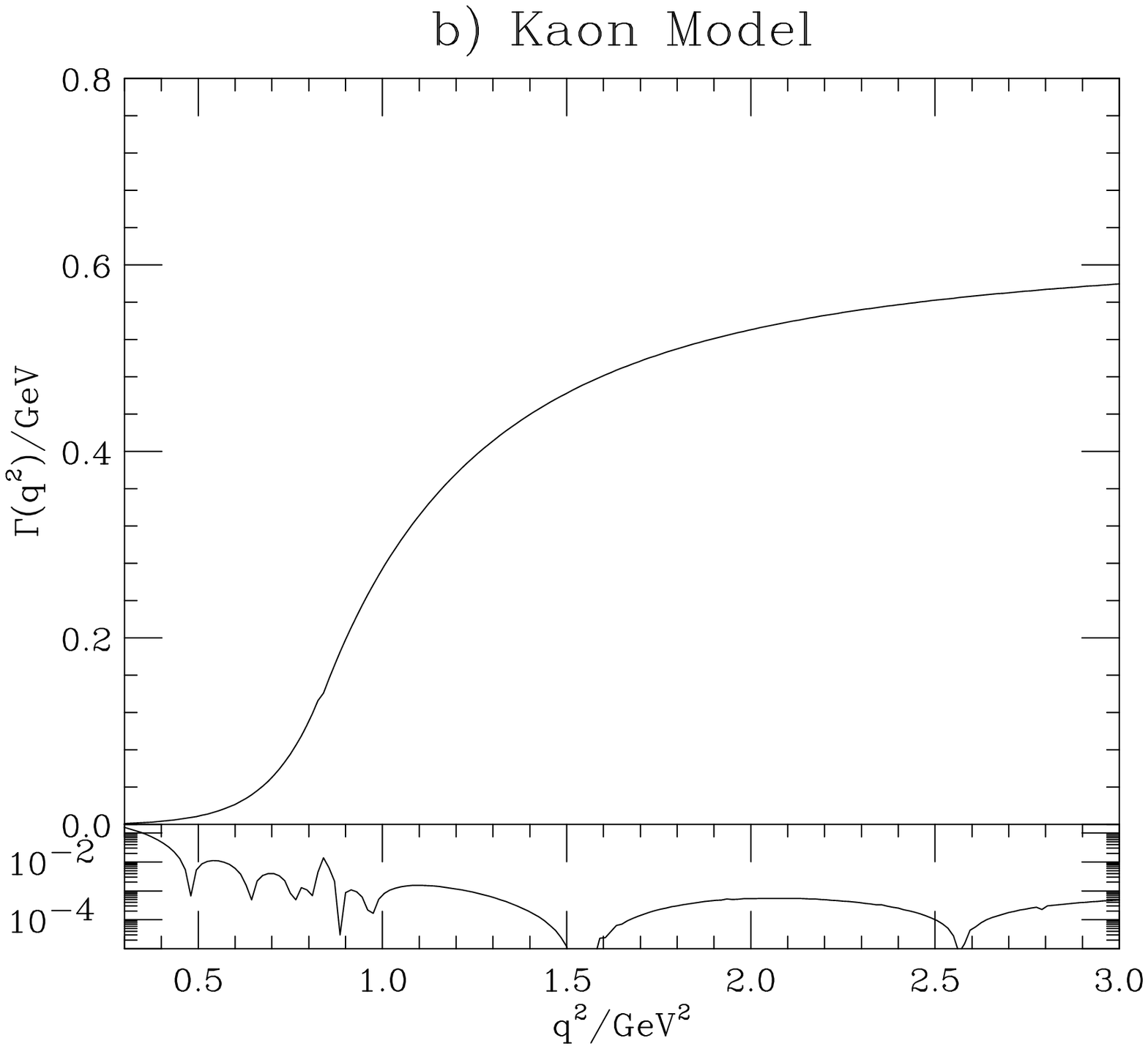}\\
\end{center}
\caption{The running $a_1$ width calculated using the model of \cite{Kuhn:1990ad}.
        a) shows the running width for the parameters used in the 
        Section~\protect\ref{sect:threemesondefaultcurrent} 
	and b) shows the running width for the
         parameters used in the Section~\protect\ref{sect:threeK}. The 
        result from numerically evaluating the matrix element in
        \HWPP\ is shown.}
\label{fig:a1width1}
\end{figure}

  This is the implementation of the model \cite{Kuhn:1990ad,Decker:1992kj,Jadach:1993hs}
  which uses the form of \cite{Kuhn:1990ad} for the $a_1$ width.
  The form factors for the different modes are given 
  in~\cite{Decker:1992kj,Jadach:1993hs}. A Breit-Wigner distribution, 
  Eqn.\,\ref{eqn:runningBW}, with a running width is used for the $a_1$. As
  the dominant decay of $a_1$ is to three pions, a more complicated form of the
  running width is used. A parameterisation of the energy dependence 
  of the running width is given \cite{Kuhn:1990ad} for the default
  parameter values for the $\rho$ form factor. Instead of using this 
  parameterisation we calculate an interpolation table for the running width 
  at initialisation with the actual $\rho$ form-factor parameters. Our calculation of
  the running width is compared with the result of \cite{Kuhn:1990ad} in 
  Fig.\,\ref{fig:a1width1};  there is good agreement between the two
  results.

\begin{figure}[!h]
\begin{center}
\includegraphics[width=0.48\textwidth,angle=90]{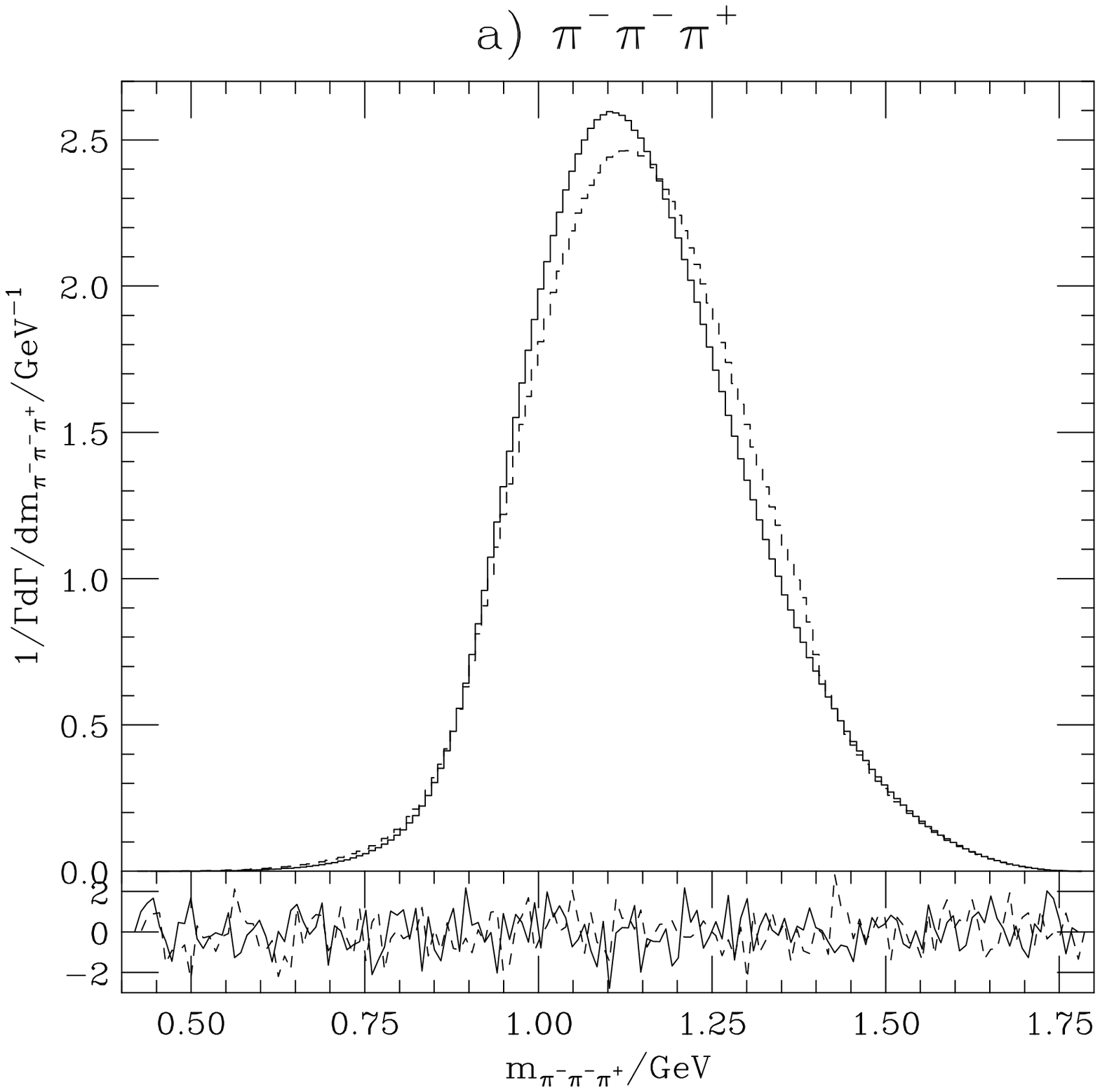}\hfill
\includegraphics[width=0.48\textwidth,angle=90]{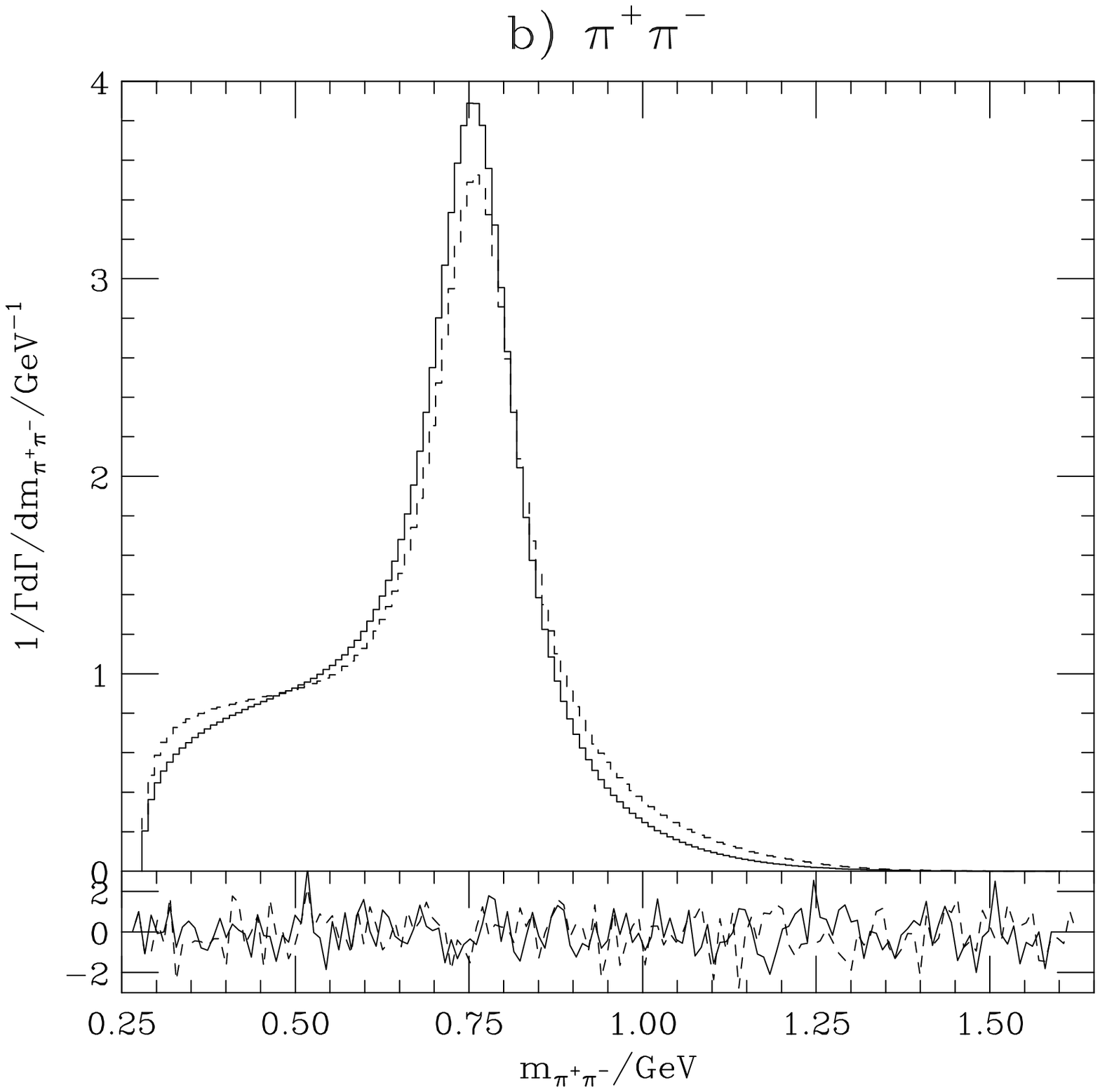}\\
\caption{Differential distribution for the mass of a) $\pi^-\pi^-\pi^+$ and
         b) $\pi^+\pi^-$ produced in the decay $\tau^-\to\pi^-\pi^-\pi^+\nu_\tau$.
        The solid line is the model of \cite{Jadach:1993hs,Kuhn:1990ad,Decker:1992kj}
        and the dashed line that of \cite{Asner:1999kj}.}
\label{fig:pippimpim}
\vspace{0.5cm}
\includegraphics[width=0.48\textwidth,angle=90]{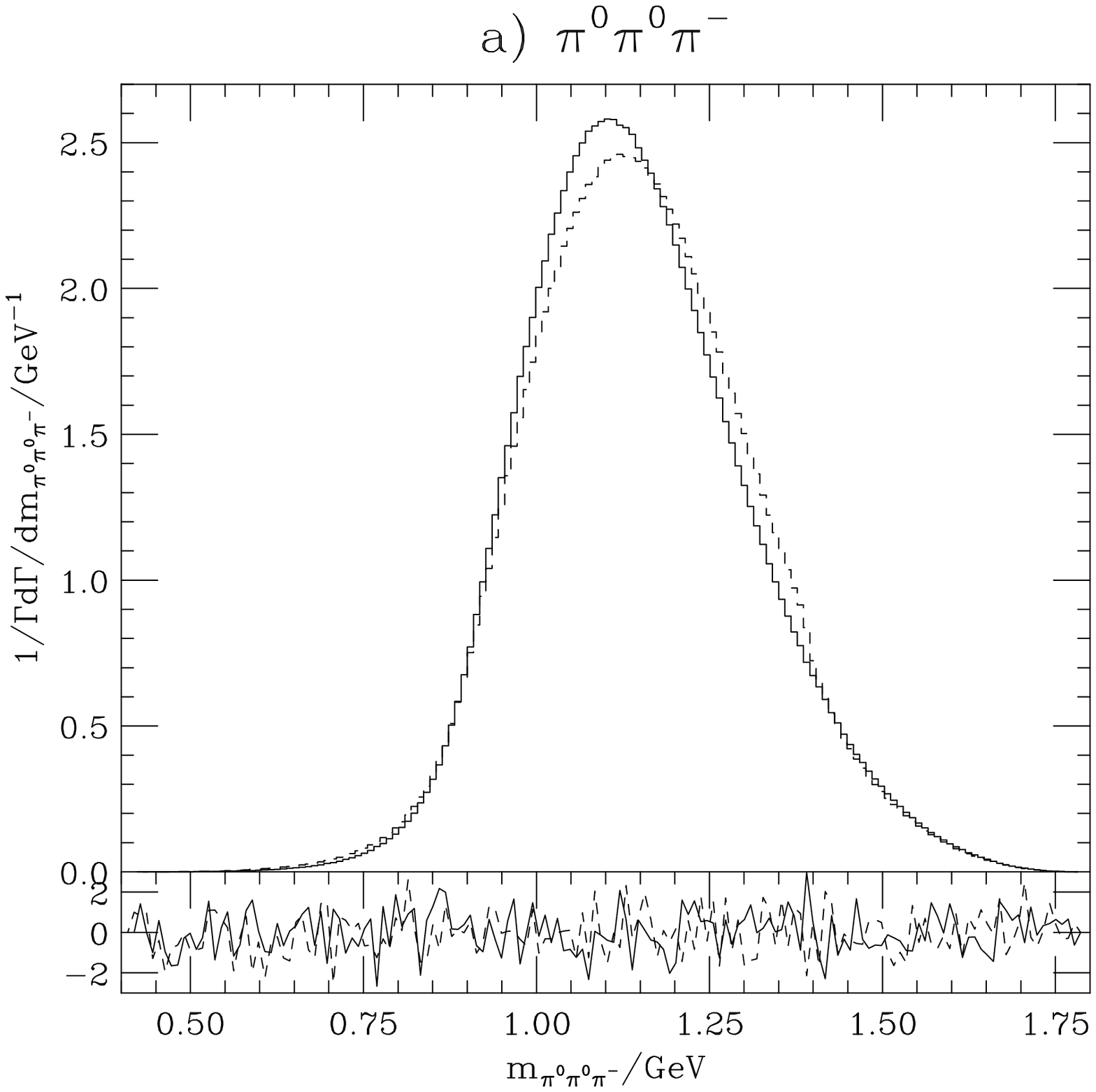}\hfill
\includegraphics[width=0.48\textwidth,angle=90]{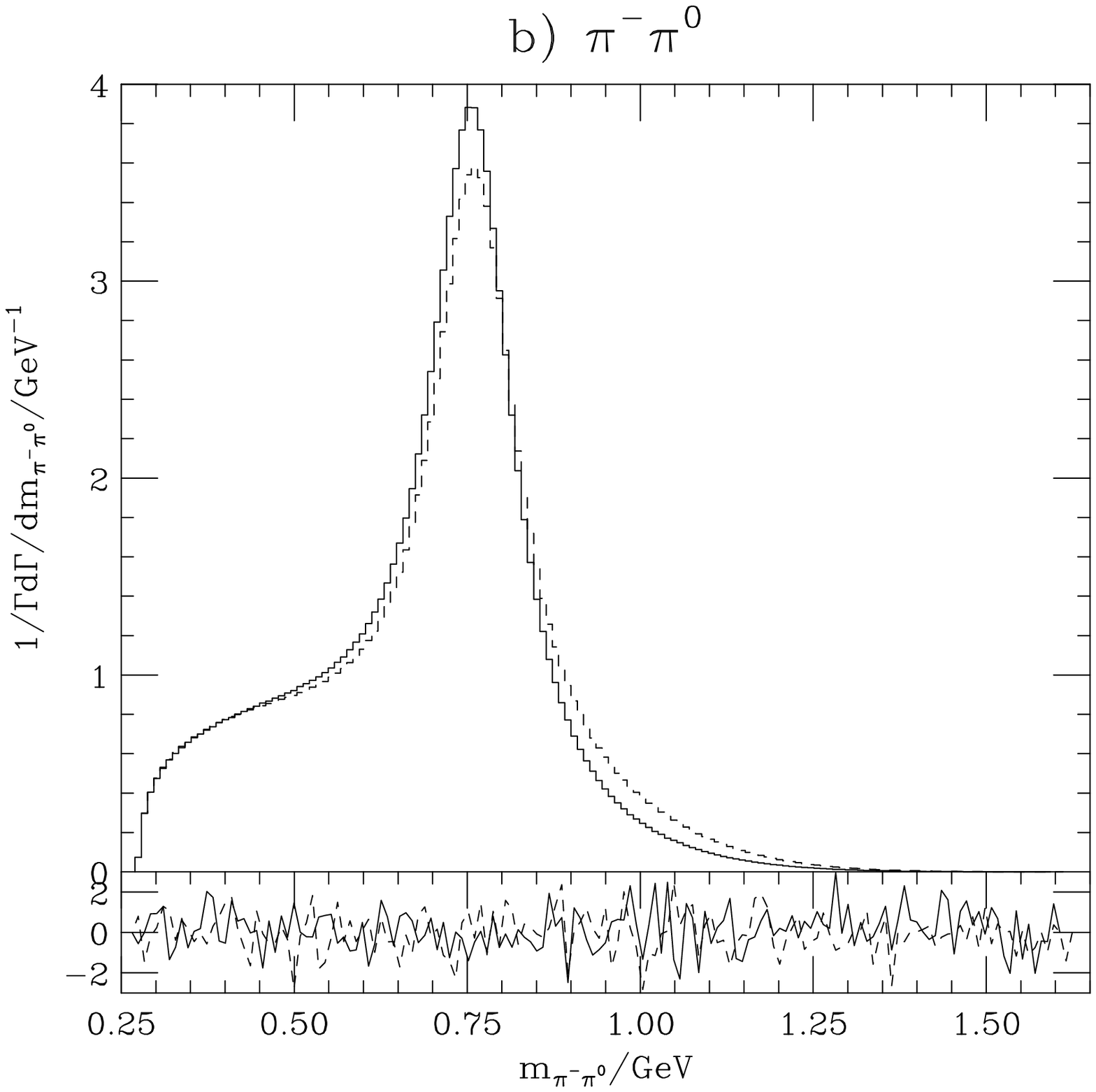}\\
\caption{Differential distribution for the mass of a) $\pi^0\pi^0\pi^-$ and
         b) $\pi^-\pi^0$ produced in the decay $\tau^-\to\pi^0\pi^0\pi^-\nu_\tau$.
        The solid line is the model of \cite{Jadach:1993hs,Kuhn:1990ad,Decker:1992kj}
        and the dashed line that of \cite{Asner:1999kj}.}
\label{fig:pi0pi0pim}
\end{center}
%\vspace{-1cm}
\end{figure}

  The results for the partial widths for all the modes are compared with the results
  from \TAUOLA\ in Table~\ref{tab:threefour} which shows good agreement between the
  two programs. The mass distributions for the hadronic system and the 
  $\pi^+\pi^-$ subsystem, which contains the $\rho$ resonance,
  for the $\tau^-\to\pi^-\pi^-\pi^+\nu_\tau$ mode are shown in
  Fig.\,\ref{fig:pippimpim}. The mass distributions of the $\pi^0\pi^0\pi^-$
  system and the $\pi^-\pi^0$ system are shown in Fig.\,\ref{fig:pi0pi0pim}, for 
  the $\tau^-\to\pi^0\pi^0\pi^-\nu_\tau$ mode.

\begin{table}
\begin{center}
\begin{tabular}{|c|c|c|c|c|}
\hline
Mode & \HWPP & \TAUOLA & Difference \\
     & $\Gamma_{\rm partial}/{\rm 10^{-13}\,GeV}$
     & $\Gamma_{\rm partial}/{\rm 10^{-13}\,GeV}$ & $/{\rm 10^{-17}\,GeV}$ \\
\hline
\multicolumn{4}{|c|}{General Model}\\
\hline
 $\pi^-\pi^-\pi^+$     & $1.4632\pm0.0001$&
                         $1.4633\pm0.0001$&
                         $-1\pm1$ \\[-1mm]
 $\pi^-\pi^0\pi^0$     & $1.4957\pm0.0001$&
                         $1.4958\pm0.0001$&
                         $-1\pm1$ \\[-1mm]
 $K^-\pi^-K^+$          & $(2.4567\pm0.0003)\times10^{-2}$&
                         $(2.4572\pm0.0001)\times10^{-2}$&
                         $-0.05\pm0.03$\\[-1mm]
 $K^0\pi^-\bar{K}^0$   & $(2.2694\pm0.0002)\times10^{-2}$&
                         $(2.2696\pm0.0001)\times10^{-2}$&
                         $-0.02\pm0.03$\\[-1mm]
 $K^-\pi^0\bar{K}^0$   & $(2.0877\pm0.0001)\times10^{-3}$&
                         $(2.0878\pm0.0001)\times10^{-3}$&
                         $-0.001\pm0.002$ \\[-1mm]
 $\pi^0\pi^0K^-$       & $(2.5421\pm0.0003)\times10^{-2}$&
                         $(2.5419\pm0.0001)\times10^{-2}$&
                         $\phantom{-}0.03\pm0.03$ \\[-1mm]
 $K^-\pi^-\pi^+$       & $0.13077\pm0.00001$&
                         $0.13078\pm0.00001$& 
                         $-0.1\pm0.2$\\[-1mm]
 $\pi^-\bar{K}^0\pi^0$ & $0.13204\pm0.00002$& 
                         $0.13201\pm0.00001$&
                         $\phantom{-}0.3\pm0.2$ \\[-1mm]
 $\pi^-\pi^0\eta$      & $(4.6175\pm0.0009)\times10^{-2}$& 
                         $(4.6160\pm0.0008)\times10^{-2}$&
                         $\phantom{-}0.15\pm0.12$ \\
\hline
\multicolumn{4}{|c|}{CLEO model for Three Pions}\\
\hline
$\pi^-\pi^-\pi^+$     & $1.2605\pm0.0001$ & $1.2604\pm0.0001$ & $1\pm1$\\[-1mm] 
$\pi^-\pi^0\pi^0$     & $1.2702\pm0.0001$ & $1.2702\pm0.0001$ & $0\pm1$\\
\hline
\multicolumn{4}{|c|}{Two Pions and a Photon}\\
\hline
$\pi^-\pi^0\gamma$    & $(1.2708\pm0.0001)\times10^{-2}$&
                           $(1.2707\pm0.0001)\times10^{-2}$& $\phantom{-}0.01\pm0.01$\\
\hline
\multicolumn{4}{|c|}{4 pions}\\
\hline
$2\pi^-\pi^0\pi^+$ & $0.95120\pm0.00009$ 
                   & $0.95112\pm0.00006$
                   & $0.8\pm1.1$\\[-1mm]
$3\pi^0\pi^-$ & $0.24453\pm0.0002$ & $0.24452\pm0.0002$ & $0.1\pm2$ \\
\hline
\multicolumn{4}{|c|}{5 pions}\\
\hline
 $\pi^-4\pi^0$    & $(3.6192\pm0.0010)\times10^{-2}$ & 
                  $(3.6185\pm0.0007)\times10^{-2}$ & 
                  $\phantom{-}0.07\pm0.12$ \\[-1mm]
 $3\pi^-2\pi^+$ & $(4.2745\pm0.0010)\times10^{-2}$&  
                  $(4.2746\pm0.0006)\times10^{-2}$&  
                  $-0.01\pm0.12$                   \\[-1mm]
 $2\pi^-2\pi^0\pi^+$ & $(1.1883\pm0.0004)\times10^{-1}$ 
                     & $(1.1884\pm0.0003)\times10^{-1}$
                     & $-0.1\pm0.5$\\                       
\hline
\end{tabular}
\end{center}
\caption{The partial widths for the three-, four- and five-meson decay modes of the
         $\tau$ calculated using \HWPP\ and \TAUOLA. In order to compare the
        results of the two programs, the masses of the decay products in \TAUOLA\
        were set to the \HWPP\ values, as was the Cabibbo angle. For the five
        pion decays the $\rho$ form factors were not included in the decay of the
        $\omega$.}
\label{tab:threefour}
\end{table}

\begin{figure}[!h]
\begin{center}
\includegraphics[width=0.48\textwidth,angle=90]{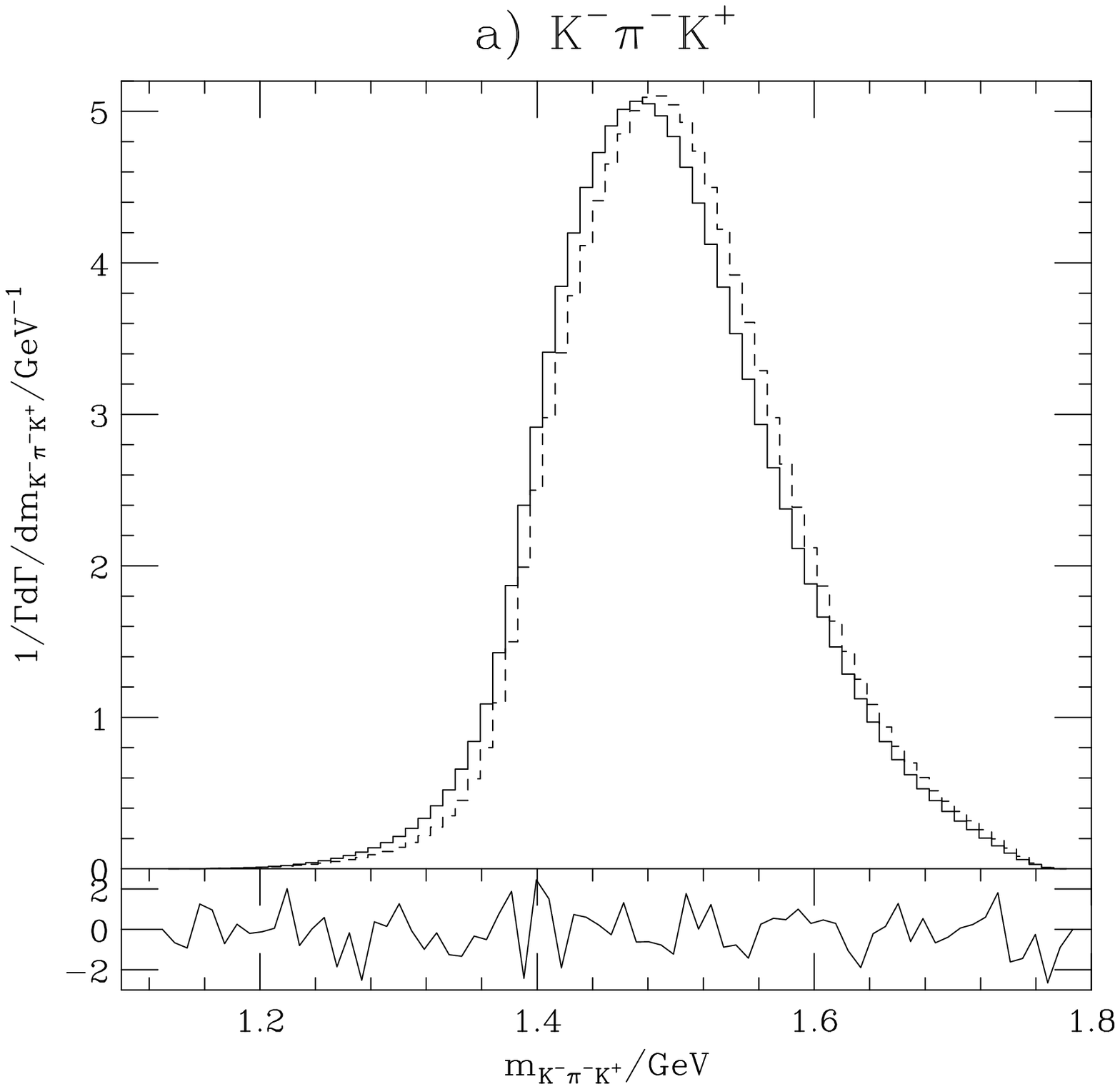}\hfill
\includegraphics[width=0.48\textwidth,angle=90]{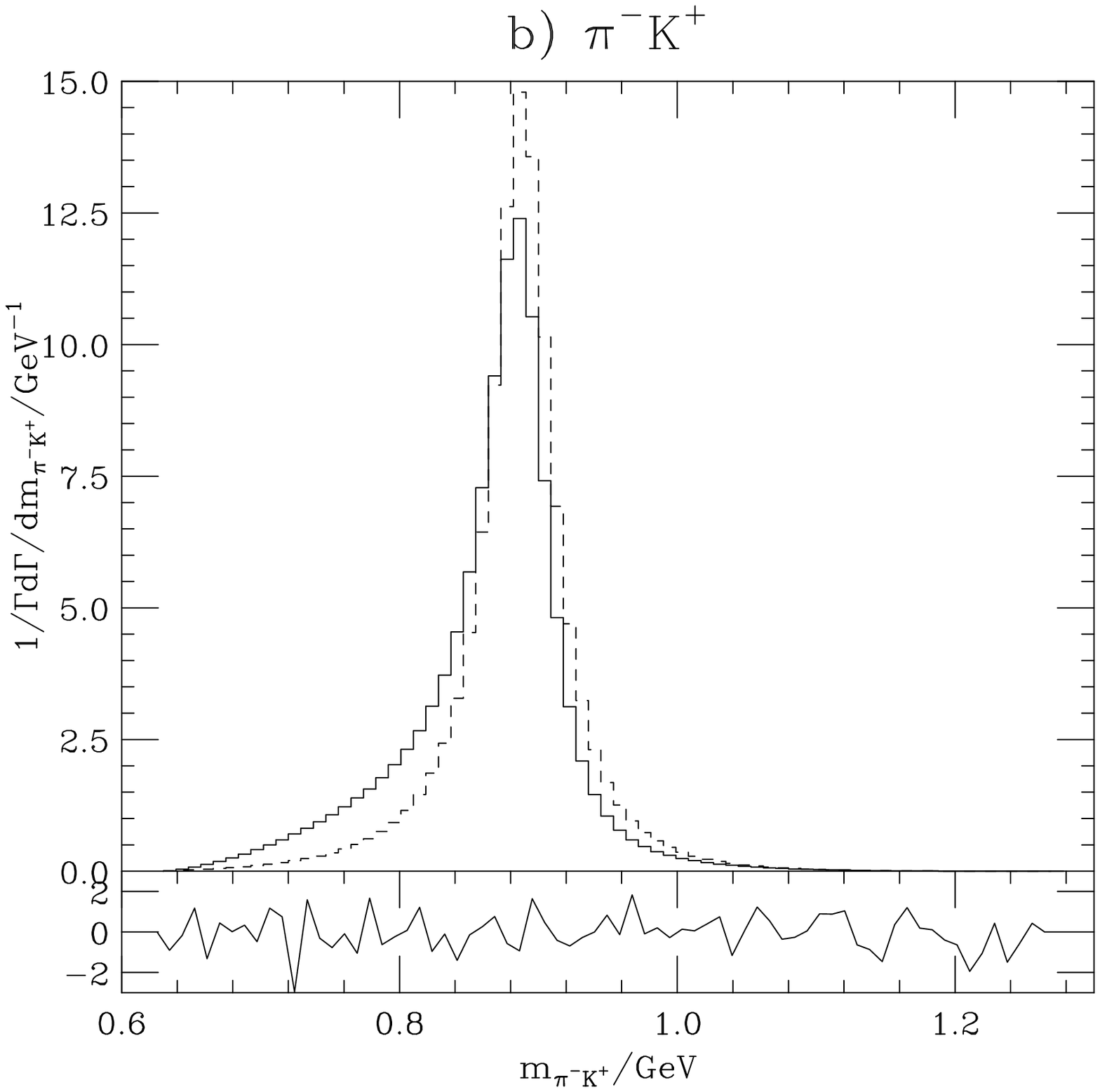}\\
\caption{Differential distribution for the mass of a) $K^-\pi^-K^+$ and
         b) $\pi^-K^+$ produced in the decay $\tau^-\to K^-\pi^-K^+\nu_\tau$.
        The solid line is the model of \cite{Jadach:1993hs,Decker:1992kj}
        and the dashed line that of \cite{Finkemeier:1995sr}.}
\label{fig:kmpimkp}
\vspace{0.5cm}
\includegraphics[width=0.48\textwidth,angle=90]{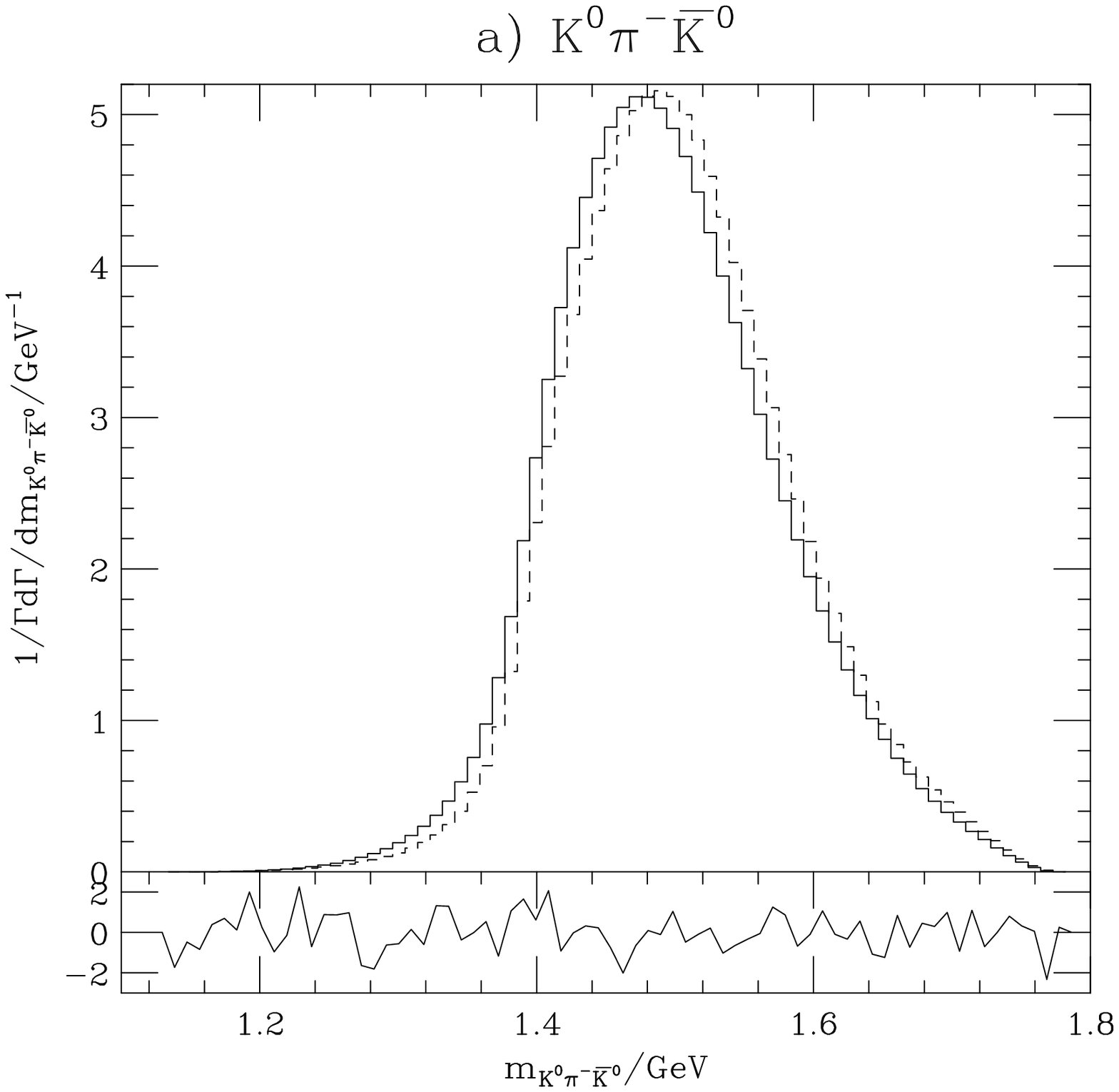}\hfill
\includegraphics[width=0.48\textwidth,angle=90]{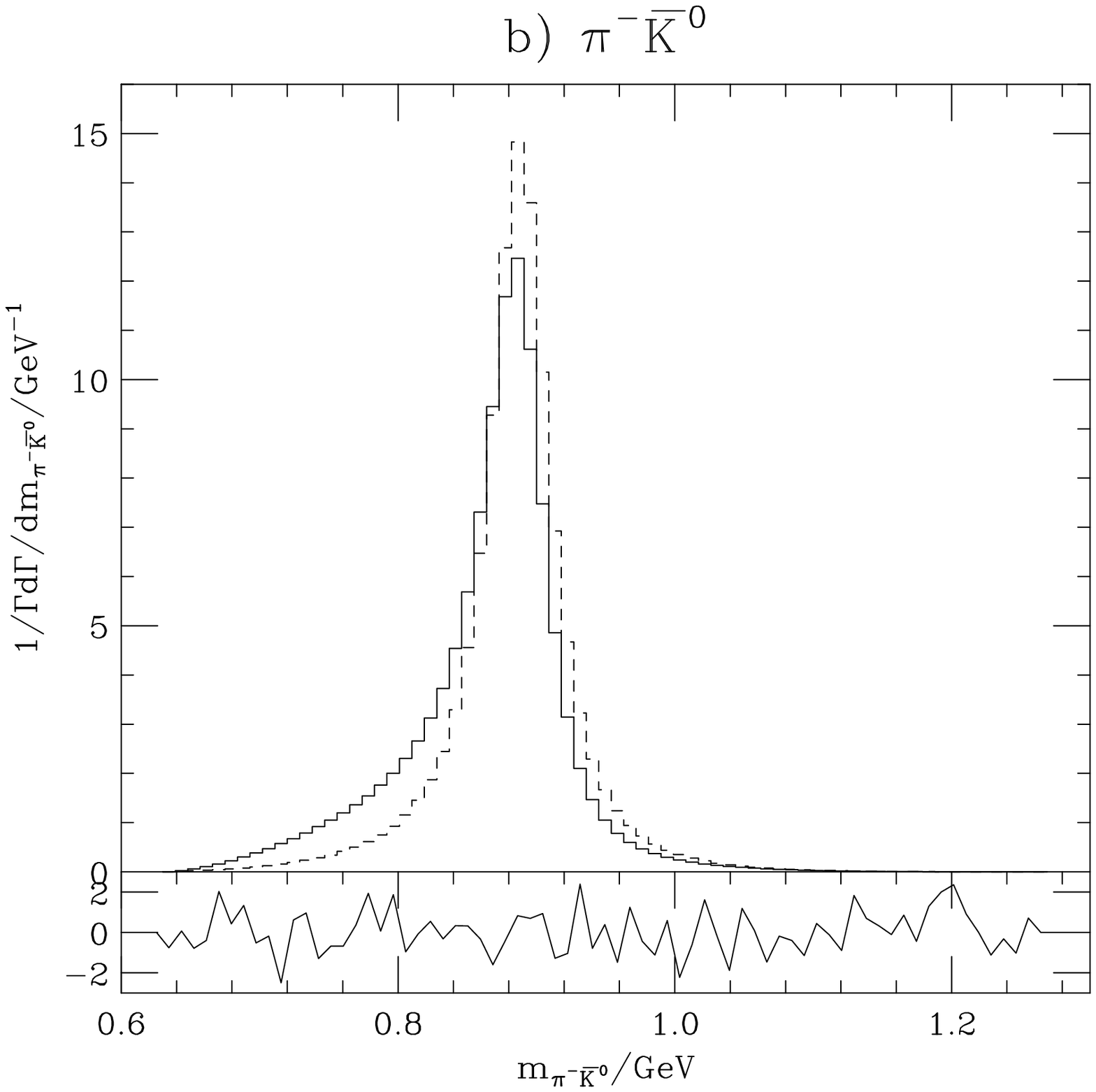}\\
\caption{Differential distribution for the mass of a) $K^0\pi^-\bar{K}^0$ and
         b) $\pi^-\bar{K}^0$ produced in the decay $\tau^-\to K^0\pi^-\bar{K}^0\nu_\tau$.
        The solid line is the model of \cite{Jadach:1993hs,Decker:1992kj}
        and the dashed line that of \cite{Finkemeier:1995sr}.}
\label{fig:k0pimkbar0}
\end{center}
%\vspace{-1cm}
\end{figure}

\begin{figure}[!h]
\begin{center}
\includegraphics[width=0.48\textwidth,angle=90]{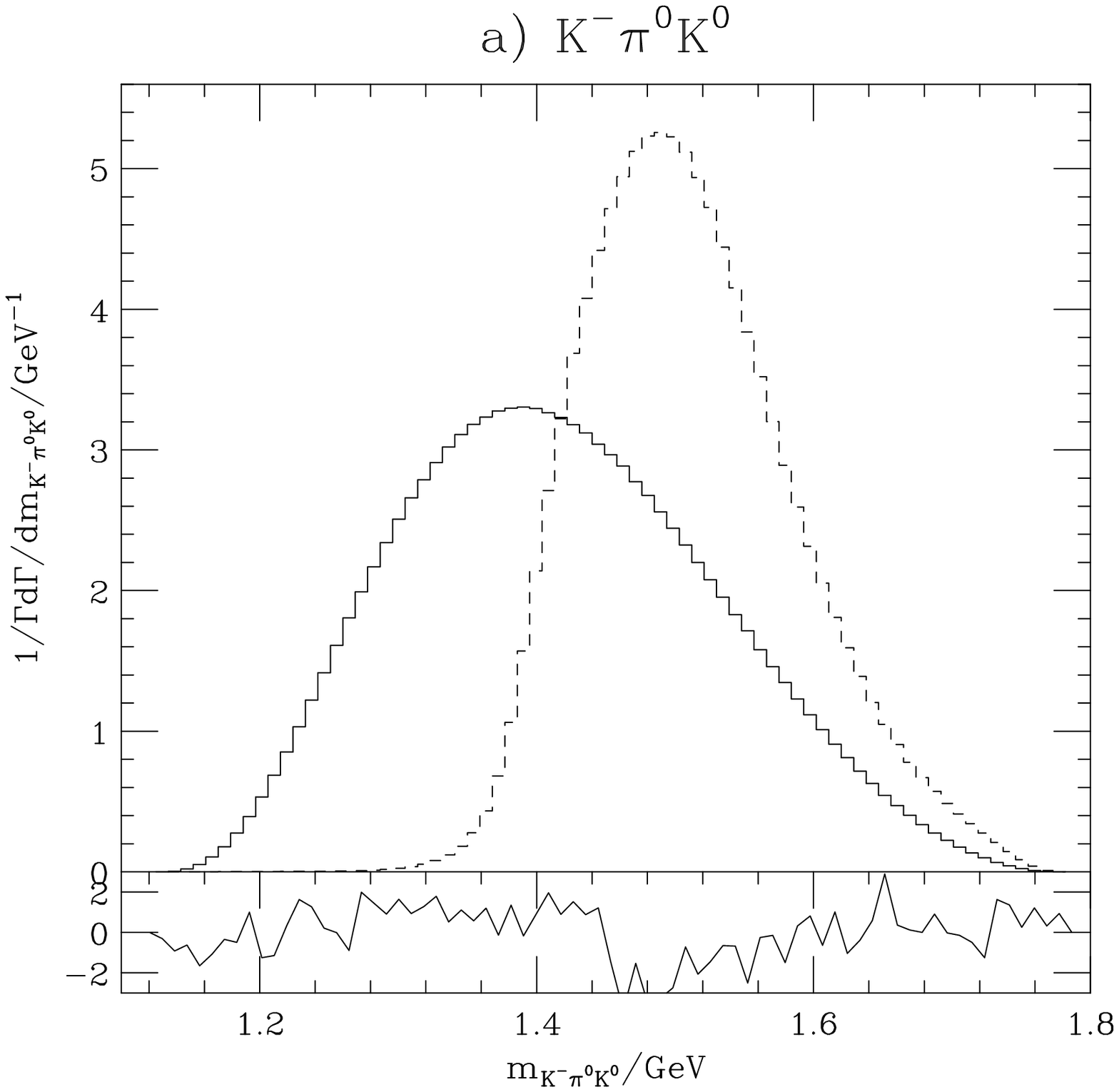}\hfill
\includegraphics[width=0.48\textwidth,angle=90]{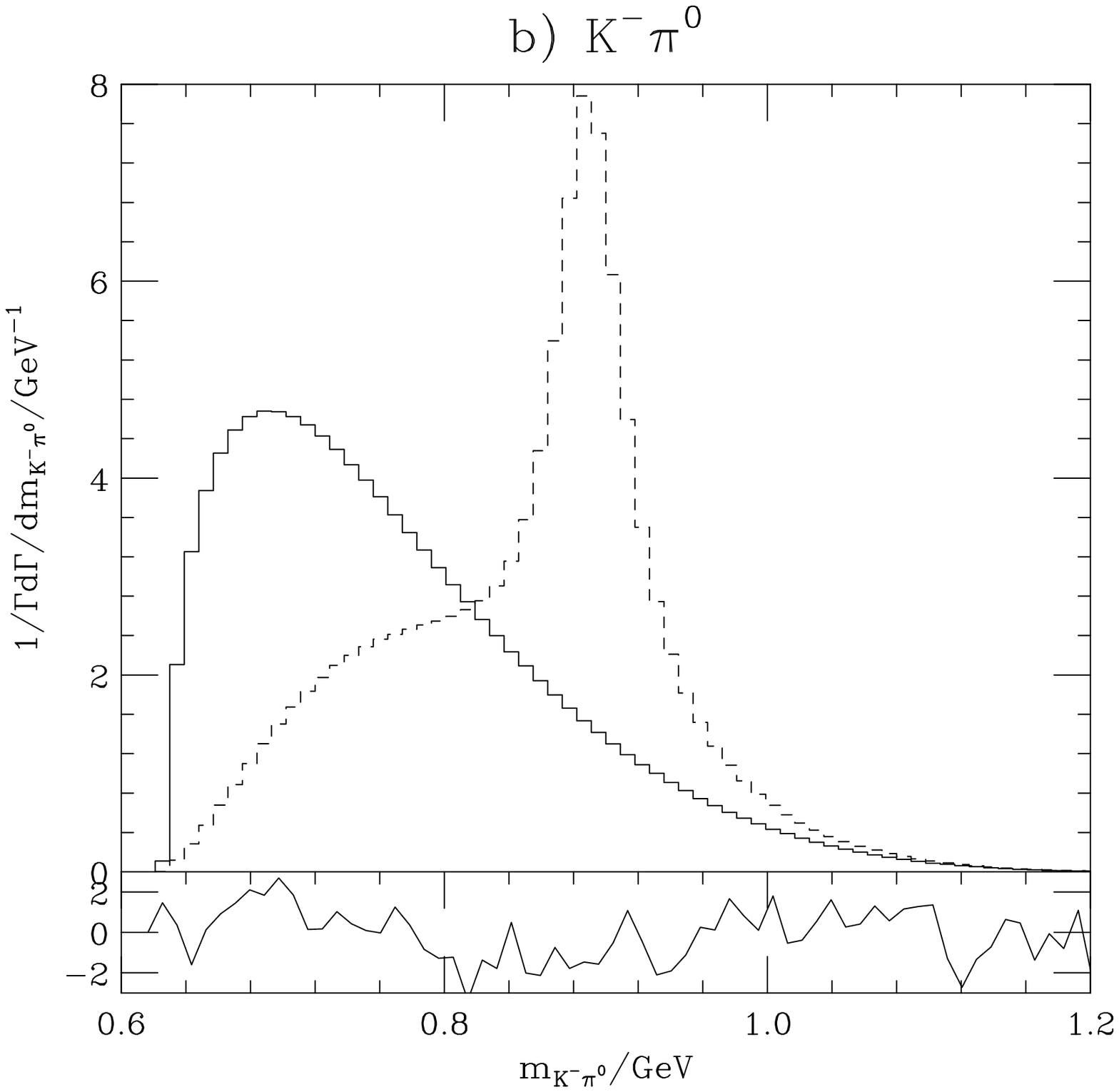}\\
\caption{Differential distribution for the mass of a) $K^-\pi^0K^0$ and
         b) $K^-\pi^0$ produced in the decay $\tau^-\to K^-\pi^0K^0\nu_\tau$.
        The solid line is the model of \cite{Jadach:1993hs,Decker:1992kj}
        and the dashed line that of \cite{Finkemeier:1995sr}.}
\label{fig:kmpi0k0}
\vspace{0.5cm}
\includegraphics[width=0.48\textwidth,angle=90]{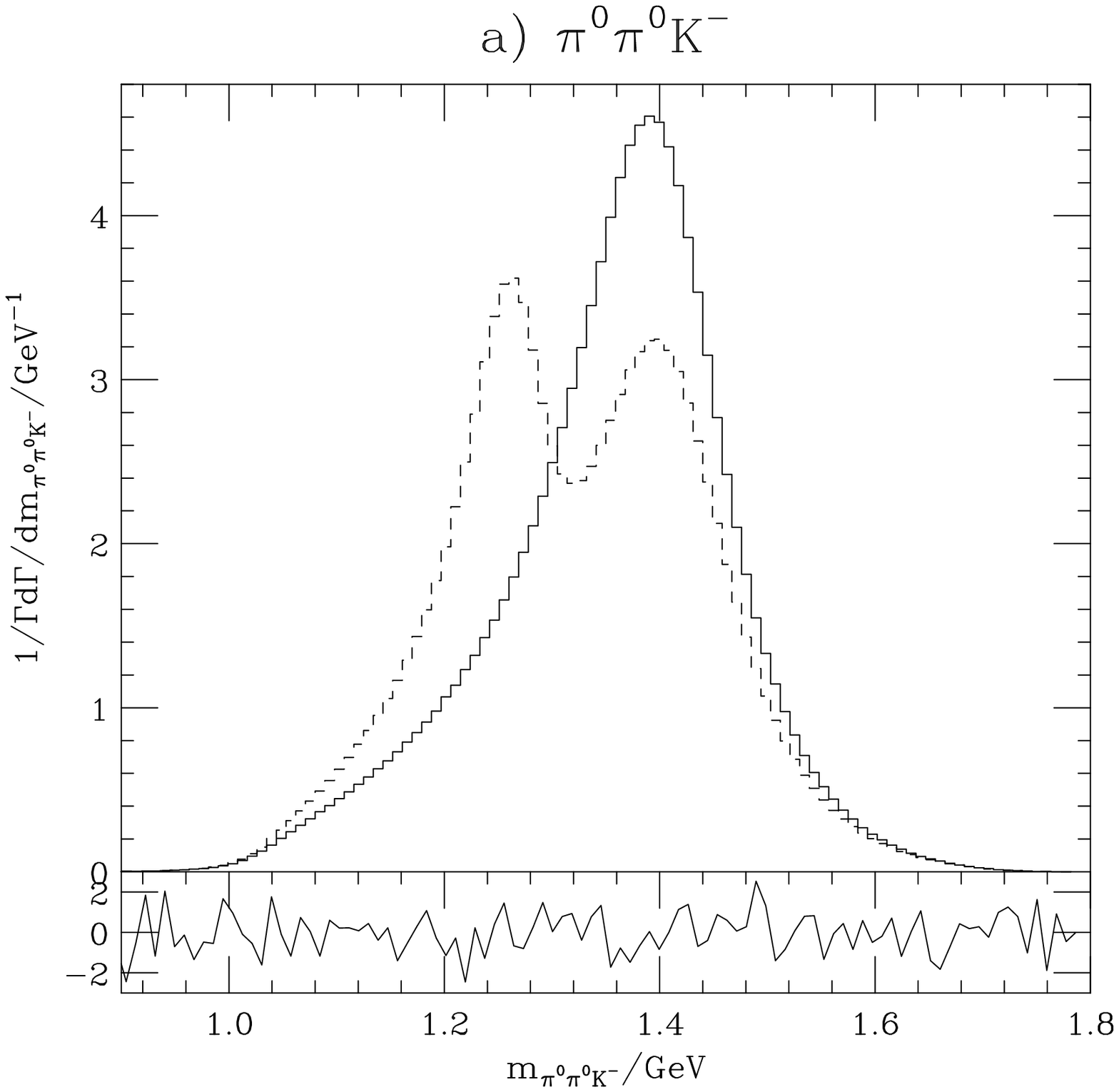}\hfill
\includegraphics[width=0.48\textwidth,angle=90]{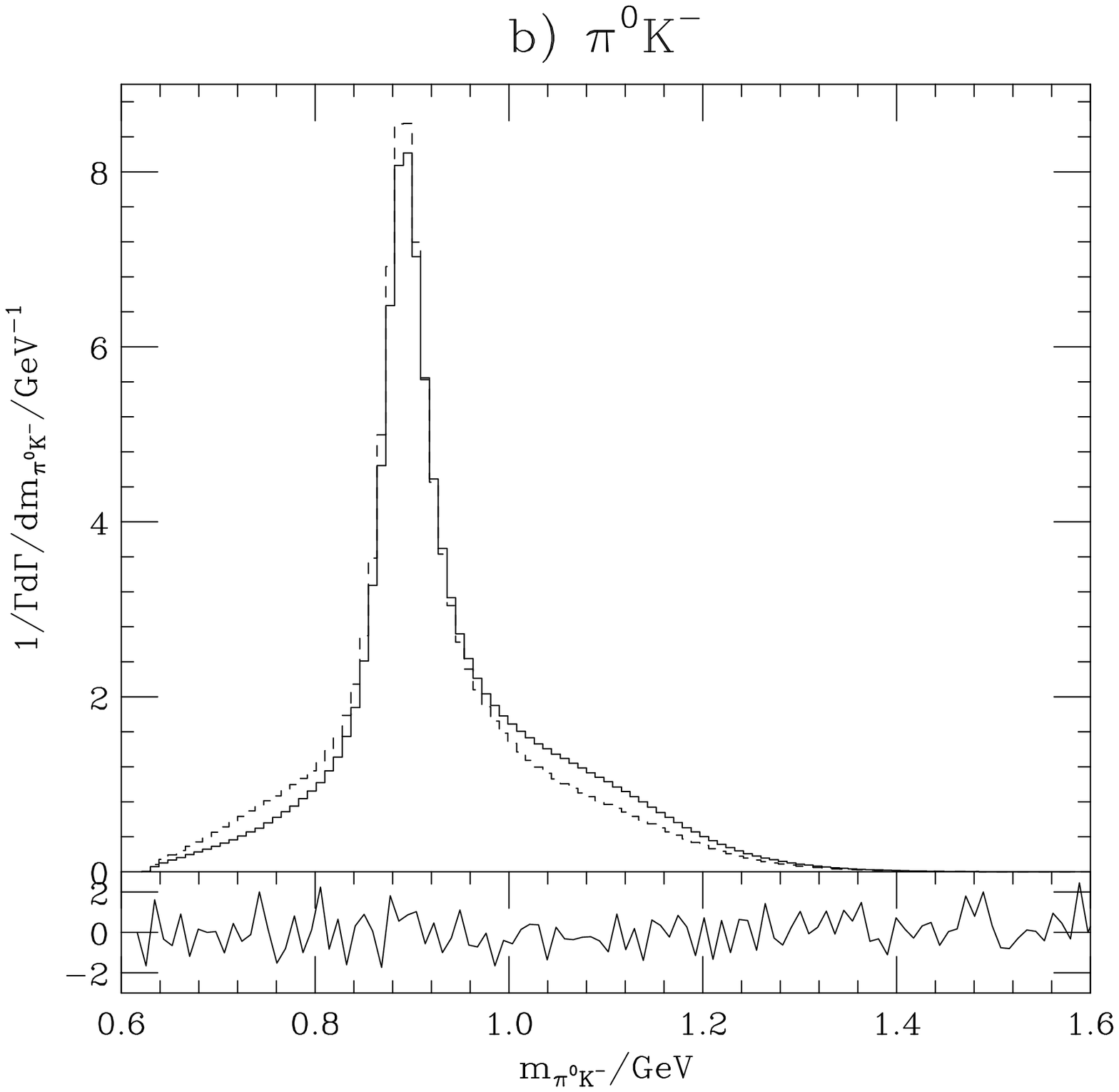}\\
\caption{Differential distribution for the mass of a) $\pi^0\pi^0K^-$ and
         b) $\pi^0K^-$ produced in the decay $\tau\to \pi^0\pi^0K^-\nu_\tau$.
        The solid line is the model of \cite{Jadach:1993hs,Decker:1992kj}
        and the dashed line that of \cite{Finkemeier:1995sr}.}
\label{fig:pi0pi0km}
\end{center}
%\vspace{-1cm}
\end{figure}

  The mass distributions 
  of the $K^-\pi^-K^+$ system and the $\pi^-K^+$ system are shown in
  Fig.\,\ref{fig:kmpimkp} for the $\tau^-\to K^-\pi^-K^+\nu_\tau$ decay mode.
  The mass distributions 
  of the $K^0\pi^-\bar{K}^0$ hadronic system and the $\pi^-\bar{K}^0$ 
  system are shown in Fig.\,\ref{fig:k0pimkbar0} for the 
  $\tau^-\to K^0\pi^-\bar{K}^0\nu_\tau$ decay mode.
  The mass distributions of the hadronic system and the $K^-\pi^0$ subsystem are shown in
  Fig.\,\ref{fig:kmpi0k0} for the $\tau^-\to K^-\pi^0K^0\nu_\tau$ decay mode.

\begin{figure}[!h]
\begin{center}
\includegraphics[width=0.48\textwidth,angle=90]{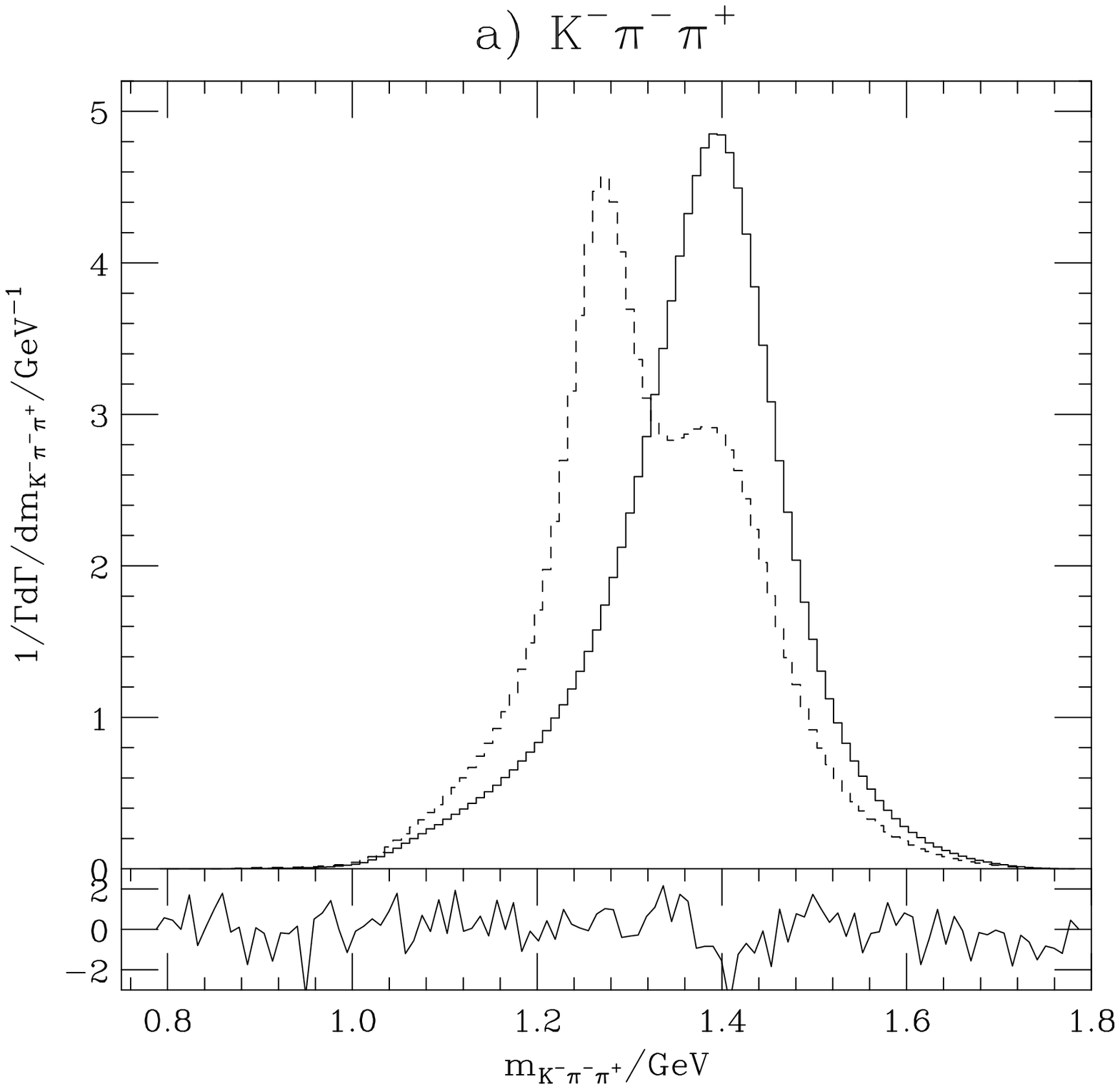}\hfill
\includegraphics[width=0.48\textwidth,angle=90]{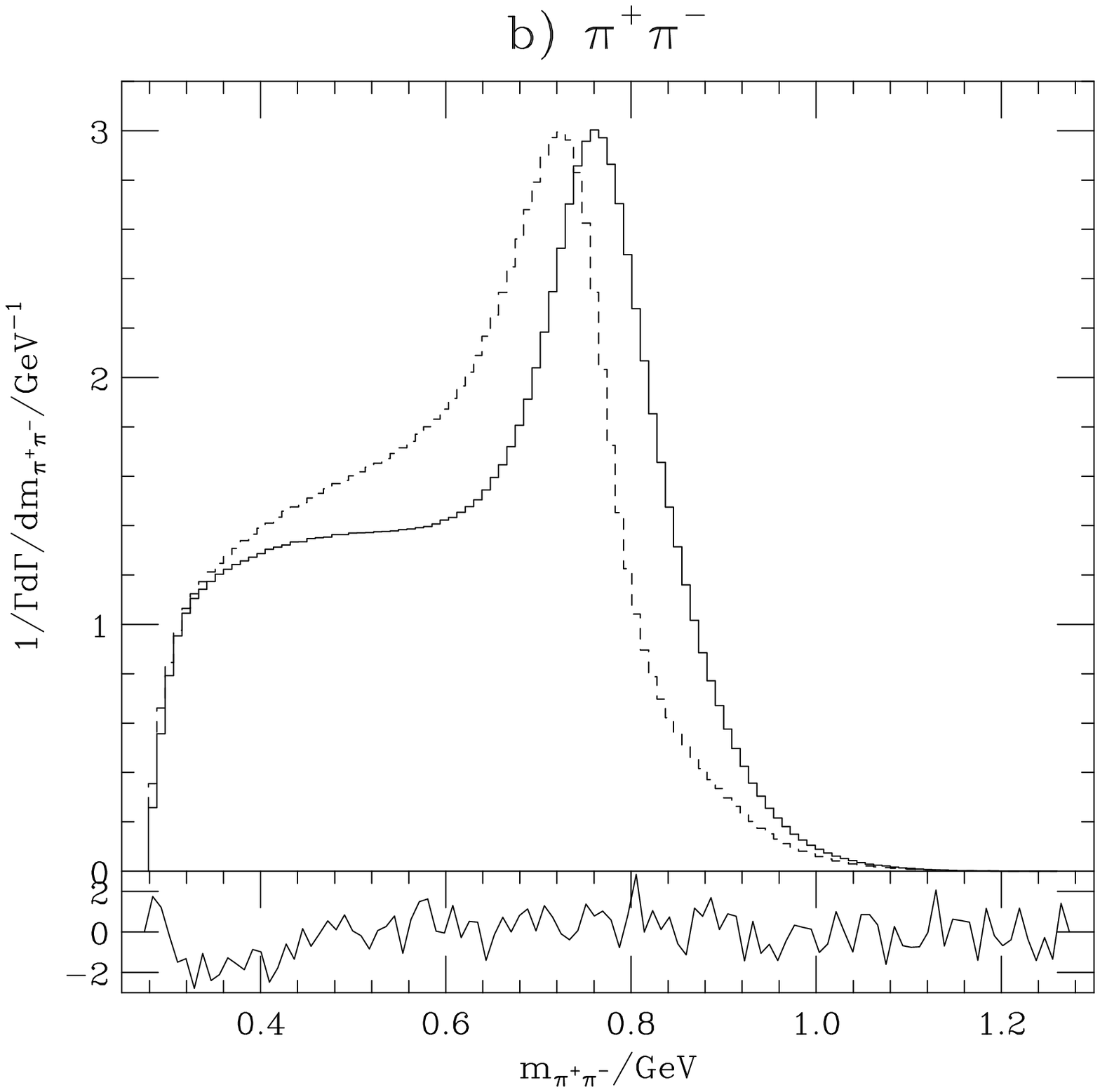}\\
\caption{Differential distribution for the mass of a) $K^-\pi^-\pi^+$ and
         b) $\pi^+\pi^-$ produced in the decay $\tau^-\to K^-\pi^-\pi^+\nu_\tau$.
        The solid line is the model of \cite{Jadach:1993hs,Decker:1992kj}
        and the dashed line that of \cite{Finkemeier:1995sr}.}
\label{fig:kmpimpip}
\vspace{0.5cm}
\includegraphics[width=0.48\textwidth,angle=90]{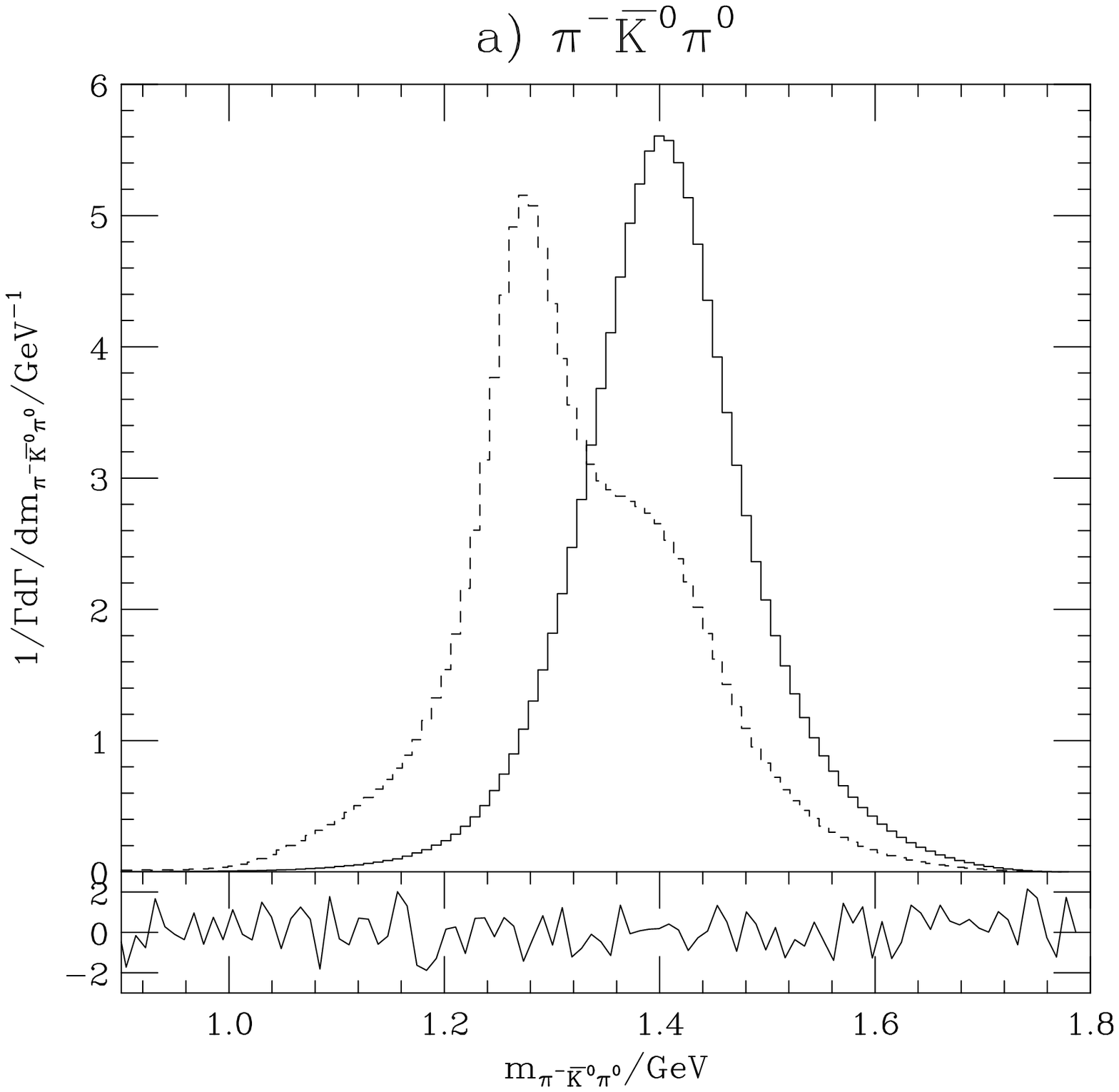}\hfill
\includegraphics[width=0.48\textwidth,angle=90]{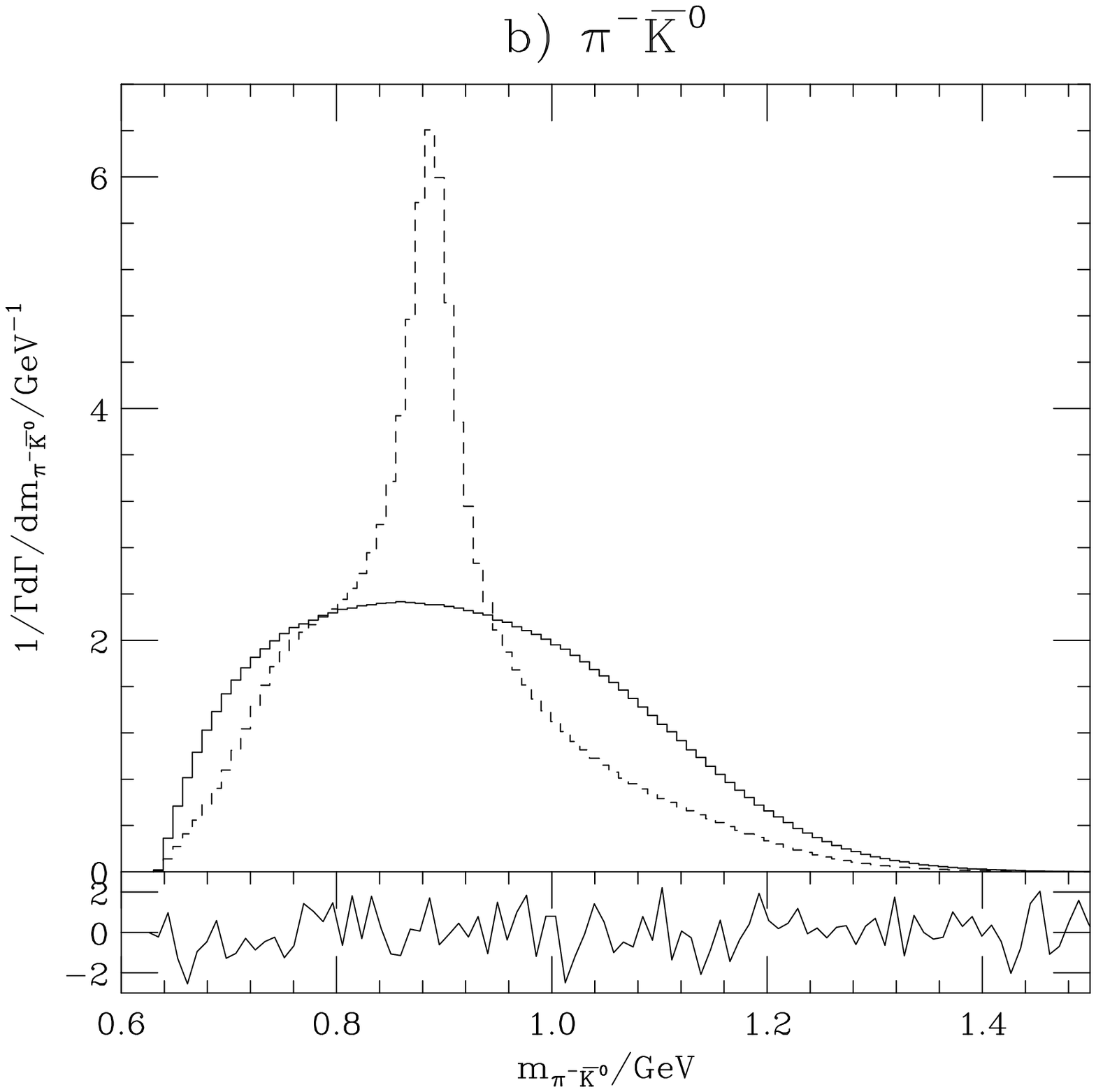}\\
\caption{Differential distribution for the mass of a) $\pi^-\bar{K}^0\pi^0$ and
         b) $\pi^-\bar{K}^0$ produced in the decay 
        $\tau^-\to \pi^-\bar{K}^0\pi^0\nu_\tau$.
        The solid line is the model of \cite{Jadach:1993hs,Decker:1992kj}
        and the dashed line that of \cite{Finkemeier:1995sr}.}
\label{fig:pimpi0kbar0}
\end{center}
%\vspace{-1cm}
\end{figure}

  The mass distributions of the hadronic system and the $\pi^0K^-$ subsystem 
  are shown in Fig.\,\ref{fig:pi0pi0km} for the decay mode
  $\tau^-\to\pi^0\pi^0 K^-\nu_\tau$.
  The mass distributions of the $K^-\pi^+\pi^-$ system and $\pi^+\pi^-$
  subsystem are shown in Fig.\,\ref{fig:kmpimpip} for the 
  decay $\tau^-\to K^-\pi^+\pi^-\nu_\tau$.
  The mass distributions of the $\pi^-\bar{K}^0\pi^0$ system and the $\pi^-\bar{K}^0$
  subsystem are shown in Fig.\,\ref{fig:pimpi0kbar0} for the decay 
  $\tau^-\to\pi^-\bar{K}^0\pi^0\nu_\tau$.

\begin{figure}
\begin{center}
\includegraphics[width=0.48\textwidth,angle=90]{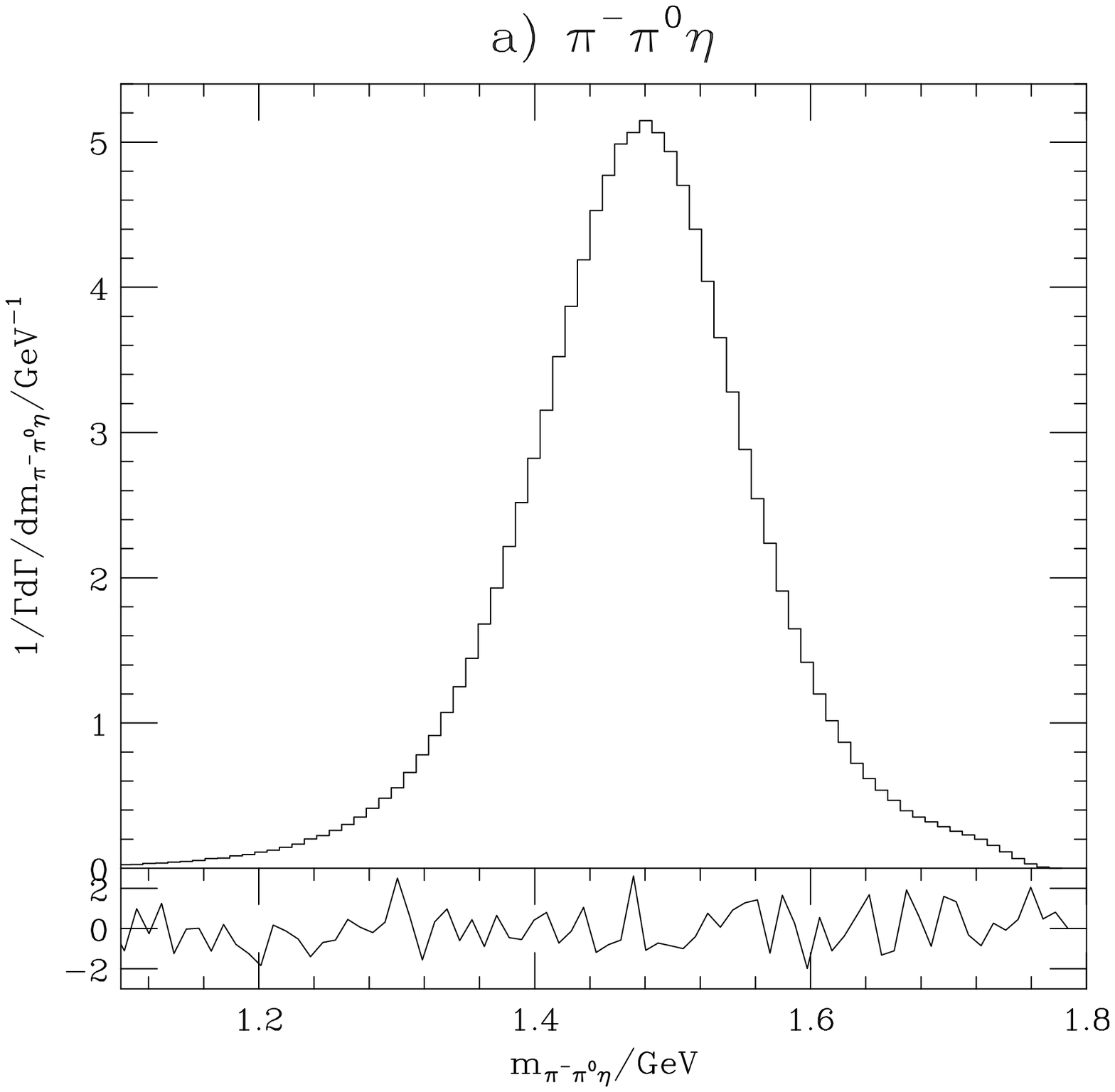}\hfill
\includegraphics[width=0.48\textwidth,angle=90]{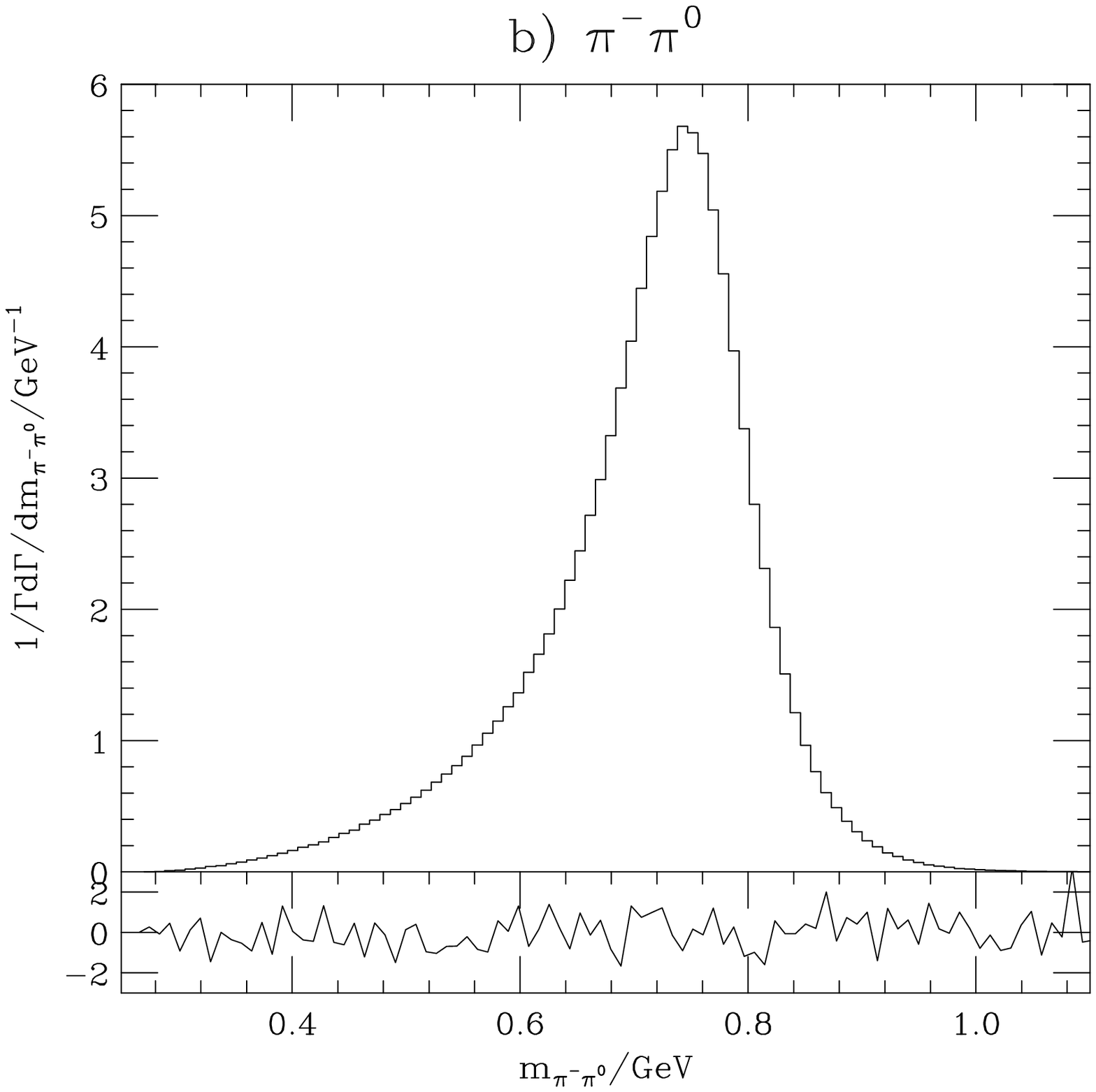}\\
\caption{Differential distribution for the mass of a) $\pi^-\pi^0\eta$ and
         b) $\pi^-\pi^0$ produced in the decay 
        $\tau^-\to \pi^-\pi^0\eta\nu_\tau$ for 
        the model of \cite{Jadach:1993hs,Decker:1992kj}.}
\label{fig:pimpi0eta}
\end{center}
%\end{figure}
%\begin{figure}
\begin{center}
\includegraphics[width=0.45\textwidth,angle=90]{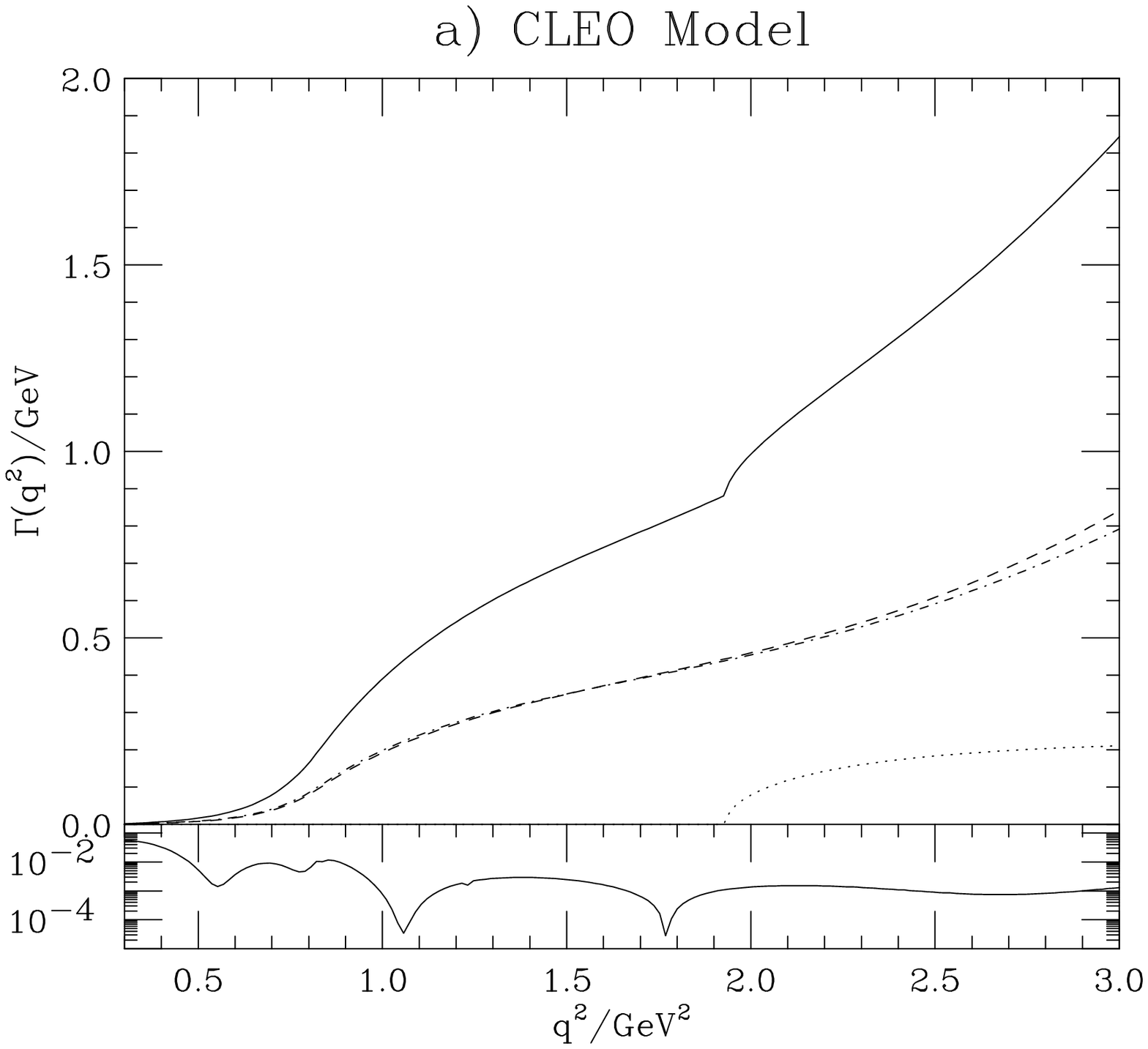}\hfill
\includegraphics[width=0.45\textwidth,angle=90]{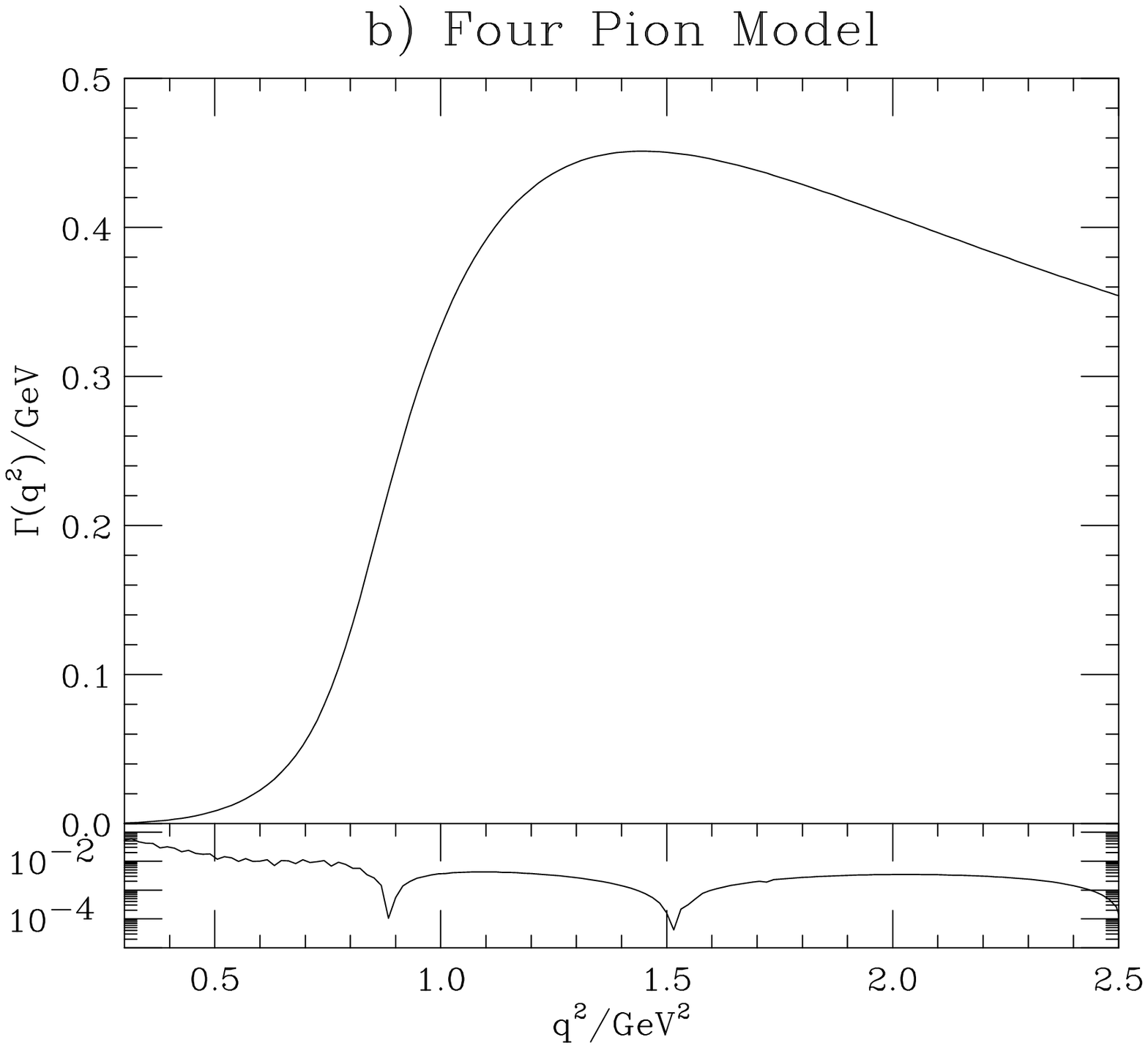}\\
\caption{Running width of the $a_1$ in the models of \cite{Asner:1999kj}
        and \cite{Bondar:2002mw}. a) shows the running width in the model of 
        \cite{Asner:1999kj} with the dotted line showing the contribution of $KK^*$,
        the dashed line showing the contribution of $\pi^-\pi^-\pi^+$, the 
        dot-dashed line showing the contribution of $\pi^0\pi^0\pi^-$ and the solid
        line the total running width.
        b) shows the running width in the model of \cite{Bondar:2002mw}.}
\label{fig:a1width2}
\end{center}
\end{figure}

  Finally the mass distributions 
  of the $\pi^-\pi^0\eta$ system and the $\pi^-\pi^0$ subsystem
  are shown in Fig.\,\ref{fig:pimpi0eta} for the $\tau^-\to\pi^-\pi^0\eta\nu_\tau$
  decay. It should be noted
  that this mode is very sensitive to the value of the $\eta$ mass used.

  In all cases there is good agreement between \HWPP\ and \TAUOLA. 

\subsubsection{CLEO Model for Three Pions}
\label{sect:3pi}

  This is the implementation of the model of~\cite{Asner:1999kj} 
  for the weak current for three pions. This model includes 
  $\rho$ mesons in both the s- and p-wave,
  the scalar $\sigma$ resonance, the tensor $f_2$ resonance and scalar $f_0(1370)$. 
  The form factors for the $\pi^0\pi^0\pi^-$
  mode are given in \cite{Asner:1999kj} and the others can be obtained by isospin
  rotation.

  In this case the running width for the $a_1$ is calculated using the current
  for the $\pi^-\pi^-\pi^0$ and $\pi^0\pi^0\pi^-$ modes of the $a_1$ together
  with an s-wave $KK^*$ contribution. The running width is shown in 
  Fig.\,\ref{fig:a1width2}a for the default parameter values and
  our calculation is in good agreement with that in \TAUOLA.

  The partial widths for the $\tau^-\to\pi^-\pi^-\pi^+\nu_\tau$ and  
  $\tau^-\to\pi^0\pi^0\pi^-\nu_\tau$ are compared with those from the
  \textsf{CLEO} version of \TAUOLA\ in Table\,\ref{tab:threefour}. 
  The mass distributions of the $\pi^-\pi^-\pi^+$ system and
  the $\pi^+\pi^-$ subsystem for the $\tau^-\to\pi^-\pi^-\pi^+\nu_\tau$ are shown in
  Fig.\,\ref{fig:pippimpim}.
  The mass distributions of the hadronic system and the $\pi^-\pi^0$ subsystem for the 
  $\tau^-\to\pi^0\pi^0\pi^-\nu_\tau$ decay are shown in Fig.\,\ref{fig:pi0pi0pim}.
  There is good agreement with \TAUOLA\ for this model which in general gives
  a higher mass for the hadronic system and a slightly broader distribution for
  the masses of the subsystems which include a rho resonance than the model of
  \cite{Jadach:1993hs,Kuhn:1990ad,Decker:1992kj}.

\pagebreak
\subsubsection{Model for modes including Kaons}
\label{sect:threeK}
     
  Like the  model of \cite{Decker:1992kj} the model of \cite{Finkemeier:1995sr}
  is designed to reproduce the correct chiral limit for tau decays to three mesons.
  However, this model makes a different choice of the resonances to use
  away from this limit for the decays involving at least one kaon and in the treatment
  of the $K_1$ resonances.

  The form factors for the different modes are given in \cite{Finkemeier:1995sr}.
  The same form of the $a_1$ Breit-Wigner is used as in 
  Section~\ref{sect:threemesondefaultcurrent}, with the mass and width
  taken from \cite{Finkemeier:1995sr}. The running width we calculate using this
  model with this choice of
  parameters is shown in Fig.\,\ref{fig:a1width1}b and is in good agreement with the
  parameterisation given in \cite{Kuhn:1990ad}.

\begin{figure}
\begin{center}
\includegraphics[width=0.48\textwidth,angle=90]{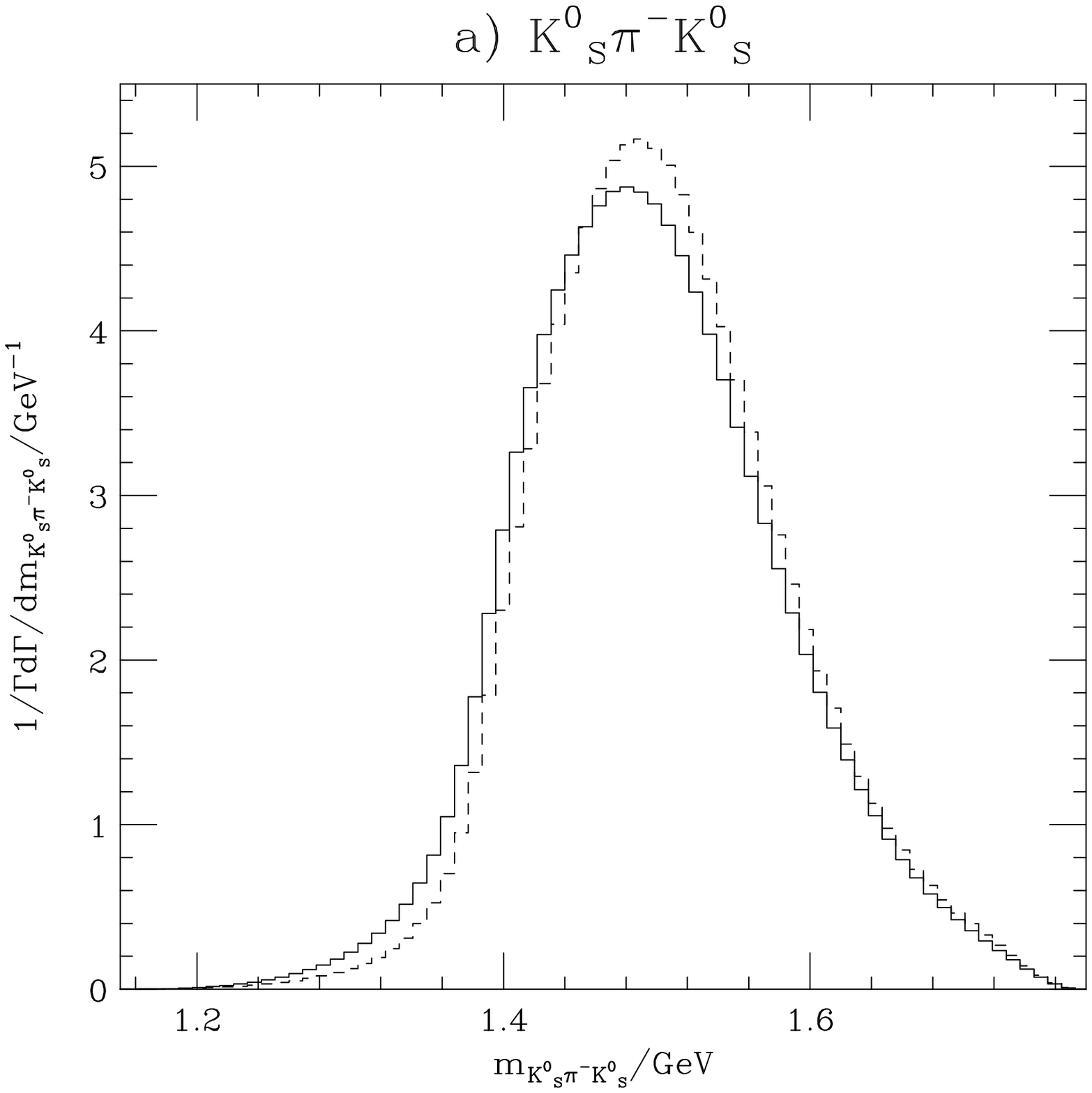}\hfill
\includegraphics[width=0.48\textwidth,angle=90]{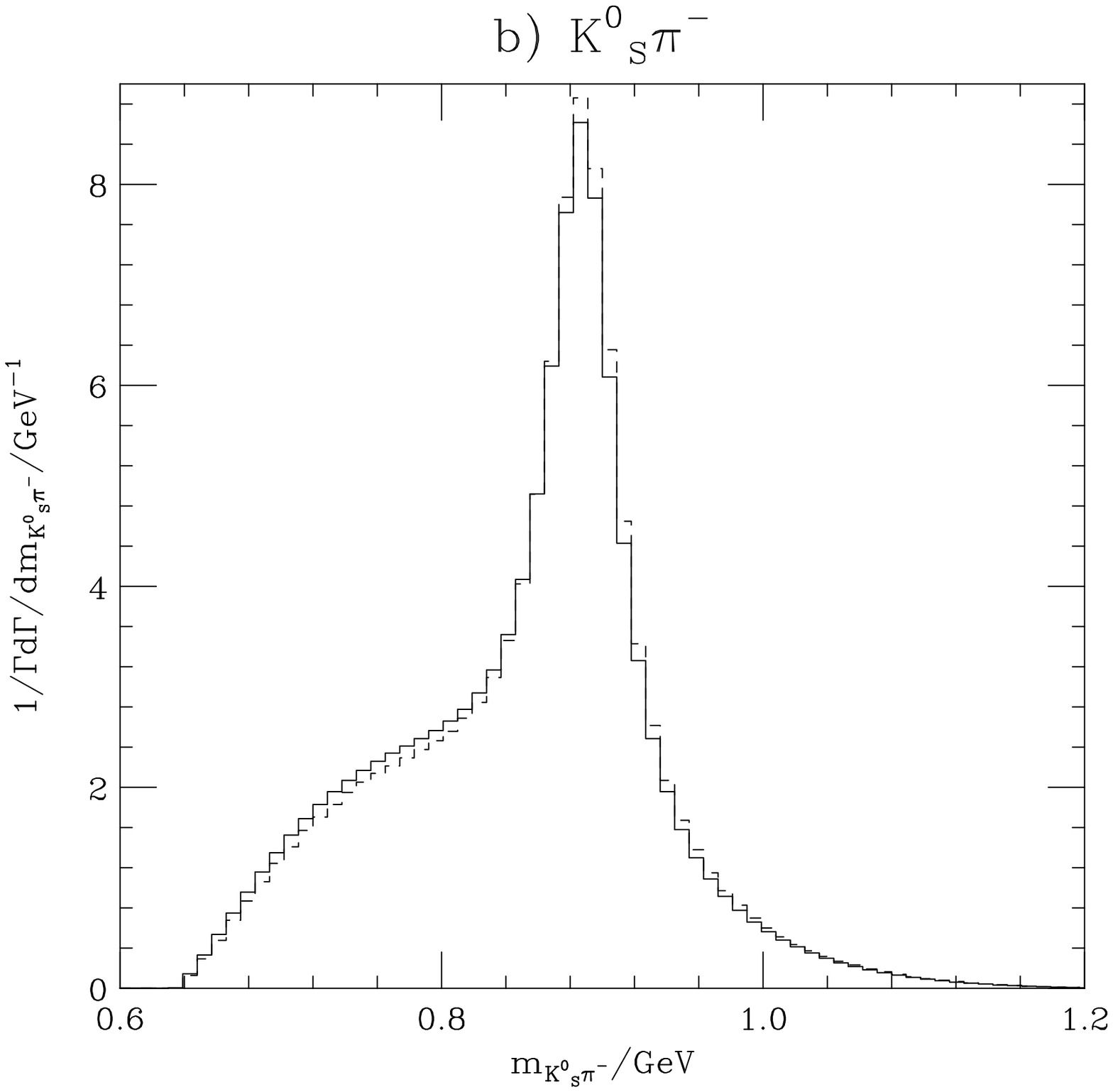}\\
\caption{Differential distribution for the mass of a) $K^0_S\pi^-K^0_S$ and
         b) $K^0_S\pi^-$ produced in the decay 
        $\tau^-\to K^0_S\pi^-K^0_S\nu_\tau$.
        The solid line is the model of \cite{Finkemeier:1995sr}
        and the dashed line is the result of the model of 
        \cite{Jadach:1993hs,Decker:1992kj}.}
\label{fig:kspimks}
\end{center}
\end{figure}
\begin{figure}
\begin{center}
\includegraphics[width=0.48\textwidth,angle=90]{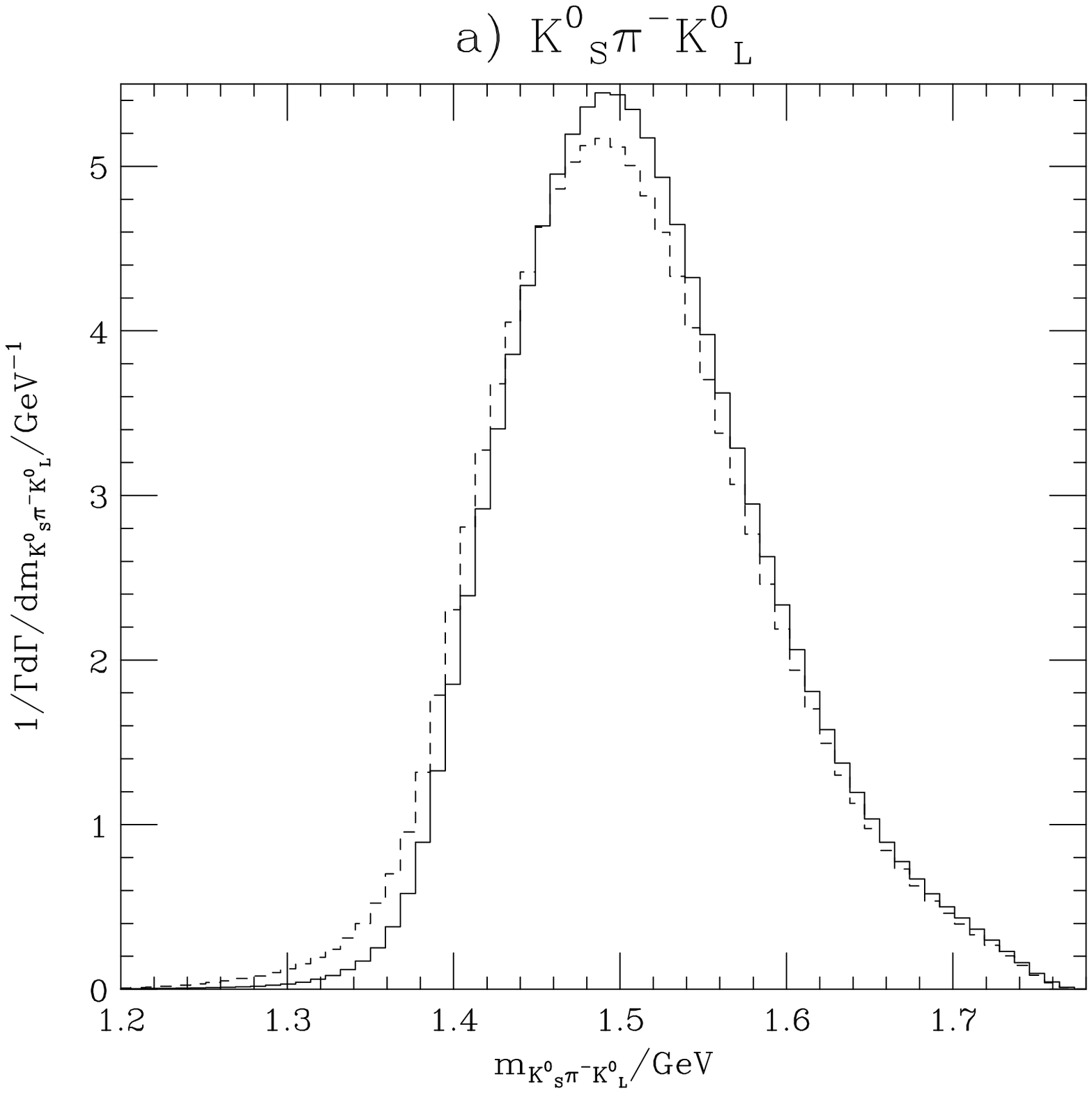}\hfill
\includegraphics[width=0.48\textwidth,angle=90]{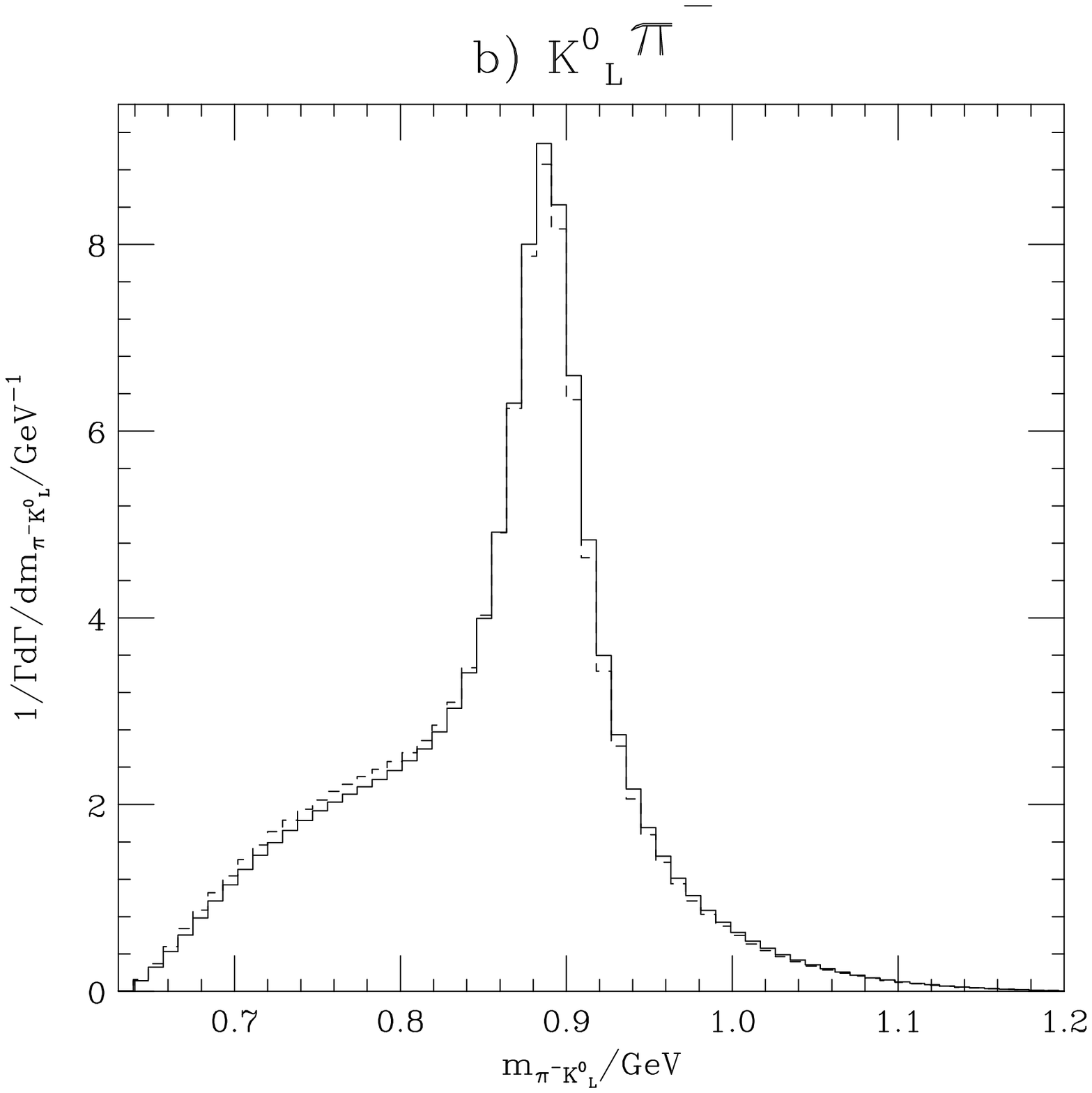}\\
\caption{Differential distribution for the mass of a) $K^0_S\pi^-K^0_L$ and
         b) $K^0_L\pi^-$ produced in the decay 
        $\tau^-\to K^0_S\pi^-K^0_L\nu_\tau$.
        The solid line is the model of \cite{Finkemeier:1995sr}
        and the dashed line is the result of the model of 
        \cite{Jadach:1993hs,Decker:1992kj}.}
\label{fig:kspimkl}
\end{center}
\end{figure}

  As well as the different choice for the resonances contributing to the various decay
  modes this model differs from that of \cite{Decker:1992kj} in the treatment of 
  the $K_1$ resonances. While the model of \cite{Decker:1992kj} assumes that
  only the $K_1(1400)$ contributes the model of \cite{Finkemeier:1995sr} assumes
  that both the $K_1(1270)$ and $K_1(1400)$ contribute with different
  relative contributions to modes including $K^*\pi$ and  
  $K\rho$ intermediate states. All the parameters are taken from
  \cite{Finkemeier:1995sr}.

  The mass distributions for the total hadronic mass and the mass of the $\pi^-K^+$
  subsystem are shown in Fig.\,\ref{fig:kmpimkp} for the decay 
  $\tau^-\to K^-\pi^-K^+\nu_\tau$. The mass distributions for the total
  hadronic mass and the mass of the $\pi^-\bar{K}^0$ system is shown in 
  Fig.\,\ref{fig:k0pimkbar0} for the decay $\tau^-\to K^0\pi^-\bar{K}^0\nu_\tau$.
  The only major difference with the model of
  \cite{Decker:1992kj} for these decay modes 
  is due to the different parameters for the $a_1$ and
  the inclusion of higher $K^*$ resonances.
  The mass distributions for the hadronic system and the $K^-\pi^0$ subsystem 
  for the decay $\tau^-\to K^-\pi^0K^0\nu_\tau$ are shown in Fig.\,\ref{fig:kmpi0k0}.
  In this case the model of \cite{Finkemeier:1995sr} includes $K^*$ resonances in
  the $K^-\pi^0$ and $\pi^0K^0$ subsystems which are not presented in the model
  of \cite{Decker:1992kj} giving different results for the masses of these systems
  and leading to a higher total hadronic mass for this decay mode.

  The mass distributions of the hadronic system and the $\pi^0K^-$ subsystem in the 
  decay $\tau^-\to \pi^0\pi^0K^-\nu_\tau$ are shown in Fig.\,\ref{fig:pi0pi0km}. In 
  this case the main difference between the models of \cite{Finkemeier:1995sr}
  and \cite{Decker:1992kj} is due to the inclusion of the $K_1(1270)$ in the model
  of \cite{Finkemeier:1995sr} which gives the two peak structure seen in the hadronic
  mass spectrum.
  The mass distributions for the hadronic system and the $\pi^+\pi^-$ subsystem for
  the decay $\tau^-\to K^-\pi^-\pi^+\nu_\tau$ are shown in Fig.\,\ref{fig:kmpimpip}.
  As with the $\pi^0\pi^0K^-$ the inclusion of the $K_1(1270)$ gives a different
  spectrum for the mass of the hadronic system when compared with that from the 
  model of \cite{Decker:1992kj}. The CLEO measurement of this mode~\cite{Asner:2000nx}
  favours a significant $K_1(1270)$ contribution but with a much larger width than
  the model of \cite{Finkemeier:1995sr} for the $K_1$.

  The mass distributions for the hadronic system and the $\pi^-\bar{K}^0$ subsystem
  for the decay $\tau^-\to \pi^-\bar{K}^0\pi^0\nu_\tau$ are shown in 
  Fig.\,\ref{fig:pimpi0kbar0}. As with the $K^-\pi^0K^0$ the presence of 
  $K^*$ resonances in the $\pi^-\bar{K}^0$ and $ \bar{K}^0\pi^0$ systems give
  a different mass distribution for these systems, which together with the 
  inclusion of the $K_1(1270)$ gives the different result for the mass of the
  hadronic system.

\begin{figure}
\begin{center}
\includegraphics[width=0.48\textwidth,angle=90]{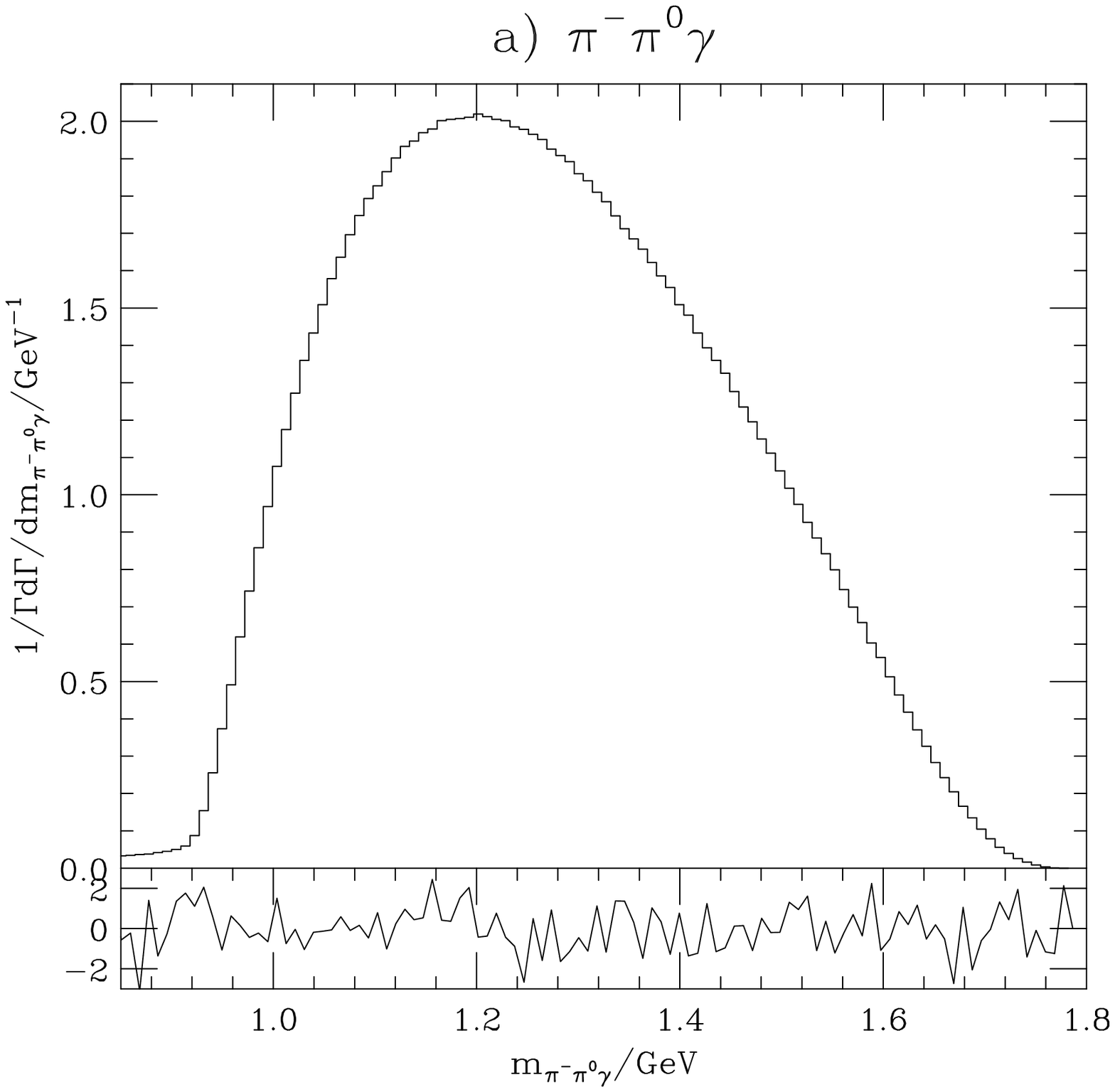}\hfill
\includegraphics[width=0.48\textwidth,angle=90]{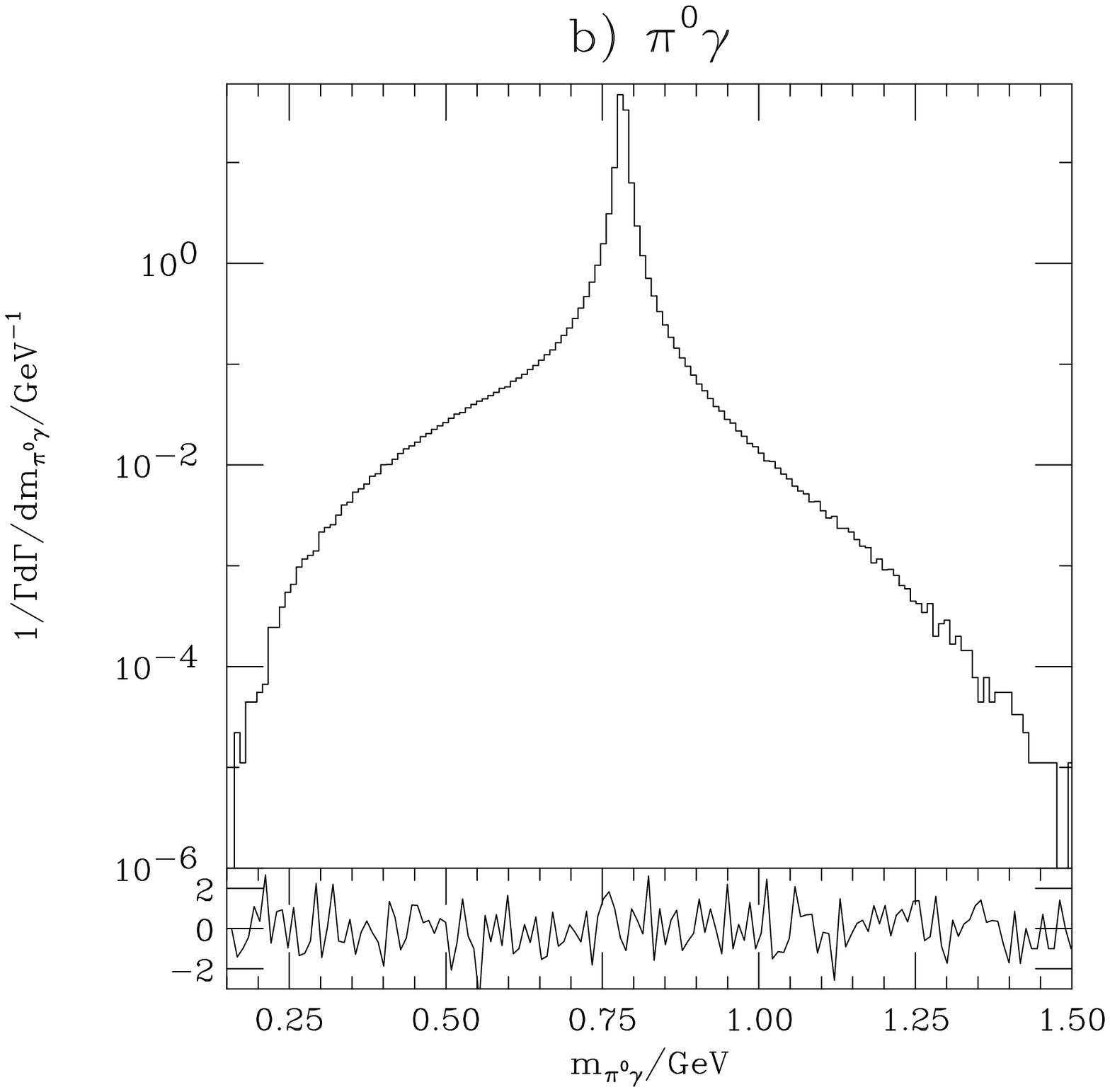}\\
\caption{Differential distribution for the mass of a) $\pi^-\pi^0\gamma$ and
         b) $\pi^-\gamma$ produced in the decay 
        $\tau^-\to \pi^-\pi^0\gamma\nu_\tau$ for  the model of \cite{Jadach:1993hs}.}
\label{fig:pimpi0gamma}
\end{center}
\end{figure}

  The mass distribution for the $K^0_S\pi^-K^0_S$ and $K^0_S\pi^-$ systems in
  the decay \mbox{$\tau^-\to K^0_S\pi^-K^0_S\nu_\tau$} are shown in Fig.\,\ref{fig:kspimks}.
  The mass distribution for the $K^0_S\pi^-K^0_L$ are $K^0_L\pi^-$ systems in the
  decay $\tau^-\to K^0_S\pi^-K^0_L\nu_\tau$ are shown in Fig.\,\ref{fig:kspimkl}.
  The results from the model of \cite{Jadach:1993hs,Decker:1992kj} are also
  shown, where the $K^0$ and $\bar{K}^0$ are allowed to decay with equal probability
  to $K^0_S$ and $K^0_L$ without changing the hadronic current.\footnote{The dashed lines in Figures~\ref{fig:kspimks} and~\ref{fig:kspimkl} are therefore identical.}
  The models are in
  reasonable agreement for the shapes of these distributions.

\subsection{Two Pions and a Photon}
\label{sect:2pigamma}

 The branching ratio for the decay $\tau^-\to\omega\pi^-\nu_\tau$ 
 is 1.95\%~\cite{Yao:2006px}. The majority of this decay is modelled as 
 an intermediate state in the four pion
 current described below. However there is a 8.90\%~\cite{Yao:2006px} branching
 ratio of the $\omega$ into $\pi^0\gamma$ which must also be modelled.
 We do this using a current for $\pi^\pm\pi^0 \gamma$ via
 an intermediate $\omega$. The hadronic current for this mode, together with
 the masses, widths and other parameters, are taken from~\cite{Jadach:1993hs}. 

  The partial width for the $\tau^-\to\pi^0\pi^-\gamma$ mode is given in
  Table.\,\ref{tab:threefour}. The mass distributions
  of the hadronic system and the $\pi^0\gamma$
  subsystem, which contains the $\omega$ resonance,
  are given in Fig.\,\ref{fig:pimpi0gamma}. There is good agreement between \HWPP\
  and \TAUOLA\ for both the partial width and the shapes of the distributions.

\begin{figure}
\begin{center}
\includegraphics[width=0.48\textwidth,angle=90]{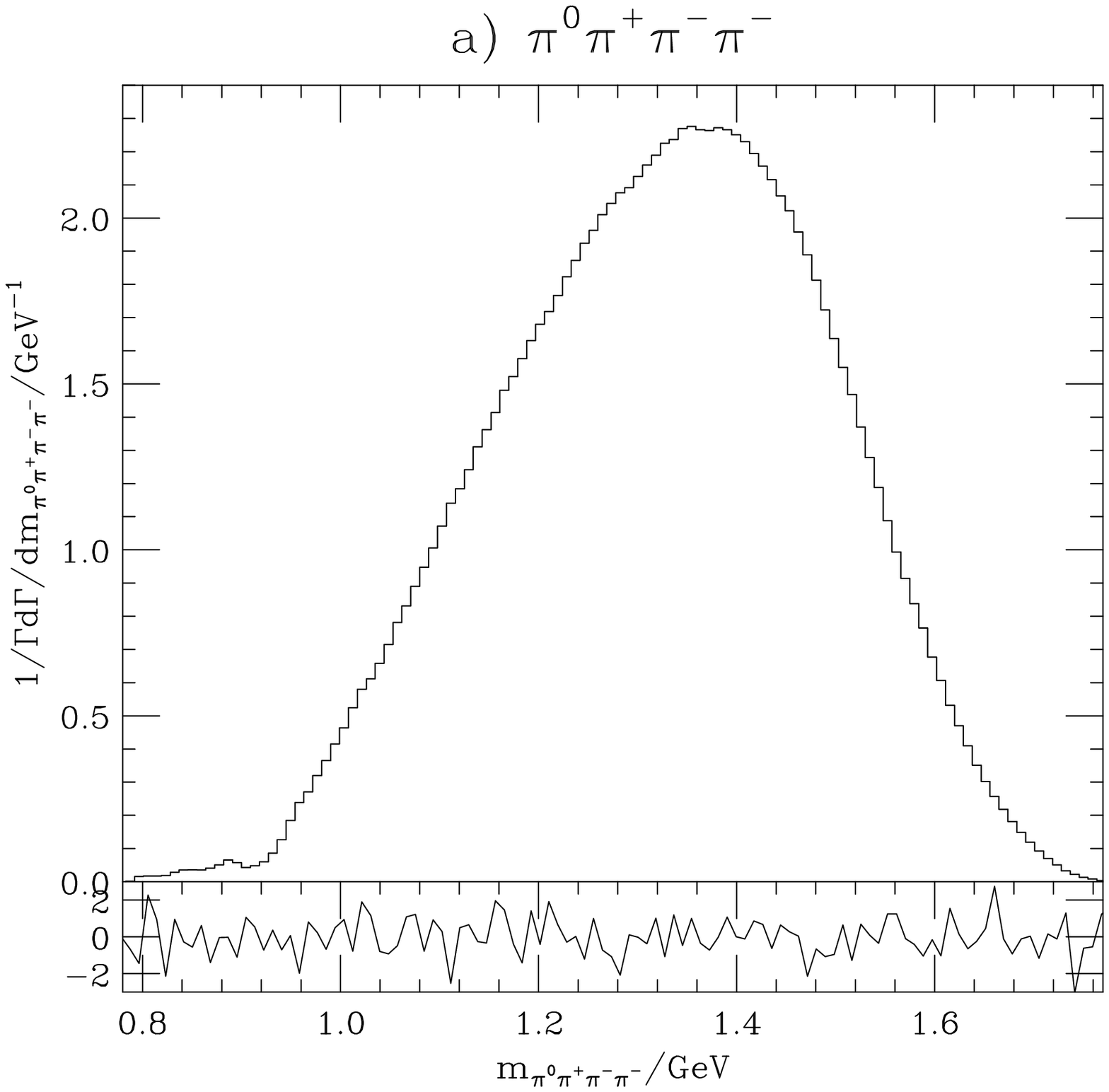}\hfill
\includegraphics[width=0.48\textwidth,angle=90]{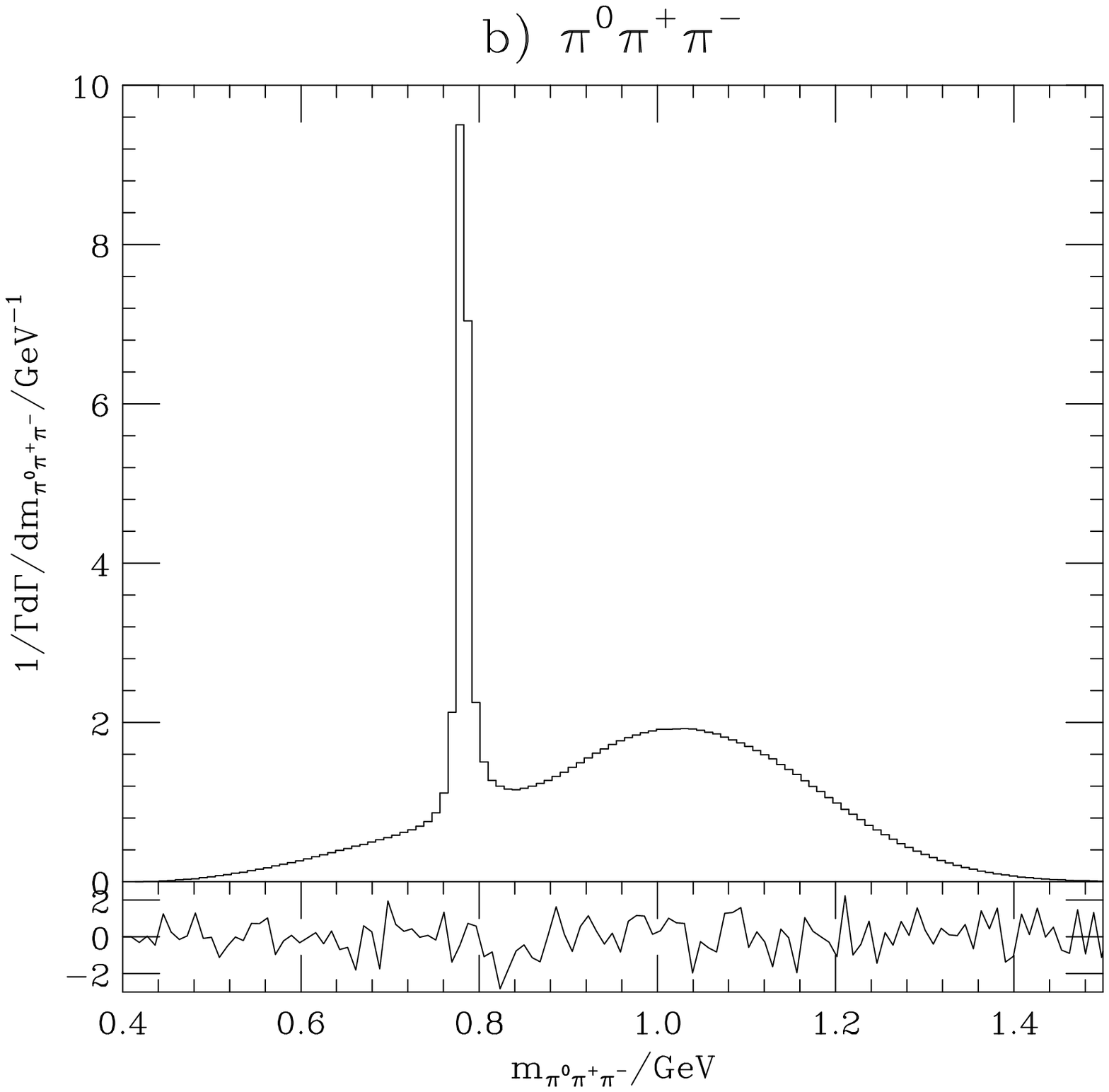}\\
\caption{Differential distribution for the mass of a) $\pi^0\pi^+\pi^-\pi^-$ and
         b) $\pi^0\pi^+\pi^-$ produced in the decay 
        $\tau^-\to \pi^0\pi^+\pi^-\pi^-\nu_\tau$ for the model of \cite{Bondar:2002mw}.}
\label{fig:pimpimpi0pip}
\end{center}
\end{figure}

\begin{figure}
\begin{center}
\includegraphics[width=0.48\textwidth,angle=90]{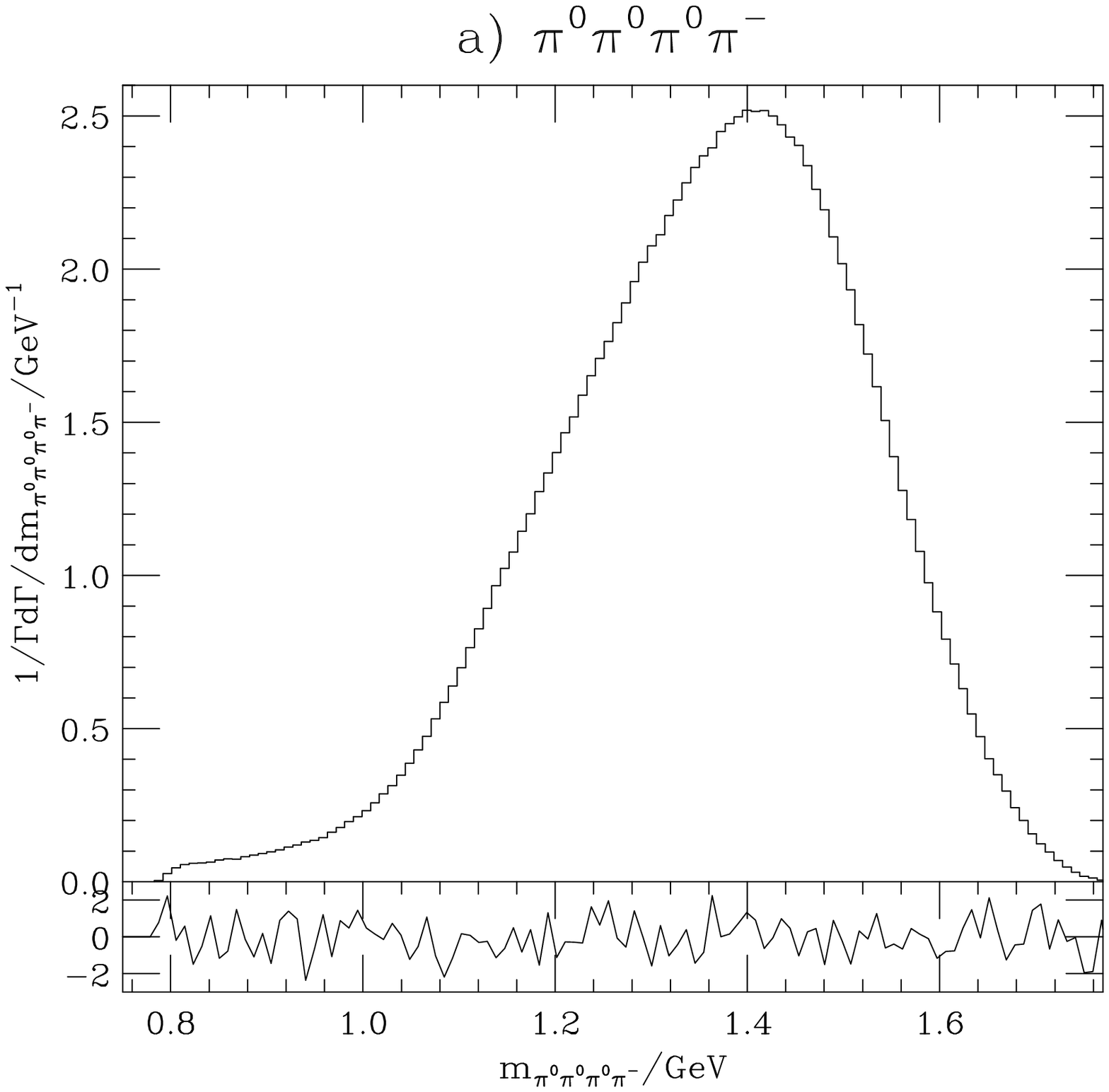}\hfill
\includegraphics[width=0.48\textwidth,angle=90]{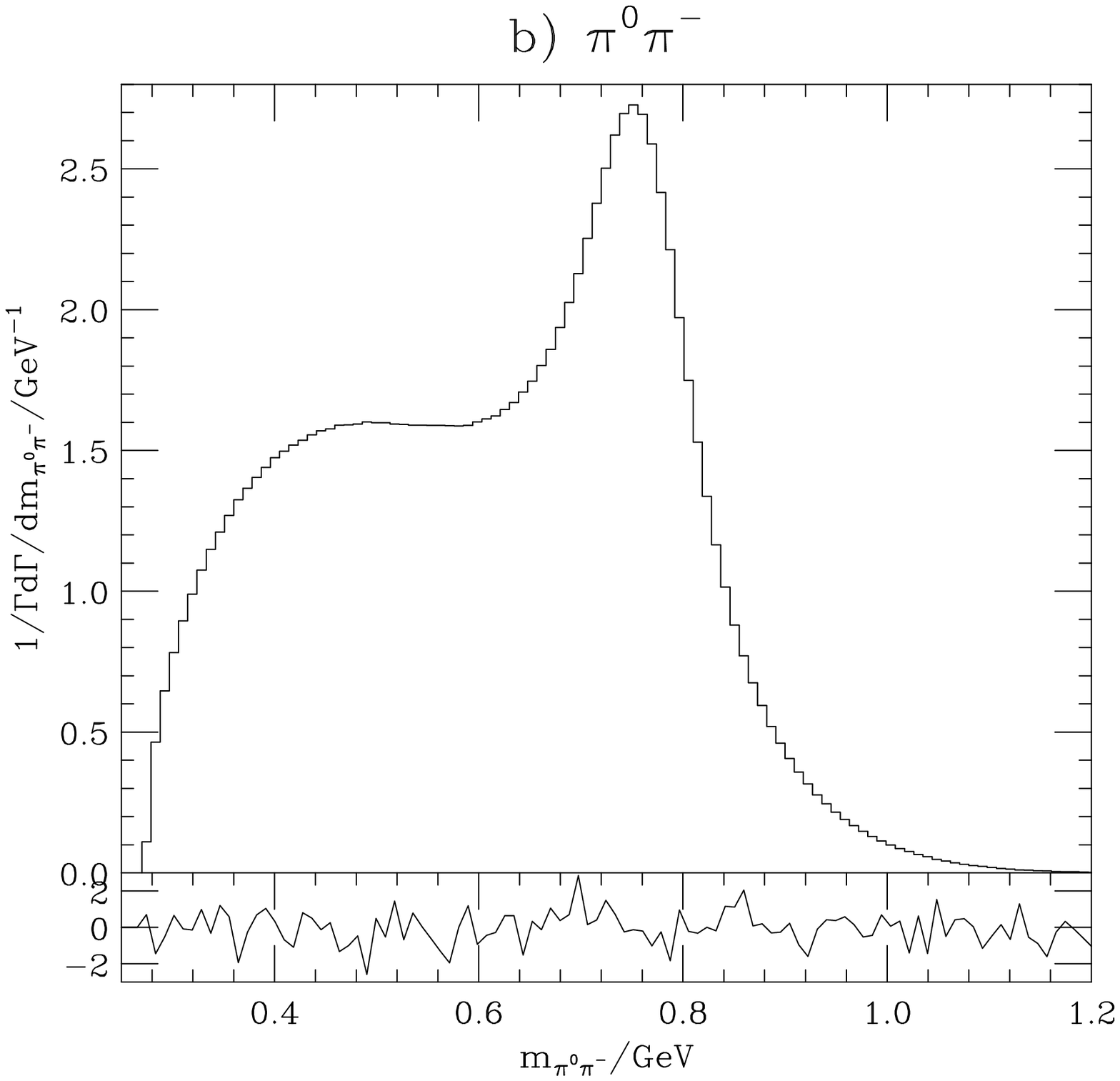}\\
\caption{Differential distribution for the mass of a) $\pi^0\pi^0\pi^0\pi^-$ and
         b) $\pi^0\pi^-$ produced in the decay 
        $\tau^-\to \pi^0\pi^0\pi^0\pi^-\nu_\tau$ for the model of \cite{Bondar:2002mw}.}
\label{fig:pi0pi0pi0pim}
\end{center}
\end{figure}
 
\subsection{Four Pions}
\label{sect:4pi}

  We use the model of \cite{Bondar:2002mw}\footnote{It should be noted that
  there were a number of mistakes in this paper which were corrected in
  \cite{Golonka:2003xt}.} to model the decay of the $\tau$ to four pions.
  The model is based on a fit to $e^+e^-$ data from Novosibirsk.
A Breit-Wigner distribution with a running width, Eqn.\,\ref{eqn:runningBW},
is used for the $\sigma$ and $\omega$ resonances with the running widths
taken from \cite{Golonka:2003xt}. The more complicated form of 
\cite{Golonka:2003xt} is used for the $\rho$ which, apart from the choice to 
normalise to $-1$ at $q^2=0$, is the same as the form of
\cite{Gounaris:1968mw} given in Eqn.\,\ref{eqn:GSBW}.

\begin{figure}
\begin{center}
\includegraphics[width=0.48\textwidth,angle=90]{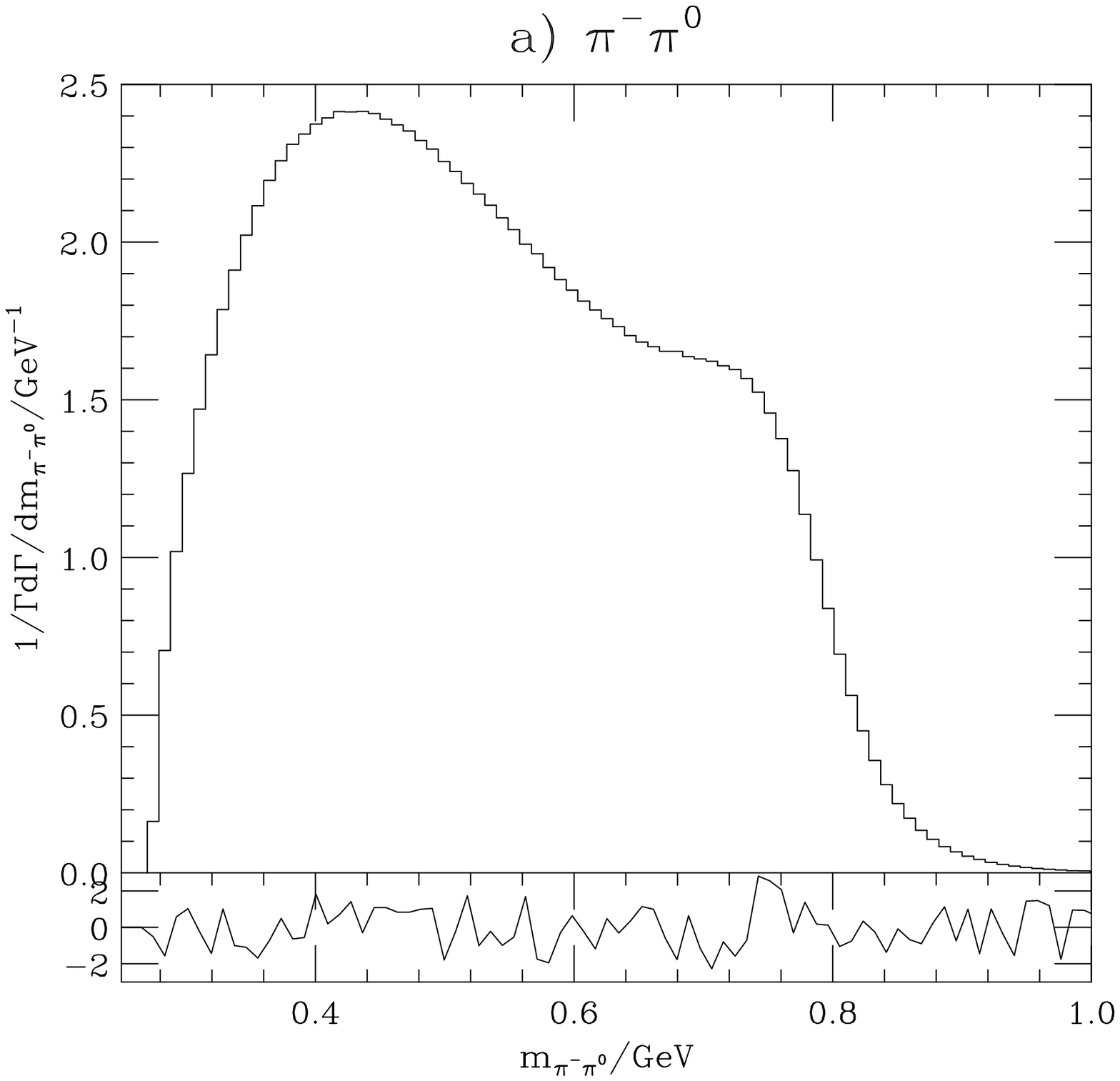}\hfill
\includegraphics[width=0.48\textwidth,angle=90]{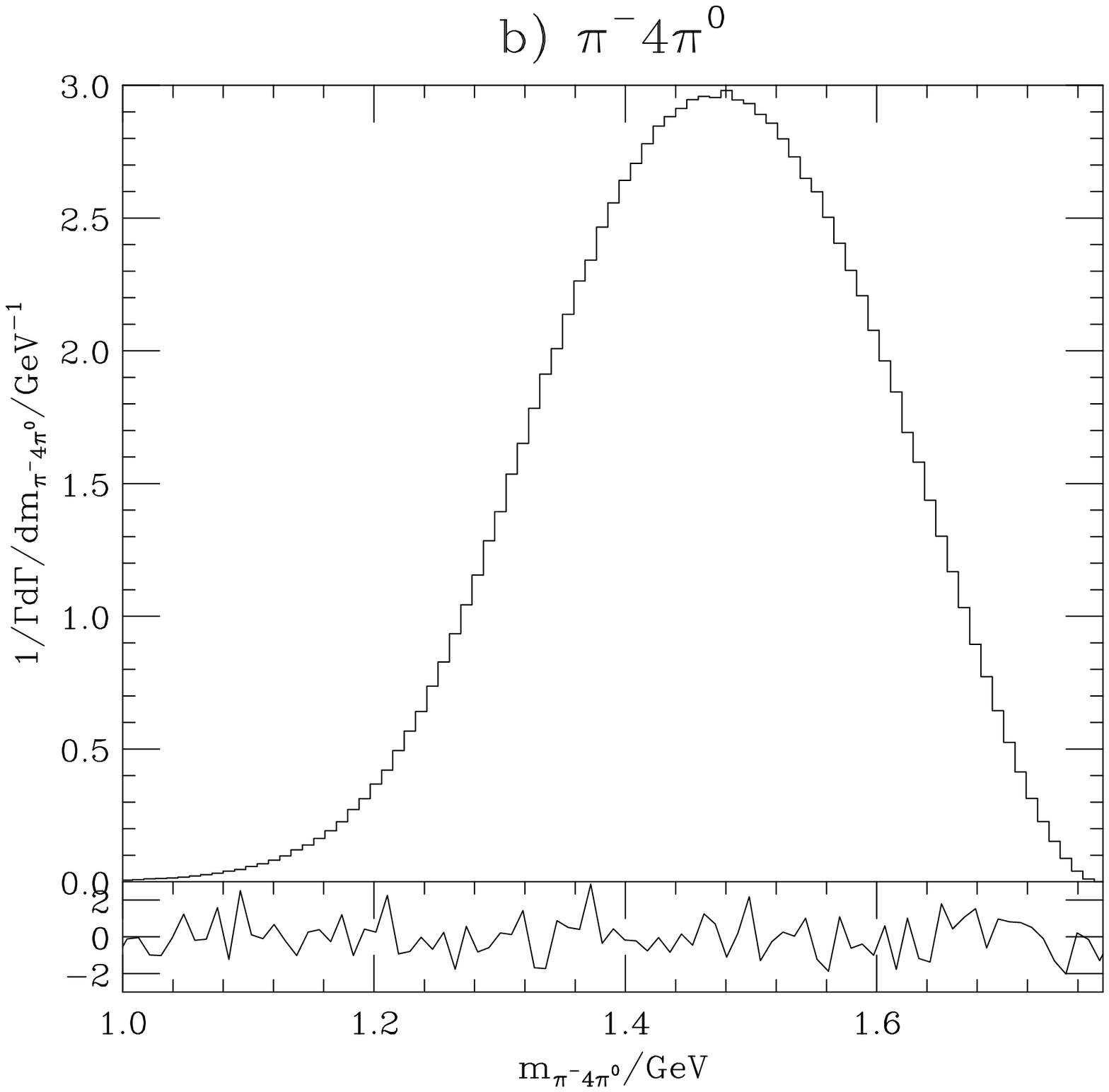}\\
\caption{Differential distribution for the mass of a) $\pi^-\pi^0$ and
         b) $\pi^-4\pi^0$ produced in the decay 
        $\tau^-\to 4\pi^0\pi^-\nu_\tau$ for the model of \cite{Kuhn:2006nw}.}
\label{fig:4pi0pim}
\vspace{5mm}
\includegraphics[width=0.48\textwidth,angle=90]{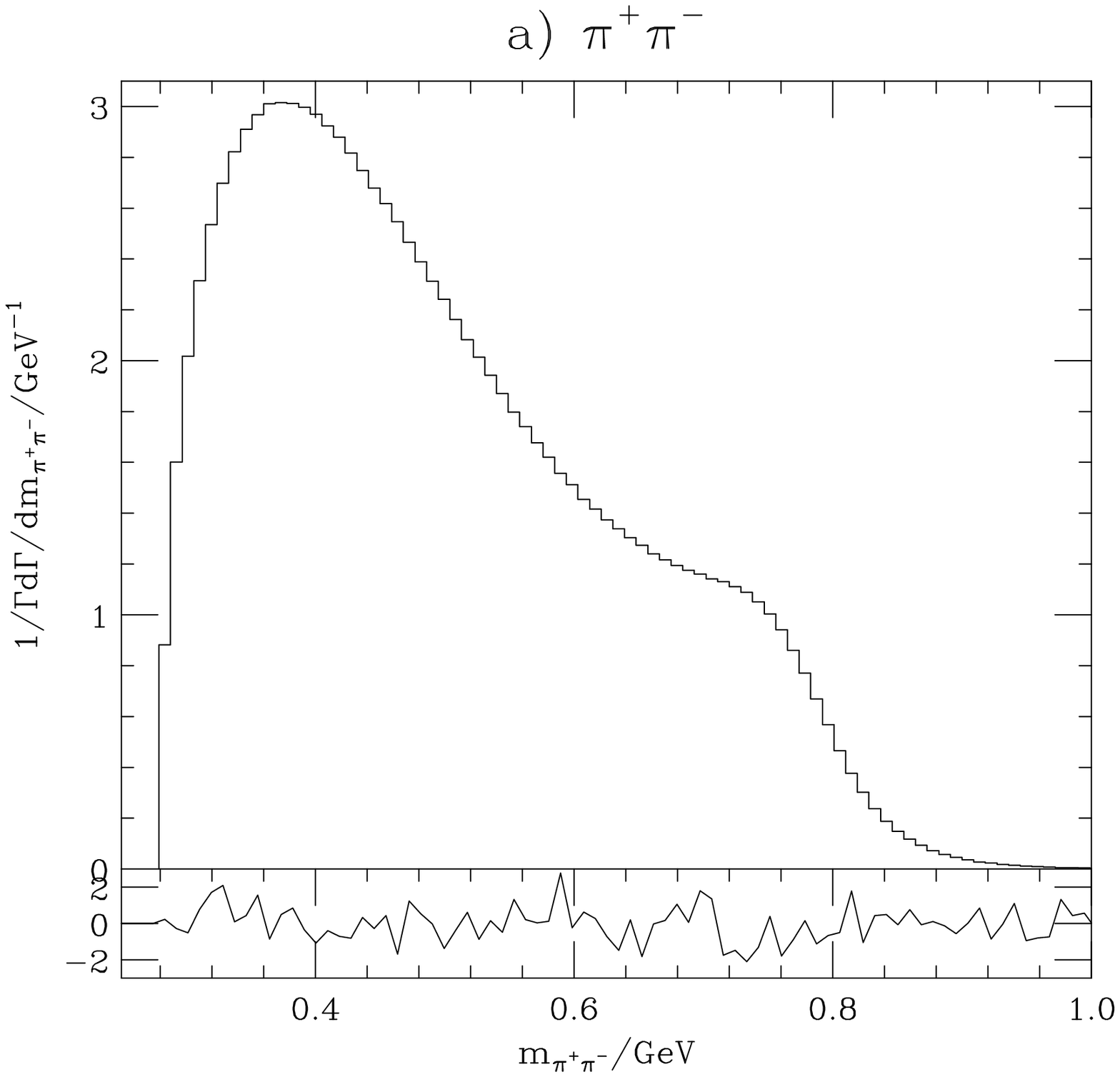}\hfill
\includegraphics[width=0.48\textwidth,angle=90]{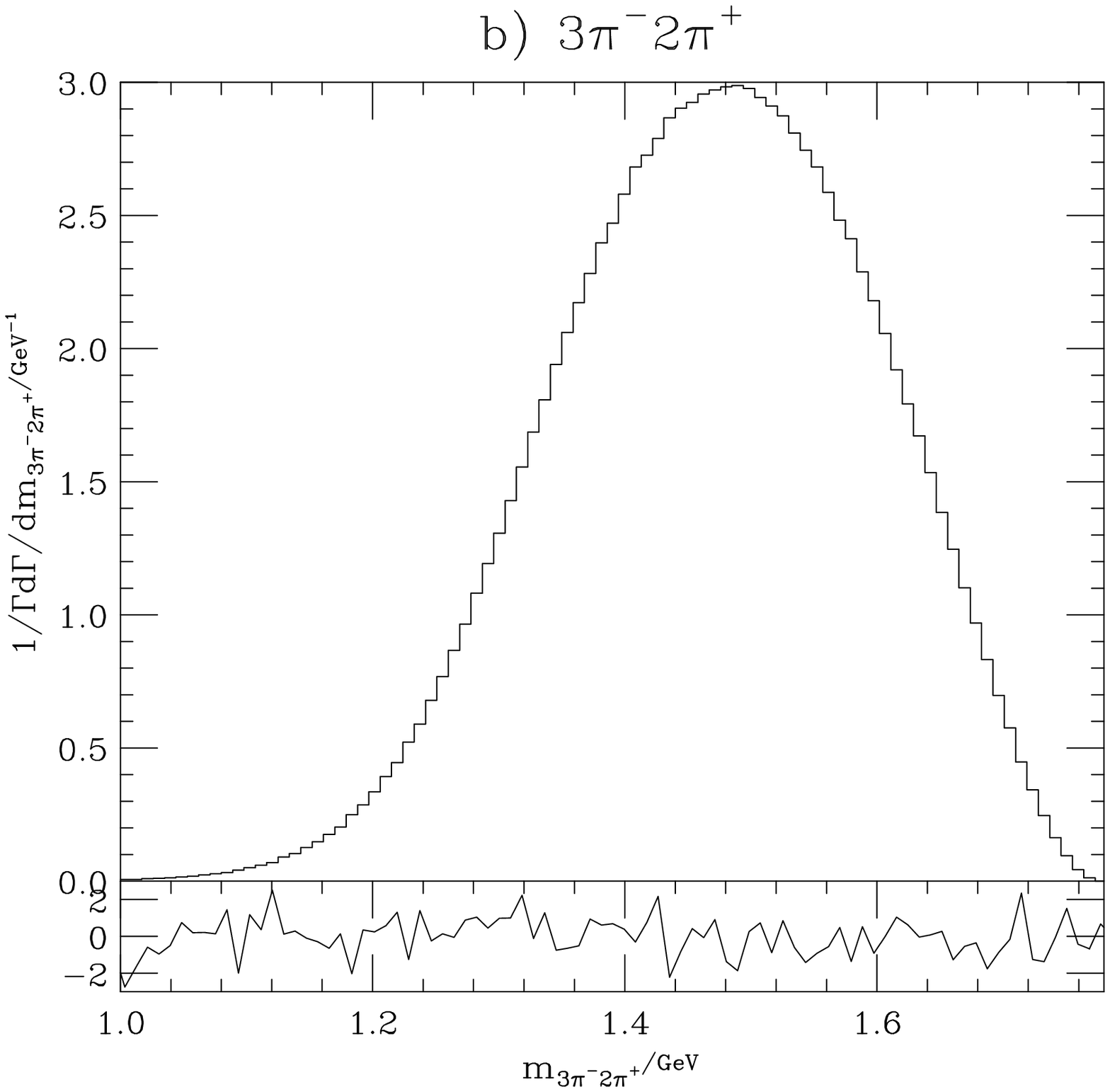}\\
\caption{Differential distribution for the mass of a) $\pi^+\pi^-$ and
         b) $3\pi^-2\pi^+$ produced in the decay 
        $\tau^-\to 3\pi^-2\pi^+\nu_\tau$ for the model of \cite{Kuhn:2006nw}.}
\label{fig:3pim2pip}
\end{center}
\end{figure}

  A Breit-Wigner distribution with a running width is used for the $a_1$ with the 
  running width calculated in \cite{Bondar:2002mw}.
The \HWPP\ calculation of the running width is compared with that from
\TAUOLA\ in Fig.\,\ref{fig:a1width2}b. There is reasonable agreement between
the two calculations apart from in the threshold region.

  The partial widths for the two modes are given in Table~\ref{tab:threefour}.
  The mass distributions for the hadronic system and the $\pi^0\pi^+\pi^-$
  subsystem, which contains the $\omega$ resonance, are shown in
  Fig.\,\ref{fig:pimpimpi0pip} for the decay $\tau^-\to\pi^0\pi^+\pi^-\pi^-\nu_\tau$.
  The mass distributions  for the hadronic system and the $\pi^0\pi^-$ subsystem are
  shown in Fig.\,\ref{fig:pi0pi0pi0pim} for the decay 
  $\tau^-\to\pi^0\pi^0\pi^0\pi^-\nu_\tau$. There is good agreement between
  \HWPP\ and \TAUOLA\ for both the partial widths and mass distributions.

\begin{figure}
\begin{center}
\includegraphics[width=0.48\textwidth,angle=90]{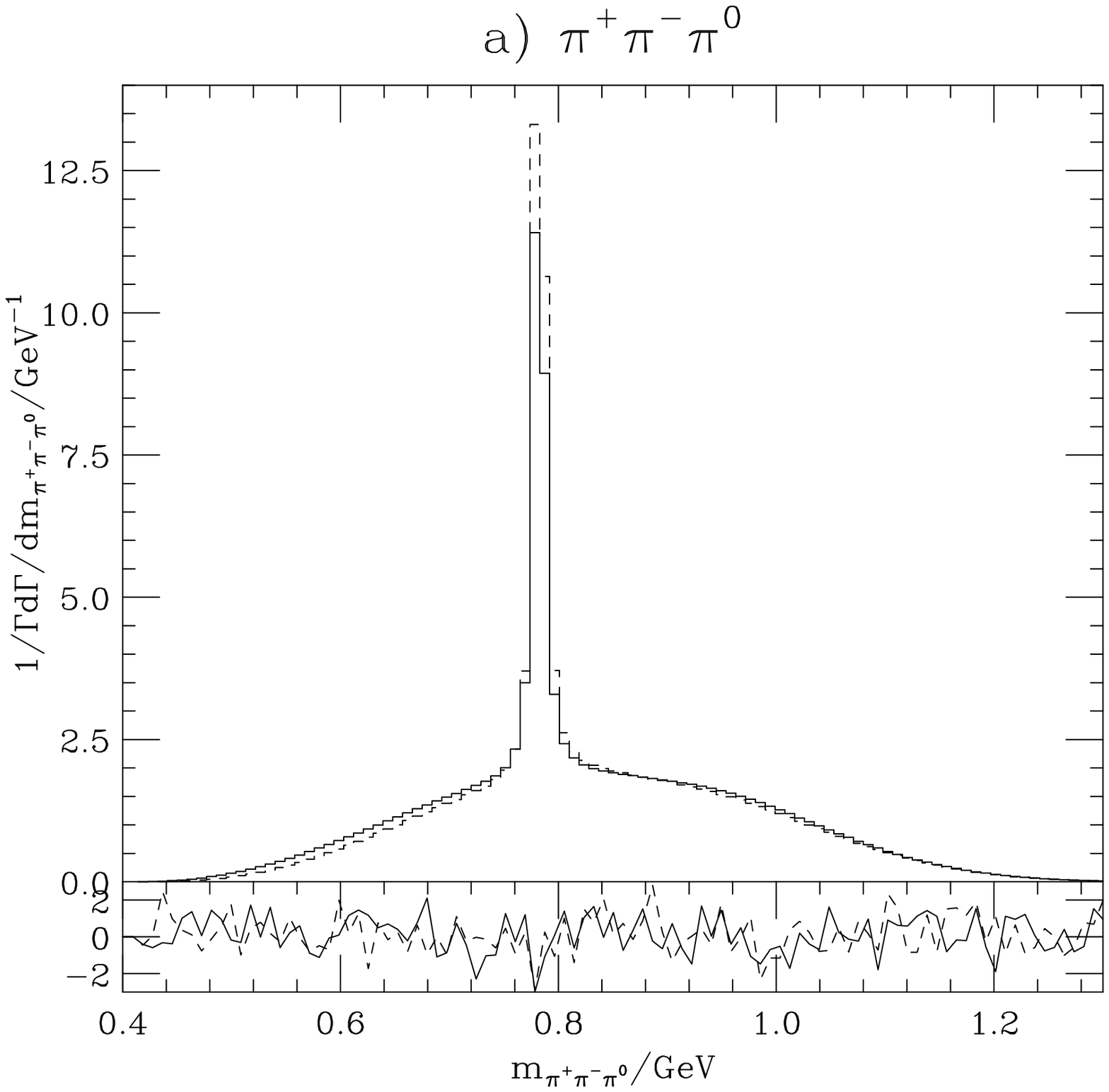}\hfill
\includegraphics[width=0.48\textwidth,angle=90]{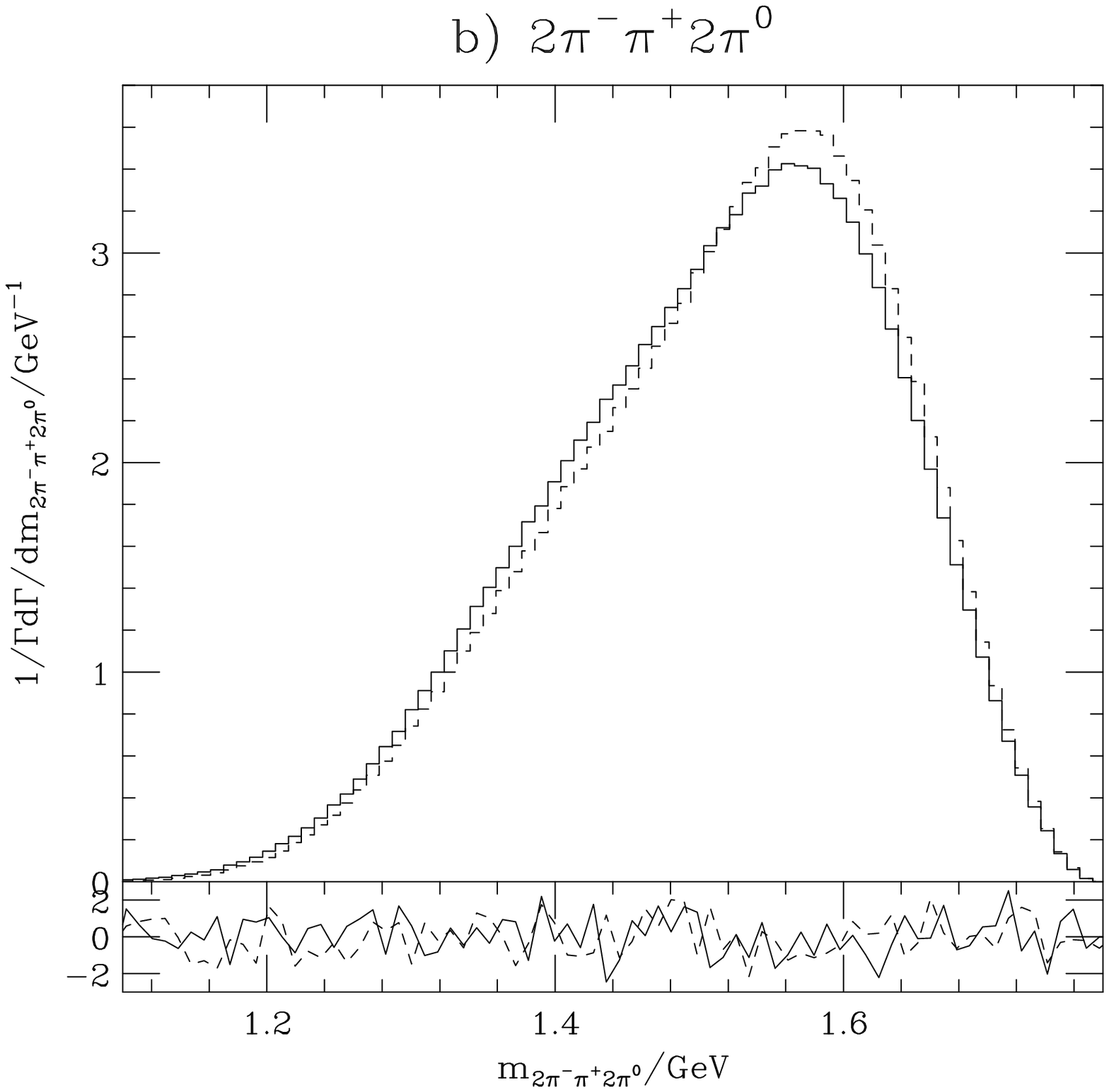}\\
\caption{Differential distribution for the mass of a) $\pi^+\pi^-\pi^0$ and
         b) $2\pi^-\pi^+2\pi^0$ produced in the decay 
        $\tau^-\to 4\pi^0\pi^-\nu_\tau$.
        The solid line is the model of \cite{Kuhn:2006nw} neglecting
        the $\rho$ propagators in the $\omega$ decay and the dashed line includes
        these propagators, by default we do not include the $\rho$ propagators.}
\label{fig:2pimpip2pi0}
\end{center}
\end{figure}

\subsection{Five pions}
\label{sect:5pi}

  We use the model of \cite{Kuhn:2006nw} which includes $\rho\omega$ and
  $\rho\sigma$ intermediate states, via the $a_1$ meson to model the 
  five pion decay modes of the $\tau$.

  The partial widths for these decay modes are compared with those from
  \TAUOLA\ in Table\,\ref{tab:threefour}. The mass distributions of the $\pi^-\pi^0$ 
  subsystem and the hadronic system are shown in Fig.\,\ref{fig:4pi0pim} for the 
  decay $\tau^-\to4\pi^0\pi^-\nu_\tau$. The mass distributions
  of the $\pi^+\pi^-$ subsystem and
  the total hadronic mass for the decay $\tau^-\to3\pi^-2\pi^+\nu_\tau$ are shown
  in Fig.\,\ref{fig:3pim2pip}. The mass distributions
  of the $\pi^+\pi^-\pi^0$ subsystem, which
  includes the $\omega$ resonance, and the total hadronic system are shown for the
  decay $\tau^-\to2\pi^-2\pi^0\pi^+\nu_\tau$ are shown in Fig.\,\ref{fig:2pimpip2pi0}.
  In all cases there is good agreement between the results of \HWPP\ and \TAUOLA\ for
  both the partial widths and the shapes of the 
  distributions.\footnote{As the $\omega\to\pi^+\pi^-\pi^0$ decay is below the threshold for
			  $\rho\pi$ production it can be modeled as either a 
		          contact interaction or via the $\rho\pi$ intermediate state.
	                  The parameters of~\cite{Kuhn:2006nw} are chosen to give the correct 
			  relative rates for the different intermediate states without the 
			  $\rho$ propagators and we take this as our default choice.}

\section{Modelling of Tau Decays}
 
  In our approach the branching ratios for a given decay mode are
  specified as input parameters  rather than
  calculated from the theoretical models used to give the distributions of the decay 
  products, which gives us the ability to adjust the branching ratios
  to the experimentally-observed values.
  In general we have taken the decay modes and branching ratios for tau
  decays from \cite{Yao:2006px}. In \cite{Yao:2006px} a set of basis
  modes is used, the branching ratios for which are constrained to sum to unity.
  In most cases we have used these modes. However, in some cases due to 
  our modelling of the decays we have combined modes. We have also
  included some of the non-basis modes, which forces us to adjust the branching
  ratios slightly from those in \cite{Yao:2006px} in order to ensure that the 
  branching ratios still sum to one. For the higher multiplicity modes where
  charged hadrons are observed, but the different rates for kaons and pions
  are not known, we assume that the hadrons are pions. The decay modes, branching
  ratios and currents used are summarized in Table\,\ref{tab:decaymodes}.
  In addition we have chosen to use the masses from~\cite{Yao:2006px} for the 
  external particles, although we use the default choice of the models we are using,
  which often come from experimental fits, for the intermediate resonances.

\begin{table}[!!p]
\begin{center}
\vspace{-0.8cm}
\begin{tabular}{|c|c|c|}
\hline
Branching & Decay Mode & Current \\
Ratio     &  &         \\ 
\hline
0.178345 & $e^-\bar{\nu}_e\nu_\tau$             & LeptonNeutrinoCurrent \\[-2mm]
0.173545 & $\mu^-\bar{\nu}_\mu\nu_\tau$         & LeptonNeutrinoCurrent \\ 
\hline
0.108924 & $\pi^-\nu_\tau$                      & ScalarMesonCurrent \\[-2mm] 
0.006885 & $K^-\nu_\tau$                        & ScalarMesonCurrent \\
\hline
0.254890 & $\pi^-\pi^0\nu_\tau$                 & TwoMesonRhoKStarCurrent \\[-2mm]
0.008957 & $\bar{K}^0\pi^-\nu_\tau$             & KPiCurrent \\[-2mm]
0.004491 & $K^-\pi^0\nu_\tau$                   & KPiCurrent \\[-2mm]
0.001513 & $K^-K^0\nu_\tau$                     & TwoMesonRhoKStarCurrent \\[-2mm]
0.000263 & $\eta K^-\nu_\tau$                   & TwoMesonRhoKStarCurrent \\
\hline
0.092370 & $\pi^-\pi^0\pi^0\nu_\tau$            & ThreePionCLEOCurrent \\[-2mm] 
0.089813 & $\pi^-\pi^+\pi^-\nu_\tau$            & ThreePionCLEOCurrent \\ 
\hline
0.003757 & $\bar{K}^0\pi^-\pi^0\nu_\tau$        & KaonThreeMesonCurrent \\[-2mm] 
0.003292 & $K^-\pi^-\pi^+\nu_\tau$              & KaonThreeMesonCurrent \\[-2mm] 
0.000555 & $K^-\pi^0\pi^0\nu_\tau$              & KaonThreeMesonCurrent \\ 
\hline
0.001519 & $K^-K^+\pi^-\nu_\tau$                & KaonThreeMesonCurrent \\[-2mm] 
0.001518 & $K^-K^0\pi^0\nu_\tau$                & KaonThreeMesonCurrent \\[-2mm] 
0.001087 & $K^0_LK^0_S\pi^-\nu_\tau$            & KaonThreeMesonCurrent \\[-2mm] 
0.000235 & $K^0_LK^0_L\pi^-\nu_\tau$            & KaonThreeMesonCurrent \\[-2mm] 
0.000235 & $K^0_SK^0_S\pi^-\nu_\tau$            & KaonThreeMesonCurrent \\ 
\hline
0.044435 & $\pi^-\pi^+\pi^-\pi^0\nu_\tau$       & FourPionNovosibirskCurrent \\[-2mm] 
0.010313 & $\pi^-\pi^0\pi^0\pi^0\nu_\tau$       & FourPionNovosibirskCurrent \\[-2mm] 
0.001762 & $\pi^-\pi^0\gamma\nu_\tau$           & TwoPionPhotonCurrent \\
\hline
0.004935 & $\pi^-\pi^-\pi^+\pi^0\pi^0\nu_\tau$  & FivePionCurrent\\[-2mm]  
0.001744 & $\eta\pi^-\pi^0\nu_\tau$             & ThreeMesonDefaultCurrent \\[-2mm] 
0.000957 & $\pi^-\pi^0\pi^0\pi^0\pi^0\nu_\tau$  & FivePionCurrent \\[-2mm] 
0.000834 & $\pi^-\pi^-\pi^-\pi^+\pi^+\nu_\tau$  & FivePionCurrent \\ 
\hline
0.000225 & $\eta\pi^-\pi^-\pi^+\nu_\tau$        & Phase Space \\[-2mm] 
0.000145 & $\eta\pi^-\pi^0\pi^0\nu_\tau$        & Phase Space \\[-2mm] 
0.000135 & $\omega\pi^-\pi^0\pi^0\nu_\tau$      & Phase Space \\[-2mm] 
0.000118 & $\omega\pi^-\pi^-\pi^+\nu_\tau$      & Phase Space \\ 
\hline
0.000400 & $K^-\omega\nu_\tau$                  & Phase Space \\[-2mm] 
0.000397 & $K^-\pi^0\pi^0\pi^0\nu_\tau$         & Phase Space \\[-2mm] 
0.000307 & $K^-\pi^+\pi^-\pi^0\nu_\tau$         & Phase Space \\[-2mm] 
0.000280 & $\eta K^{*-}\nu_\tau$                & Phase Space \\[-2mm] 
0.000238 & $\bar{K}^0\pi^-\pi^0\pi^0\nu_\tau$   & Phase Space \\[-2mm] 
0.000225 & $\bar{K}^0\pi^-\pi^-\pi^+\nu_\tau$   & Phase Space \\ 
\hline
0.000297 & $K^0\bar{K}^0\pi^-\pi^0\nu_\tau$     & Phase Space \\[-2mm] 
0.000059 & $K^-K^+\pi^-\pi^0\nu_\tau$           & Phase Space \\ 
\hline
\end{tabular}
\end{center}
\vspace{-0.3cm}
\caption{Decay modes and branching ratios used for $\tau$ decays in \HWPP\ together
         with the model of the hadronic current used for the decay.}
\label{tab:decaymodes}
\vspace{-1cm}
\end{table}

  The branching ratios for the leptonic modes are taken from \cite{Yao:2006px} 
  and the distribution of the decay products is
  modelled using the \textsf{LeptonNeutrinoCurrent} described in
  Section~\ref{sect:leptonneutrinocurrent}. The decays to a single
  charged meson are modelled using the \textsf{ScalarMesonCurrent}, described in 
  Section~\ref{sect:scalarmesoncurrent}, with 
  the branching ratios taken from \cite{Yao:2006px}.
  The $K\pi$ modes are modelled using \textsf{KPiCurrent}, described in Section~\ref{sect:kpicurrent},
  and the parameters from
  the fit of \cite{Epifanov:2007rf}.
  The remaining two meson modes are modelled using the \textsf{TwoMesonRhoKStarCurrent},
  described in Section~\ref{sect:twomeson},
  and the branching ratios from \cite{Yao:2006px}, the resonance
  parameters for the different modes are described in Section~\ref{sect:twomeson}.

  The three-pion modes are modelled using the \textsf{ThreePionCLEOCurrent}, as described in 
  Section~\ref{sect:3pi}, together
  with the branching ratios from \cite{Yao:2006px}.
  We take the branching ratios for modes with three mesons, at least one of which 
  is a kaon, from \cite{Yao:2006px} and use the \textsf{KaonThreeMesonCurrent},
  as described in Section~\ref{sect:threeK},
  to model the distribution of the decay products.

  We split the observed $\tau^-\to\omega\pi\nu_\tau$ rate into its dominant pieces,
  i.e. \mbox{$\omega\to\pi^+\pi^-\pi^0$} and $\omega\to\pi^0\gamma$. The three pion
  mode is included with the rest of the four pion tau decays and is modelled using
  the \textsf{FourPionNovosibirskCurrent}, described in Section~\ref{sect:4pi},
  whereas the $\pi^0\gamma$ is modelled
  using the \textsf{TwoPionPhotonCurrent}, described in Section~\ref{sect:2pigamma}.
  The branching ratios are taken from
  \cite{Yao:2006px}. In general the treatment of non-dominant modes
  is often a problem when we include resonances 
  as intermediate particles in the hadronic currents. However, in the other cases,
  such as the $\omega$ component to the five pion decays, the contribution of
  the other modes is smaller and comparable with other neglected tau decay modes and
  is not included.

  The five pion decays for the tau are modelled using the \textsf{FivePionCurrent},
  described in Section~\ref{sect:5pi}.
  As with the four-pion modes we include the $\omega$ contribution with the rest of
  the five-pion modes as the current includes modelling of the intermediate $\omega$
  contribution. In this case we neglect the sub-dominant $\omega$ modes due to their
  smaller contribution. The branching ratios are taken from \cite{Yao:2006px}.

  The inclusion of the $\eta3\pi$ and $\omega3\pi$ modes is sufficient to 
  saturate the observed six pion rates from \cite{Yao:2006px} and we therefore
  include these modes using a phase-space distribution for the decay products.

  We also include a number of modes with one kaon and three pions, or two kaons
  and two pions and $K^*\eta$ using a phase-space distribution for the decay products
  and the branching ratios from \cite{Yao:2006px}. The largest branching ratio
  for one of these modes is $K\omega$ with a branching ratio of $0.0004$ and the sum of
  the branching ratios for these modes is less than three per mille.
  This means that they make a 
  relatively small contribution and if necessary the simulation can be improved by 
  implementing hadronic currents for these modes.

  As the decay modes, the decay models and branching ratios are specified in a data
  file these can easily be changed given new experimental results or better
  modelling of the hadronic currents.

\section{Other Applications}

  The hadronic currents obtained from $\tau$ decay can be used to describe
  other decays which occur via the weak current. It is an important test of the
  new structure of decays and weak currents in \HWPP\ that the currents can be used
  to simulate these decays.

  One obvious application is
  to use the currents, together with the na\"{\i}ve factorization approximation, to simulate
  the weak hadronic decays of bottom and charm 
  hadrons.\footnote{We will use this approach in the simulation of hadronic decays
                    in \HWPP~\cite{Mesondecays}.}
  Here we will consider another possible use of these currents, the simulation
  of the weak decay of BSM particles where there is a small mass difference between
  two of the particles. One such example of this is the decay of the lighter chargino,
  which is almost mass degenerate with the lightest neutralino, in AMSB models.
  In most studies of these models only the leptonic
  and single-pion modes are included, in some more sophisticated studies the 
  two- and three-pion modes were also 
  considered~\cite{Chen:1996ap,Barr:2002ex}. 

\begin{figure}
\begin{center}
\includegraphics[width=0.45\textwidth,angle=90]{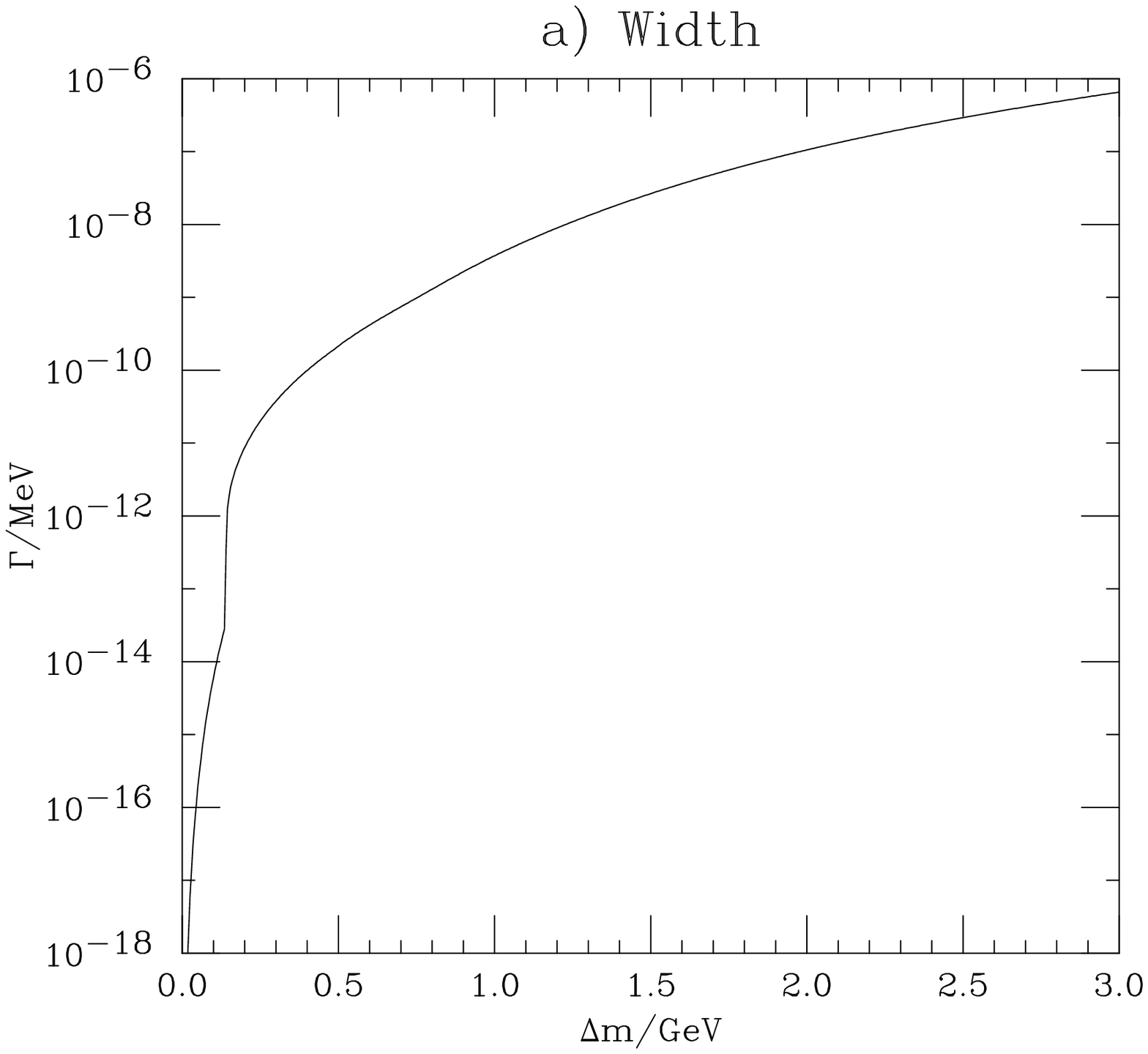}\hfill
\includegraphics[width=0.45\textwidth,angle=90]{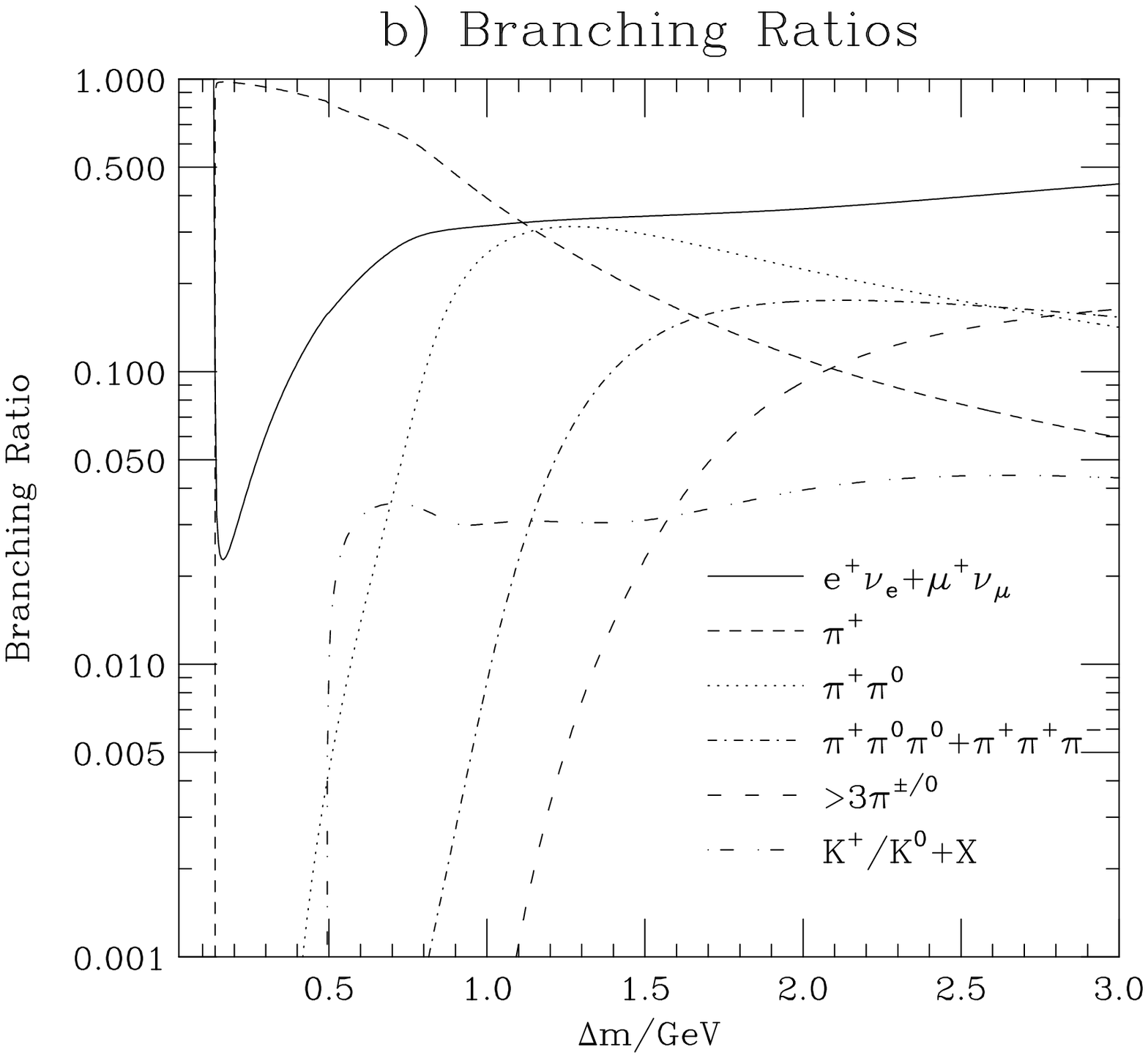}\\
\caption{The a) width and b) branching ratios of the chargino as a function
         of the mass difference between the chargino and lightest neutralino.
         The solid line shows the sum of the leptonic modes, 
         \ie $\chi^+_1\to\chi^0_1e^+\nu_e$ and $\chi^+_1\to\chi^0_1\mu^+\nu_\mu$. 
         The dashed line shows the pion mode $\chi^+_1\to\chi^0_1\pi^+$ and the dotted
         shows the $\chi^+_1\to\chi^0_1\pi^+\pi^0$ mode. The dot-dashed line shows
         the sum of the $\chi^+_1\to\chi^0_13\pi$ modes. The long dashed line
        shows the sum of $\chi^+_1\to\chi^0_1\{4,5\}\pi$,
        $\chi^+_1\to\chi^0_1\pi^+\pi^0\gamma$, and
        $\chi^+_1\to\chi^0_1\pi^+\pi^0\eta$ modes. The long dot-dashed line
        shows the sum of the modes involving kaons.
}
\label{fig:chargino}
\end{center}
\end{figure}

  Using the hadronic currents from tau decays it is easy to extend the simulation
  of BSM decays in \cite{Gigg:2007cr} to automatically calculate the 
  partial widths and branching ratios for decays with $X\to Y W^\pm$ where the
  $W^\pm$ is highly off-shell. These decays are calculated by using the 
  \textsf{Vertices} for the BSM model, which encode the Feynman rules, and the hadronic
  current. The partial widths are normalised in order to give the correct branching
  ratios for the $\tau$ with the modes which are modelled using a phase-space
  distribution neglected. This is a reasonable approximation given the relatively
  small contribution of these modes. As the partial widths depend on both the overall mass scale,
  as well as the mass difference, the branching ratios, while often close to, are 
  not necessarily the same as those of the $\tau$ when the mass difference is equal to 
  the tau mass. In practice, this approach is normally only used
  for small values of the mass splitting and matched to the perturbative calculation
  with outgoing quarks at a mass difference below the charm threshold.

  We illustrate this by considering the decay $\chi^+_1\to\chi^0_1X$ at the AMSB parameter point 
  SPS9~\cite{Allanach:2002nj}. The SUSY
  parameters were generated using \textsf{SOFTSUSY}~\cite{Allanach:2001kg}.
  The mass difference between the lightest neutralino and chargino is very 
  sensitive to higher-order corrections, this can be seen by the large
  difference between the values obtained using \textsf{SOFTSUSY}~($1.167$\,GeV)
  and \textsf{ISAJET}7.58~\cite{Baer:1999sp}~($0.167$\,GeV), which was used 
  in~\cite{Allanach:2002nj}.
  Given the uncertainty in the mass difference, and to show dependence
  of the results upon it, we present the width and branching ratios 
  as a function of the mass difference, Fig.\,\ref{fig:chargino},
  between the neutralino and chargino by
  keeping all the other SUSY parameters fixed to the values given by~\textsf{SOFTSUSY}\ 
  and varying the chargino mass. As can be seen, for larger values of the mass difference the multi-pion 
  modes~($>3$ pions) dominate, although at lower values they typically make a small
  contribution.

\section{Conclusion}
\label{sect:conclude}

  We have described the simulation of tau decays in the \HWPP\ event generator.
  A number of tests of the simulation have been performed giving us confidence
  in the implementation of the various hadronic currents and accuracy of the 
  simulation. 

  This new simulation makes use of the factorization of the matrix element and 
  the inheritance mechanism of C++ to produce a simulation which is easy to 
  extend by implementing new hadronic currents. The interface feature of the 
  \textsf{ThePEG} framework \cite{Bertini:2000uh},
  on which \HWPP\ is based, allows easy access
  to all the parameters of the hadronic currents, allowing different experimental
  fits to be used without changing any code.

  The range of hadronic currents that are included together with the simulation
  of spin correlation effects
  in all tau decays gives a state-of-the-art description of tau decays. The new
  code structure will make this simulation easy to maintain and develop in the
  future.

\section*{Acknowledgments}

  We are grateful to the other members of the \HWPP\ collaboration for both 
  their contributions to the development of the program which underlies this work
  and many useful discussions. This work was supported by Science and Technology
  Facilities Council, formally the Particle Physics and Astronomy Research Council,
  and the European Union Marie Curie Research Training Network MCnet under
  contract MRTN-CT-2006-035606.

\appendix

\section{Breit-Wigner Distributions}

  A number of different forms of the Breit-Wigner distribution are used in the 
  various hadronic currents.
  The simplest form of the distribution, which is rarely used, is a fixed width
  Breit-Wigner distribution
\begin{equation}
{\rm B}(s) = \frac{M^2-iM\Gamma}{M^2-s-iM\Gamma},
\label{eqn:fixedBW}
\end{equation}
  where $s$ is the virtual mass squared of the resonance, $M$ is the mass of the 
  resonance and $\Gamma$ is the width. As with many of the forms we use, this
  is designed so that ${\rm B}(0)=1$.

  In most cases we make use of a more sophisticated form including a running
  width for the particle, $\Gamma(s)$:
\begin{equation}
{\rm BW}(s) = \frac{M^2}{M^2-s-i\sqrt{s}\,\Gamma(s)},\label{eqn:runningBW}
\end{equation}
  The details of this
  form of the Breit-Wigner distribution depend on the choice of the running
  width. If the decays of the resonance are dominated by a simple two-body decay,
  the running width is given by
\begin{subequations}
\begin{eqnarray}
\Gamma^{\rm s-wave}(s) &=& \Gamma\,\frac{M^2}s\frac{p(s)}{p(M^2)},\label{eqn:swavewidth}\\
\Gamma^{\rm p-wave}(s) &=& \Gamma\,\frac{M^2}s\left(\frac{p(s)}{p(M^2)}\right)^3,\label{eqn:pwavewidth}\\
\Gamma^{\rm d-wave}(s) &=& \Gamma\,\frac{M^2}s\left(\frac{p(s)}{p(M^2)}\right)^5,\label{eqn:dwavewidth}
\end{eqnarray}
\end{subequations}
  where $\Gamma$ is the physical width and $p(s)$ is the momentum for the 2-body
  decay of a particle of virtual mass $s$ in the rest frame of the particle. The angular momentum of
  the decay products is denoted as s-, p-, and d-wave for spin 0, 1, and 2 systems.
  For particles where the dominant decay is a three-body mode, for example the $a_1$
  meson, a more complicated form of the running width is often used. In this case 
  we still have ${\rm BW}(0)=1$, as the running width is zero below the threshold
  for the decay modes.

  In some cases for the $\rho$ meson we use the form of
  Gounaris and Sakurai~\cite{Gounaris:1968mw}, which includes
  a running mass correction derived from an effective range formula
  for the p-wave $\pi\pi$ scattering phase,
\begin{equation}
{\rm BW}(s) = \frac{M^2+dM\Gamma}
{M^2-s+\Gamma \frac{M^2}{k_M^3}\left\{k^2_s\left[h(s)-h(M^2)\right]+k^2_Mh'(M^2)(M^2-s)\right\}-iM\Gamma\left(\frac{k_s}{k_M}\right)^3\frac{M}{\sqrt{s}}}\label{eqn:GSBW}
\end{equation}
 where $M$ is the mass of the $\rho$ and $\Gamma$ is its width.
 The function $k(s)$ is defined to be 
\begin{subequations}
\begin{eqnarray}
k(s) &=& \sqrt{\frac14s-m^2_\pi} \ \ \ \ \ \ \ s\geq4m_\pi^2 \\
     &=& i\sqrt{m^2_\pi-\frac14s}\ \ \ \ \ \, \,  s<4m_\pi^2,
\end{eqnarray}
\end{subequations}
where we have used $k_M=k(M^2)$ and $k_s=k(s)$ in the definition of the Breit-Wigner
distribution.
The function $h(s)$ is taken to be
\begin{subequations}
\begin{eqnarray}
h(s)&=&\frac2\pi\frac{k}{\sqrt{s}}\ln\left(\frac{\sqrt{s}+2k}{2m_\pi}\right) \ \ \ \ \ \ \ \ \ \ \ s\geq4m_\pi^2 \\
&=&i\frac2\pi\frac{k}{\sqrt{s}}\cot^{-1}\sqrt{\frac{s}{4m_\pi^2-s}}\ \ \ \ \ \ \ \,  s<4m_\pi^2,
\end{eqnarray}
\end{subequations}
and $h'(s) = \frac{{\rm d}h}{{\rm d}s}$.
The constant $d$
\begin{equation}
d = \frac3\pi\frac{m^2_\pi}{k^2_M}\ln\left(\frac{M+2k_M}{2m_\pi}\right)
+\frac{M}{2\pi k_M}-\frac{m^2_\pi M}{\pi k_M^3}
\end{equation} 
is defined to give ${\rm BW}(0)=1$.

%\bibliography{Herwig++}

\begin{thebibliography}{10}

\bibitem{Richardson:2001df}
P. Richardson,
\newblock JHEP 11 (2001) 029, hep-ph/0110108.
%%CITATION = HEP-PH 0110108;%%

\bibitem{Catani:2001cc}
S. Catani et~al.,
\newblock JHEP 11 (2001) 063, hep-ph/0109231.
%%CITATION = HEP-PH 0109231;%%

\bibitem{Frixione:2002ik}
S. Frixione and B.R. Webber,
\newblock JHEP 06 (2002) 029, hep-ph/0204244.
%%CITATION = HEP-PH 0204244;%%

\bibitem{Gieseke:2003rz}
S. Gieseke, P. Stephens and B. Webber,
\newblock JHEP 12 (2003) 045, hep-ph/0310083.
%%CITATION = HEP-PH 0310083;%%

\bibitem{Nason:2004rx}
P. Nason,
\newblock JHEP 11 (2004) 040, hep-ph/0409146.
%%CITATION = HEP-PH/0409146;%%

\bibitem{Bertini:2000uh}
M. Bertini, L. L{\"{o}}nnblad and T. Sj{\"{o}}strand,
\newblock Comput. Phys. Commun. 134 (2001) 365, hep-ph/0006152.
%%CITATION = HEP-PH 0006152;%%
%\bibitem{Lonnblad:2003wz}
L. L{\"{o}}nnblad,
\newblock Nucl. Instrum. Meth. A502 (2003) 549.
%%CITATION = NUIMA,A502,549;%%

\bibitem{Gieseke:2003hm}
S. Gieseke et~al.,
\newblock JHEP 02 (2004) 005, hep-ph/0311208.
%%CITATION = HEP-PH 0311208;%%
%\bibitem{Gieseke:2006ga}
S. Gieseke et~al.,
\newblock (2006), hep-ph/0609306.
%%CITATION = HEP-PH 0609306;%%
%\bibitem{Gigg:2007vr}
M. Gigg and P. Richardson,
\newblock (2007), arXiv:0706.2921 [hep-ph].
%%CITATION = ARXIV:0706.2921;%%

\bibitem{Gleisberg:2003xi}
T. Gleisberg et~al.,
\newblock JHEP 02 (2004) 056, hep-ph/0311263.
%%CITATION = HEP-PH 0311263;%%

\bibitem{Jadach:1993hs}
S. Jadach et~al.,
\newblock Comput. Phys. Commun. 76 (1993) 361.
%%CITATION = CPHCB,76,361;%%

\bibitem{Golonka:2003xt}
P. Golonka et~al.,
\newblock (2003), hep-ph/0312240.
%%CITATION = HEP-PH 0312240;%%

\bibitem{Knowles:1988vs}
I.G. Knowles,
\newblock Nucl. Phys. B310 (1988) 571.
%%CITATION = NUPHA,B310,571;%%

\bibitem{Knowles:1988hu}
I.G. Knowles,
\newblock Comput. Phys. Commun. 58 (1990) 271.
%%CITATION = CPHCB,58,271;%%

\bibitem{Collins:1987cp}
J.C. Collins,
\newblock Nucl. Phys. B304 (1988) 794.
%%CITATION = NUPHA,B304,794;%%

\bibitem{Was:2002gv}
Z. W\c{a}s and M. Worek,
\newblock Acta Phys. Polon. B33 (2002) 1875, hep-ph/0202007.
%%CITATION = HEP-PH 0202007;%%

\bibitem{Gigg:2007cr}
M. Gigg and P. Richardson,
\newblock (2007), hep-ph/0703199.
%%CITATION = HEP-PH/0703199;%%

\bibitem{Desch:2003mw}
K. Desch, Z. W\c{a}s and M. Worek,
\newblock Eur. Phys. J. C29 (2003) 491, hep-ph/0302046.
%%CITATION = HEP-PH 0302046;%%

\bibitem{Hamilton:2006xz}
K. Hamilton and P. Richardson,
\newblock JHEP 07 (2006) 010, hep-ph/0603034.
%%CITATION = HEP-PH/0603034;%%

\bibitem{Yennie:1961ad}
D.R. Yennie, S.C. Frautschi and H. Suura,
\newblock Ann. Phys. 13 (1961) 379.
%%CITATION = APNYA,13,379;%%

\bibitem{Anderson:1999ui}
CLEO, S. Anderson et~al.,
\newblock Phys. Rev. D61 (2000) 112002, hep-ex/9910046.
%%CITATION = HEP-EX 9910046;%%

\bibitem{Abe:2005ur}
Belle, K. Abe et~al.,
\newblock (2005), hep-ex/0512071.
%%CITATION = HEP-EX 0512071;%%

\bibitem{Schael:2005am}
ALEPH, S. Schael et~al.,
\newblock Phys. Rept. 421 (2005) 191, hep-ex/0506072.
%%CITATION = HEP-EX 0506072;%%

\bibitem{Kuhn:1990ad}
J.H. K{\"{u}}hn and A. Santamaria,
\newblock Z. Phys. C48 (1990) 445.
%%CITATION = ZEPYA,C48,445;%%

\bibitem{Gounaris:1968mw}
G.J. Gounaris and J.J. Sakurai,
\newblock Phys. Rev. Lett. 21 (1968) 244.
%%CITATION = PRLTA,21,244;%%

\bibitem{Epifanov:2007rf}
Belle, D. Epifanov et~al.,
\newblock (2007), arXiv:0706.2231 [hep-ex].
%%CITATION = ARXIV:0706.2231;%%

\bibitem{Finkemeier:1995sr}
M. Finkemeier and E. Mirkes,
\newblock Z. Phys. C69 (1996) 243, hep-ph/9503474.
%%CITATION = HEP-PH 9503474;%%

\bibitem{Pich:1987qq}
A. Pich,
\newblock Phys. Lett. B196 (1987) 561.
%%CITATION = PHLTA,B196,561;%%

\bibitem{Barate:1999hj}
{ALEPH}, R. Barate et~al.,
\newblock Eur. Phys. J. C11 (1999) 599, hep-ex/9903015.
%%CITATION = HEP-EX 9903015;%%

\bibitem{Lyon:2004sp}
A.J. Lyon,
\newblock SLAC-R-785.

\bibitem{Finkemeier:1996dh}
M. Finkemeier and E. Mirkes,
\newblock Z. Phys. C72 (1996) 619, hep-ph/9601275.
%%CITATION = HEP-PH 9601275;%%

\bibitem{Decker:1992kj}
R. Decker et~al.,
\newblock Z. Phys. C58 (1993) 445.
%%CITATION = ZEPYA,C58,445;%%

\bibitem{Asner:1999kj}
CLEO, D.M. Asner et~al.,
\newblock Phys. Rev. D61 (2000) 012002, hep-ex/9902022.
%%CITATION = HEP-EX 9902022;%%

\bibitem{Bondar:2002mw}
A.E. Bondar et~al.,
\newblock Comput. Phys. Commun. 146 (2002) 139, hep-ph/0201149.
%%CITATION = HEP-PH 0201149;%%

\bibitem{Asner:2000nx}
CLEO, D.M. Asner et~al.,
\newblock Phys. Rev. D62 (2000) 072006, hep-ex/0004002.
%%CITATION = HEP-EX/0004002;%%

\bibitem{Yao:2006px}
Particle Data Group, W.M. Yao et~al.,
\newblock J. Phys. G33 (2006) 1.
%%CITATION = JPHGB,G33,1;%%

\bibitem{Kuhn:2006nw}
J.H. K{\"{u}}hn and Z. W\c{a}s,
\newblock (2006), hep-ph/0602162.
%%CITATION = HEP-PH 0602162;%%

\bibitem{Mesondecays}
D. Grellscheid, K. Hamilton and P. Richardson,
\newblock {S}imulation of {M}eson {D}ecays in the {H}erwig++ {E}vent
  {G}enerator,
\newblock in preparation.

\bibitem{Chen:1996ap}
C.H. Chen, M. Drees and J.F. Gunion,
\newblock Phys. Rev. D55 (1997) 330, hep-ph/9607421.
%%CITATION = HEP-PH/9607421;%%
%\bibitem{Chen:1999yf}
C.H. Chen, M. Drees and J.F. Gunion,
\newblock (1999), hep-ph/9902309.
%%CITATION = HEP-PH/9902309;%%

\bibitem{Barr:2002ex}
A.J. Barr et~al.,
\newblock JHEP 03 (2003) 045, hep-ph/0208214.
%%CITATION = HEP-PH/0208214;%%

\bibitem{Allanach:2002nj}
B.C. Allanach et~al.,
\newblock (2002), hep-ph/0202233.
%%CITATION = HEP-PH/0202233;%%

\bibitem{Allanach:2001kg}
B.C. Allanach,
\newblock Comput. Phys. Commun. 143 (2002) 305, hep-ph/0104145.
%%CITATION = HEP-PH/0104145;%%

\bibitem{Baer:1999sp}
H. Baer et~al.,
\newblock (1999), hep-ph/0001086.
%%CITATION = HEP-PH/0001086;%%

\end{thebibliography}

\end{document}